\documentclass{elsart}
\usepackage{epsfig}
\usepackage{amssymb}
\usepackage{lscape}
\usepackage{amsmath}
\usepackage{bm}
\usepackage{graphicx}
\usepackage{graphics}
\usepackage{color}
\usepackage{pstricks}
\usepackage{cite}
\usepackage[leftbars]{changebar}
\newcommand{\bOmega}{\bm \Omega}
\newcommand{\bomega}{\bm \omega}

\newcommand{\bnabla}{{\bm \nabla}}

\newcommand{\dd}{\text{d}}
\newcommand{\bM}{{\mathbf M}}
\newcommand{\bC}{{\mathbf C}}

\newcommand{\bD}{{\mathbf D}}

\newcommand{\cH}{{\mathcal H}}

\newcommand{\cV}{{\mathcal V}}

\newcommand{\ddp}[2]{\frac{\partial #1}{\partial #2}}

\newcommand{\eg}{\text{e.g.}}
\newcommand{\ie}{\text{i.e.}}

\def \AA {\mathbf A}

\newcommand{\CC}{{\mathbf C}}

\newcommand{\uu}{{\mathbf u}}

\newcommand{\Rey}{\textrm{Re}}

\journal{Computers and Fluids}
\begin{document}

\begin{frontmatter}
\title{Transitional cylindrical swirling flow in presence of a flat free surface}
\author[EPFL,MIT]{Roland Bouffanais\corauthref{cor}\thanksref{FNRS}},
\thanks[FNRS]{Supported by a Swiss National Science Foundation Grant No. 200020--101707}
\corauth[cor]{Corresponding author.}
\ead{bouffana@mit.edu}
\author[MONASH]{David Lo Jacono}
\ead{david.lojacono@eng.monash.edu.au} \address[EPFL]{Laboratory of
  Computational Engineering,\\ \'Ecole Polytechnique F\'ed\'erale de
  Lausanne,\\ STI--IGM--LIN, Station 9,\\ CH--1015 Lausanne,
  Switzerland}
\address[MIT]{Present address: Massachusetts Institute of
  Technology,\\ Department of Mechanical Engineering,\\ 77
  Massachusetts Avenue, Bldg 5--326,\\ Cambridge, MA 02139}
\address[MONASH]{Department of Mechanical and Aerospace Engineering,\\
  Monash University,\\ Victoria, 3800 Melbourne, Australia}
\begin{abstract}
  This article is devoted to the study of an incompressible viscous
  flow of a fluid partly enclosed in a cylindrical container with an
  open top surface and driven by the constant rotation of the bottom
  wall. Such type of flows belongs to a group of recirculating
  lid-driven cavity flows with geometrical axisymmetry and of the
  prescribed boundary conditions of Dirichlet type---no-slip on the
  cavity walls. The top surface of the cylindrical cavity is left open
  with an imposed stress-free boundary condition, while a no-slip
  condition with a prescribed rotational velocity is imposed on the
  bottom wall. The Reynolds regime corresponds to transitional flows
  with some incursions in the fully laminar regime.
  The approach taken here revealed new flow states that were
  investigated based on a fully three-dimensional solution of the
  Navier--Stokes equations for the free-surface cylindrical swirling
  flow, without resorting to any symmetry property unlike all other
  results available in the literature. Theses solutions are obtained through direct numerical simulations based on a Legendre spectral element method.
\end{abstract}
\end{frontmatter}
%
%
\section{Introduction}\label{sec:introduction}
Besides the differences in terms of geometry, the lid-driven cubical
cavity flow \cite{shankar00:_fluid_mechan_driven_cavit} and the
cylindrical swirling flow investigated in this article, present
similar features typical of shear-driven recirculating flows such as
intense wall-jets, shear layers in the vicinity of the driven wall,
and secondary recirculating flows, all of which are very dependent on
the flow parameters. Nevertheless, the geometry---cubical on one hand
and cylindrical on the other hand---dramatically influences the nature
and structure of these secondary flows: corner eddies for the cubical
cavity and recirculation bubbles or vortex breakdown in the
cylindrical case.
%
%
\subsection{General considerations} \label{sec:general-considerations}
Following the pioneering work of Bogatyrev \& Gorin
\cite{bogatyrev78:_end} and Koseff \& Street
\cite{koseff84,koseff84:_visual}, it was shown that contrary to
intuition, the lid-driven \textit{cubical} cavity flow is essentially
three-dimensional, even when considering large aspect ratio. It is
only recently that the three-dimensionality of the lid-driven
\textit{cylindrical} cavity flow was confirmed numerically by
Blackburn \& Lopez \cite{blackburn00:_symmet,blackburn02:_modul} after
it was suggested but not fully proved experimentally by S{\o}rensen
\cite{soerensen92:_visual}, Spohn \etal\ \cite{spohn98:_exper},
Sotiropoulos \& Ventikos \cite{sotiropoulos98:_trans}, and Pereira \&
Sousa \cite{pereira99:_confin}. In 2001, Sotiropoulos \& Ventikos
\cite{sotiropoulos01} gave full experimental evidence of the
three-dimensional character of the flow with the onset of
non-axisymmetric modes. The three-dimensional nature of these driven
cavity flows therefore appears as a general characteristic of internal
recirculating shear-driven flows.

In the sequel, we will only consider the cylindrical lid-driven cavity
flow also referred to as ``swirling'' flow without any additional
precision. The first experiments by Vogel
\cite{vogel68:_exper_ergeb_str_geh} and later Ronnenberg
\cite{ronnenberg77:_kompon_laser_vergeic_unter_drehs_r} showed that
Ekman suction and pumping, induced by the Ekman layers on the rotating
and stationary disks, lead to the formation of a concentrated vortex
core along the axis in the closed cavity case. The two dimensionless
numbers characterizing this flow are the height-to-radius aspect ratio
$\Lambda=H/R$ and the Reynolds number $\Rey=R^2\Omega/\nu$,
where $H$ and $R$ are the height and radius of the cylinder
respectively, $\Omega$ the constant angular velocity of the bottom
end-wall, and $\nu$ the kinematic viscosity of the Newtonian
fluid. For specific values of the aspect ratio $\Lambda$, and above a
critical Reynolds number, the vortex core breaks down in the form of
one or more recirculation bubbles which are on-axis in the closed
cavity case and on- or off-axis in the open cavity case. Owing to the
enormous extent of work in the area of vortex breakdown (VB), (see
reviews by Hall \cite{hall72:_vortex}, Leibovich
\cite{leibovich78:_struc}, Shtern \& Hussain \cite{shtern99:_collap},
Kerswell \cite{kerswell02:_ellip}, and Arndt \cite{arndt02:_cavit}), we
will only briefly recall the central features of VB. As defined by
Leibovich in its review on the structure of vortex breakdown
\cite{leibovich78:_struc}, the term ``vortex breakdown'' refers to a
disturbance characterized by the formation of an internal stagnation
point on the vortex axis, followed by reversed flow in a region of
limited axial extent. Two forms of vortex breakdown predominate, one
called ``near-axisymmetric'' (sometimes ``axisymmetric'' or
``bubble-like''), and the other called ``spiral''.

The practical importance of vortex breakdown lies mainly in the field
of aeronautics, where they can be observed over wings---mainly delta
wings---with highly swept leading edges when the angle of incidence
exceeds a critical value. Vortex breakdown can be a limiting factor on
the operating altitude of slender-winged flying vehicles. Moreover, the occurrence of VB in
the wake of a large aircraft is relevant to the safety of flight in
dense air-traffic, which is becoming more and more frequent with the
constant increase in air-traffic over the years. VB is also important
in other fields for example it has been observed in the swirling flows
through nozzles and diffusers \cite{faler77:_disrup}, and in the field
of reactive flows, in combustion chambers. Besides the tremendous
importance of VB in engineering applications, it is also a
prototypical phenomenon allowing to elucidate the fundamental aspects
of the bubble mode.
%
%
\subsection{The lid-driven cylindrical cavity flow}\label{sec:closed-cavity}
The first comprehensive experimental study of the closed cylindrical
container case was undertaken by Escudier \cite{escudier84:_obser},
and Escudier \& Keller \cite{escudier85:_recir}, who extended the
earlier results of Vogel \cite{vogel68:_exper_ergeb_str_geh} and
Ronnenberg \cite{ronnenberg77:_kompon_laser_vergeic_unter_drehs_r} to
obtain the first map of VB transitions with respect to the aspect
ratio $\Lambda$ and the Reynolds number. Escudier
\cite{escudier84:_obser} revealed flow states with one, two or even
three successive breakdowns, as well as a transition from steadiness
to unsteadiness. S{\o}rensen \cite{soerensen92:_visual} extended to a
broader range of Reynolds number in the same experiment as Escudier
\cite{escudier84:_obser} for the closed container, and inferred that
above a critical Reynolds number in the unsteady flow regime, the
meridional flow becomes highly asymmetric. The first experimental
study of the open cylindrical container case with a free surface on
the top, was undertaken by Spohn \etal\ \cite{spohn93:_obser}, who
highlighted the significant change in the structure, the occurrence
and the location of the breakdown bubbles. 
In the steady closed
cylinder case, Hourigan \etal\ \cite{hourigan95:_spiral} investigated
the asymmetric spiraling effects along the cylinder axis prior to the
first vortex breakdown. They argued that the observed asymmetry was
purely an experimental artifact and not an evidence of the
three-dimensional nature of the flow. 
Spohn \etal\ \cite{spohn98:_exper} were the first to investigate
experimentally the origin of possible asymmetric features of the
instabilities at their onset. The steady breakdown bubbles reported by
Spohn \etal\ \cite{spohn98:_exper} showcase asymmetric features
comparable to earlier measurements, and also to unsteady bubbles
observed in circular diffusers by Faler \& Leibovich
\cite{faler77:_disrup}. As a matter of fact, the work of Spohn \etal\
\cite{spohn98:_exper} is really a pioneering work in the acceptance of
the axisymmetry breaking, amongst fluid experimentalists, see Br{\o}ns \etal~\cite{broens08:_dye_visual_near_three_dimen_stagn_point}. It is
noteworthy at this point, that the complex physics associated with
these intricate phenomena occurring in closed/open rotating
cylindrical container is still not clearly understood.

Like for the lid-driven cubical cavity flow, and in relation with the
simple geometry of the flow, the rotating cylindrical cavity flow has
been extensively studied using direct numerical simulations. It is
important to note that since the early seventies, the method of choice
has consisted in solving the streamfunction-vorticity formulation of
the axisymmetric incompressible Navier--Stokes equations. Without
being exhaustive, the following list of references gives an overview
of the numerical simulation of the closed lid-driven cylindrical flow
over three decades: Pao \cite{pao70}, Lugt \& Haussling
\cite{lugt73:_devel,lugt82:_axisy}, Dijkstra \& van Heijst
\cite{dijkstra83}, Lugt \& Abboud \cite{lugt87:_axisy}, Neitzel
\cite{neitzel88:_streak}, Daube \& S{\o}rensen \cite{daube89:_numer},
Lopez \cite{lopez90:_axisy}, Brown \& Lopez \cite{brown90:_axisy},
Lopez \& Perry \cite{lopez92:_axisy}, S{\o}rensen \& Christensen
\cite{soerensen95:_direc}, Watson \& Neitzel \cite{watson96:_numer},
Gelfgat \etal\ \cite{gelfgat96:_stead,gelfgat96:_stabil}, Tsitverblit
\& Kit \cite{tsitverblit98}, and Br{\o}ns \etal\
\cite{brons99:_stream}. All these works were able to reproduce with a
reasonable accuracy, the basic features observed experimentally and
reported earlier including the size, shape and number of recirculation
bubbles. The onset of vortex breakdown was to some extent captured by
several of these numerical simulations. Lopez \cite{lopez90:_axisy},
and Brown \& Lopez \cite{brown90:_axisy} suggested a physical
mechanism for the intricate phenomena observed. They prove the
existence of a standing centrifugal wave, whose amplitude increases
with the Reynolds number and which can create a stagnation point on
the cylinder axis, initiating the breakdown process. It is worth
recalling that the streamfunction-vorticity formulation is adequate
and appropriate only for the study of flow dynamics preserving the
property of axisymmetry. 
At the inception of any instability breaking the axisymmetry of the
flow, a three-dimensional solution of the Navier--Stokes equations is
required, thereby increasing considerably the complexity of the
task. The last remark justifies the observed changes in terms of
numerical modeling of Lopez' group and S{\o}rensen's group, to allow
them to investigate axisymmetry breaking in the closed cylinder case
\cite{blackburn00:_symmet,blackburn02:_modul,blackburn03,
  bisgaard06:_vortex,shen06:_numer}. Therefore, three-dimensional flow
structures have started being simulated more recently, see Gelfgat
\etal\ \cite{gelfgat01:_three}, Sotiropoulos \& Ventikos
\cite{sotiropoulos01}, Sotiropoulos \etal\
\cite{sotiropoulos01:_chaot}, Marques \& Lopez
\cite{marques01:_preces}, Blackburn \& Lopez
\cite{blackburn00:_symmet,blackburn02:_modul}, Serre \& Bontoux
\cite{serre02:_vortex}, Blackburn \cite{blackburn02:_three}, and Lopez
\cite{lopez06:_rotat}.

Apart from the canonical case with a single driving lid in rotation at
a constant angular velocity, different variations of the problem have
been extensively studied in the past years: \eg\ cylinder with co- and
counter-rotating end-covers by Br{\o}ns \etal\ \cite{brons99:_stream},
steady axisymmetric flow in an open cylindrical container with a
partially rotating bottom wall by Piva \& Meiburg
\cite{piva05:_stead}, vortex scenario and bubble generation in a
cylindrical cavity with rotating top and bottom by Okulov \etal\
\cite{okulov05:_vortex}. Mullin
\etal\ \cite{mullin00} also included a rod at the axis to control the
breakdown, and Pereira \& Sousa \cite{pereira99:_confin} significantly
changed the configuration by replacing the flat rotating bottom cover
by a cone. As noted by Br{\o}ns \etal\ \cite{brons01:_topol}, all
these studies show a large set of flow structures which are quite
sensitive to variations of external parameters. Mununga \etal\ \cite{mununga04:_confin_flow_vortex_break_contr} and Lo~Jacono \etal\ \cite{lo08:_contr_of_vortex_break_in} investigated different strategies for the control of vortex breakdown.

\subsection{Open swirling flow}\label{sec:open-swirling-flow}
The focus in the present article is on the canonical problem of a
cylinder with a rotating bottom end-wall but replacing the stationary
solid top end-wall by a free surface. The flow associated with this
problem was first studied experimentally by Spohn \etal\
\cite{spohn93:_obser,spohn98:_exper}. They observed the influence of
the top free surface---assuredly clean of surfactants---on the onset,
structure, nature and number of recirculating bubbles. Their central
observations are that breakdown bubbles still appear, but are off-axis
and may be attached to the free surface, depending on the aspect ratio
$\Lambda$ and the Reynolds number. Of course, such structures could
not be observed in the closed case because of the no-slip condition
imposed on the top wall. All the past simulations of free-surface
swirling flows rely on the central assumptions that the free surface
is flat and clean, which means that the Froude number is very small
and that surface tension effects are negligible. With these assumptions, the flow is identical to the flow in the lower half part of a cylinder with two solid covers in co-rotation, \ie\
rotating at the same angular velocity. Br{\o}ns \etal\ reported a wide
range of topologies of vortex breakdown bubbles in a bottom-driven
cylinder with a free surface.
Valentine \& Jahnke \cite{valentine94:_flows}, observed in their
simulations the existence of one or two toroidal-like types of
recirculation bubble having their stagnation lines attached to the
free surface, depending on the value of the Reynolds number. Their
study was complemented by the work of Lopez \cite{lopez95:_unstead}
for oscillating unsteady flows. Information relevant to the present
problem with a free surface all indicate consistent flow behavior at
small aspect ratio \ie, $0.5\leq \Lambda \leq 1.0$ in that stagnation
occurs off-axis and associated secondary flow creates a toroidal
recirculation bubble. Steady free-surface flows have been computed by
Iwatsu \cite{iwatsu05:_numer,iwatsu04:_analy} providing flow state
classifications with new flow patterns not revealed in the previous
studies.
%
%
\subsection{Motivations and objectives}\label{sec:Motivations_objectives}
The present study is motivated by several factors. Firstly, compared
to the closed cylinder case, only some limited aspects of the open
swirling flow have been investigated so far. The study of this
intricate problem is relatively new and consequently the body of
knowledge in some $(\Lambda , \Rey)$-parameter regions appears
fairly limited. Secondly, most of the past studies involving numerical
simulations of this free-surface swirling flow, used axisymmetric
streamfunction-vorticity formulations: Br{\o}ns \etal\
\cite{brons01:_topol}, Iwatsu \cite{iwatsu05:_numer,iwatsu04:_analy},
and Piva \& Meiburg \cite{piva05:_stead}. To our knowledge, the only
fully three-dimensional numerical simulations is due to Lopez \etal\
\cite{lopez04:_symmet}, who investigated mainly symmetry breaking
issues.

In the present article, new flow states are investigated based on a
fully three-dimensional solution of the Navier--Stokes equations
without the need to resort to symmetry properties by doubling the
computational domain and enforcing co-rotation of both end-walls. To
our knowledge, the present study provides the most general available
results for this flat-free-surface problem. Both, steady and unsteady
flows are considered for different sets of governing parameters
$(\Lambda, \Rey)$. A Legendre spectral element method is used
to provide an accurate solution of the governing equations, while the
stress-free boundary condition is naturally enforced into the weak
formulation of the problem.

The mathematical model and the problem formulation are detailed in
Sec.~\ref{sec:formulation}, while the original computational approach
of this study is presented in
Sec.~\ref{sec:computational}. Subsequently, Sec.~\ref{sec:numerical}
contains all the numerical results corresponding to various physical
situations and flow states. Finally, the article ends with
Sec.~\ref{sec:conclusions} providing summary and conclusions on the
present work.
%
\section{Mathematical model and problem formulation} \label{sec:formulation}
\subsection{Mathematical description of the problem}
\begin{figure}[htbp]
  \centering
  \includegraphics[width=0.55\textwidth]{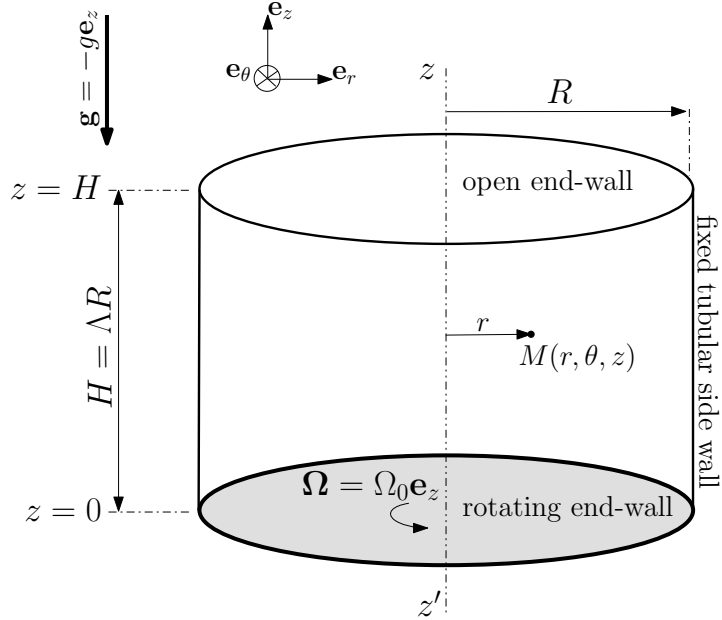}
  \caption{Schematic of the geometry studied with the set of
    coordinates employed.}\label{fig:sketch}
\end{figure}
The fluid enclosed in the cylindrical cavity is assumed to be
incompressible, Newtonian with uniform density and temperature. The
flow is governed by the Navier--Stokes equations
\begin{align}
  \frac{\partial u_i}{\partial t} + u_j\frac{\partial u_i}{\partial
    x_j} & =
  \frac{\partial \sigma_{ij}^*}{\partial x_j} + g_i, \label{eq:NS-1}\\
  \frac{\partial u_j}{\partial x_j} & = 0, \label{eq:NS-2}
\end{align}
where $\sigma_{ij}^* = -p \delta_{ij}+2 \nu D_{ij}$ is the reduced
Cauchy stress tensor of the fluid, $p$ the static or reduced pressure,
$D_{ij}$ the rate-of-deformation tensor, $\nu$ the assuredly constant
and uniform kinematic viscosity, and $g_i$ the components of the
acceleration of gravity ($g_1=g_2=0$ and $g_3=-g$). Inside the fluid
domain denoted by $\cV$, no-slip boundary conditions are imposed on
all cavity walls: the tubular side-wall, the bottom end-wall in steady
rotation, see Fig.~\ref{fig:sketch}. The mathematical expression of
the no-slip condition on the tubular side-wall simply reads
\begin{equation}\label{eq:no-slip-tubular}
  u(r=R,\theta,z,t) = v(r=R,\theta,z,t) = w(r=R,\theta,z,t) = 0, 
  \quad 0 \leq \theta \leq 2\pi, \ 0\leq z \leq H.
\end{equation}
The flow is driven by imposing a prescribed angular velocity
distribution of the bottom end-wall, which transfers its kinetic
energy to the fluid above. The details regarding the imposition of
this Dirichlet boundary condition for the velocity field at the lid
are discussed in the next section~\ref{sec:lid}. The top surface is
left open and is modeled as a flat, fixed and clean free surface. The
details regarding the imposition of this stress-free condition on the
free surface are discussed in Sec.~\ref{sec:free-surface}. As
mentioned in the Introduction, Sec.~\ref{sec:general-considerations},
two parameters that determine completely the flow state are the
height-to-radius aspect ratio $\Lambda=H/R$ and the Reynolds number
$\Rey=R^2\Omega_0/\nu$, based on the maximal prescribed angular
velocity $\Omega_0$ of the bottom end-wall.

In the sequel, the length, time, velocity, vorticity, helicity,
streamfunction, (reduced) pressure and kinetic energy, and enstrophy
are non-dimensionalized with respect to the reference scales $R$,
$\Omega_0^{-1}$, $R\Omega_0$, $\Omega_0$, $R\Omega_0^2$,
$R^3\Omega_0$, $R^2\Omega_0^2$, $\Omega_0^2$ respectively.

\subsection{Angular velocity distribution} \label{sec:lid} 
As already mentioned in Bouffanais \etal\ \cite{bouffanais06:_large}
for the study of the lid-driven cubical cavity flow, imposing a given
angular velocity distribution on the bottom end-wall of a cavity is
not an easy task numerically. Indeed, imposing a constant angular
velocity profile leads to a singularity (discontinuous behavior in the
velocity boundary conditions) at the circular edge between the bottom
end-wall and the tubular side-wall, see Fig.~\ref{fig:sketch}. Without
adequate treatment, this discontinuous behavior will undermine the
convergence and the accuracy of any numerical method in the vicinity
of the lid. The same remedy as in the lid-driven cubical cavity
problem in \cite{bouffanais06:_large} is used here for the same
reasons and with analogous justifications. A regularized angular
velocity profile is employed by prescribing the following high-order
polynomial expansion which vanishes along its first derivatives on the
circular edge
\begin{equation} 
  \label{eq:angular-velocity-distribution}
  \bOmega (r,\theta,z=0,t) = \Omega_0 \left[ 1-\left(\frac{r}{R}
    \right)^{16} \right]^2 \, \mathbf{e}_z,
\end{equation}
which leads to the following expressions in Cartesian coordinates of
the components of the prescribed velocity field on the bottom end-wall
\begin{align} 
  u(x,y,z=0,t) = u_x(x,y,z=0,t) & = -y \Omega_0 \left[ 1-\left(
      \sqrt{x^2+y^2}/R\right)^{16} \right]^2,\label{eq:velocity-distribution-1}\\
  v(x,y,z=0,t) = u_y(x,y,z=0,t) & = +x \Omega_0 \left[ 1-\left(
      \sqrt{x^2+y^2}/R\right)^{16} \right]^2,\label{eq:velocity-distribution-2}\\
  w(x,y,z=0,t) = u_z(x,y,z=0,t) & =
  0,\label{eq:velocity-distribution-3}
\end{align}
where $x=(r,\theta=0)$ and $y=(r,\theta=\pi/2)$. This profile flattens very quickly near the circular edge
$(r/R=1,\theta,z/H=0)$ while away from it, it grows rapidly to a
constant value $\Omega_0$ of the angular velocity over a short
distance. The highest polynomial degree of this distribution is
32. Such high-order polynomial expansions lead to steep velocity
gradients in the vicinity of the circular edge of the bottom
end-wall. The grid refinement, in terms of spectral element
distribution near the disk will be presented in greater details in
Sec.~\ref{sec:computational}. One of the constraint in the grid design
is to ensure the proper resolution of the lid velocity distribution by
the spectral element decomposition.
%
%
\subsection{Free-surface modeling} \label{sec:free-surface}
The analysis of a two-phase flow is based on the coupled hydrodynamics
interactions between adjacent layers over a broad range of space and
time scales. This analysis can be significantly simplified if the
dynamics of the interface is almost entirely dependent---from the
hydrodynamics and physico-chemistry viewpoints---on one phase, \eg\ a
liquid phase, and almost independent of the dynamics of the second
phase, \eg\ a gas phase. Based on this hypothesis, the surface is said
to be \emph{free}. Consequently, the two fluid phases can only exert
constant normal stresses. Sarpkaya in his review entitled ``Vorticity,
free surface and surfactants'' \cite{sarpkaya96:_vortic} gives a clear
characterization and definition of a free surface: ``Although, the
exterior of a free surface is free from externally imposed shear, the
interior is not necessarily free from the shear generated
internally. In fact, surface deformations and contaminants give rise
to surface-gradients and tangential stresses in the internal side of
the bounding interface. From a mathematical viewpoint, a free surface
means that the density and the viscosity of the upper fluid are zero
and that the existence of a continuum above the interface is
inconsequential. From a practical point of view, the free surface
means that the dynamics of the continuum above the interface has
negligible influence on the lower phase, \ie\ a free surface is a
simplifying approximation for an \emph{almost free surface}.''

In the present study, the modeling of the interface between the fluids
in the cylindrical cavity as a free surface is supplemented by an
additional simplifying approximation: the free surface is supposed to
remain flat and fixed all along the dynamic range of investigation. In
general, the dynamics of the free surface depends on the
non-dimensional Froude number defined here as
\begin{equation} \label{eq:froude-number}
  \textrm{Fr}=\frac{R^2\Omega_0^2}{gH},
\end{equation} 
which measures the relative importance of the inertial effects
compared to the stabilizing gravitational effects. Therefore, assuming
a flat free surface corresponds mathematically to a Froude number
identically zero. As a consequence and in consistency with the latter
assumption, the axial component of the velocity $w=u_z$ needs to
vanish at the free surface $z/H=1$
\begin{equation}\label{eq:w}
  w(x,y,z=H,t) = 0,\qquad x^2+y^2 \leq R^2,
\end{equation}
thereby expressing the kinematic boundary condition at the free
surface. This latter condition on the axial velocity $w$ is to be
supplemented with the stress-free condition at the free surface
\begin{equation}\label{eq:stress-free}
  \sigma^*_{ij}\hat{n}_j = -p \delta_{ij}\hat{n}_j + 2\nu D_{ij}\hat{n}_j = 0,
\end{equation}
where $\hat{\mathbf{n}}$ is the local outward unit vector normal at
the free surface, which in the present particular situation is the
unit normal vector $\mathbf{e}_z$ in the $z$-direction. Consequently,
the stress-free condition \eqref{eq:stress-free} reduces to
$\sigma^*_{i3}=\sigma^*_{iz}=0$, $i=1,2,3$ at $z=H$, and is explicitly
stated as
\begin{align}
  \sigma^*_{13} = 2 \nu D_{13}   &= 0 ,\\
  \sigma^*_{23} = 2 \nu D_{23}   &= 0 ,\\
  \sigma^*_{33} = -p+2\nu D_{33} &= 0 ,
\end{align}
which under the zero-deformation condition \eqref{eq:w}, simplifies to
\begin{align}
  \left(\frac{\partial u}{\partial z} + \frac{\partial w}{\partial x}
  \right) = \frac{\partial u}{\partial z}&=0,\label{eq:stress-free-1}\\
  \left(\frac{\partial v}{\partial z} + \frac{\partial w}{\partial y}
  \right) = \frac{\partial v}{\partial z}&=0,\label{eq:stress-free-2}\\
  -p+2\nu \frac{\partial w}{\partial z} &=0.\label{eq:stress-free-3}
\end{align}
%
%
\section{Computational approach}\label{sec:computational}
\subsection{Space discretization} \label{sec:space-discretization} 
\begin{figure}[htbp]
\centering
\includegraphics[width=0.32\textwidth]{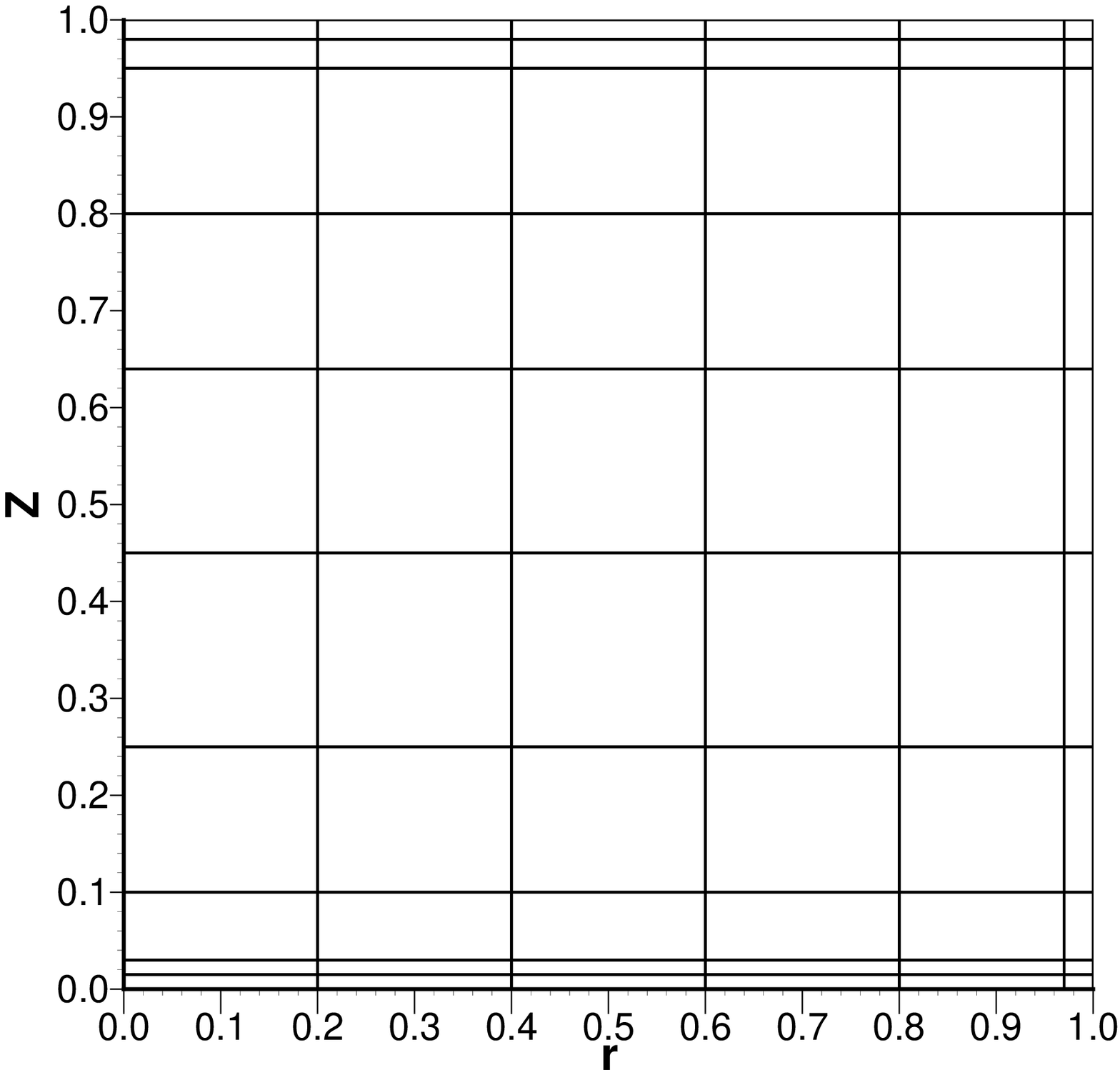}
\includegraphics[width=0.32\textwidth]{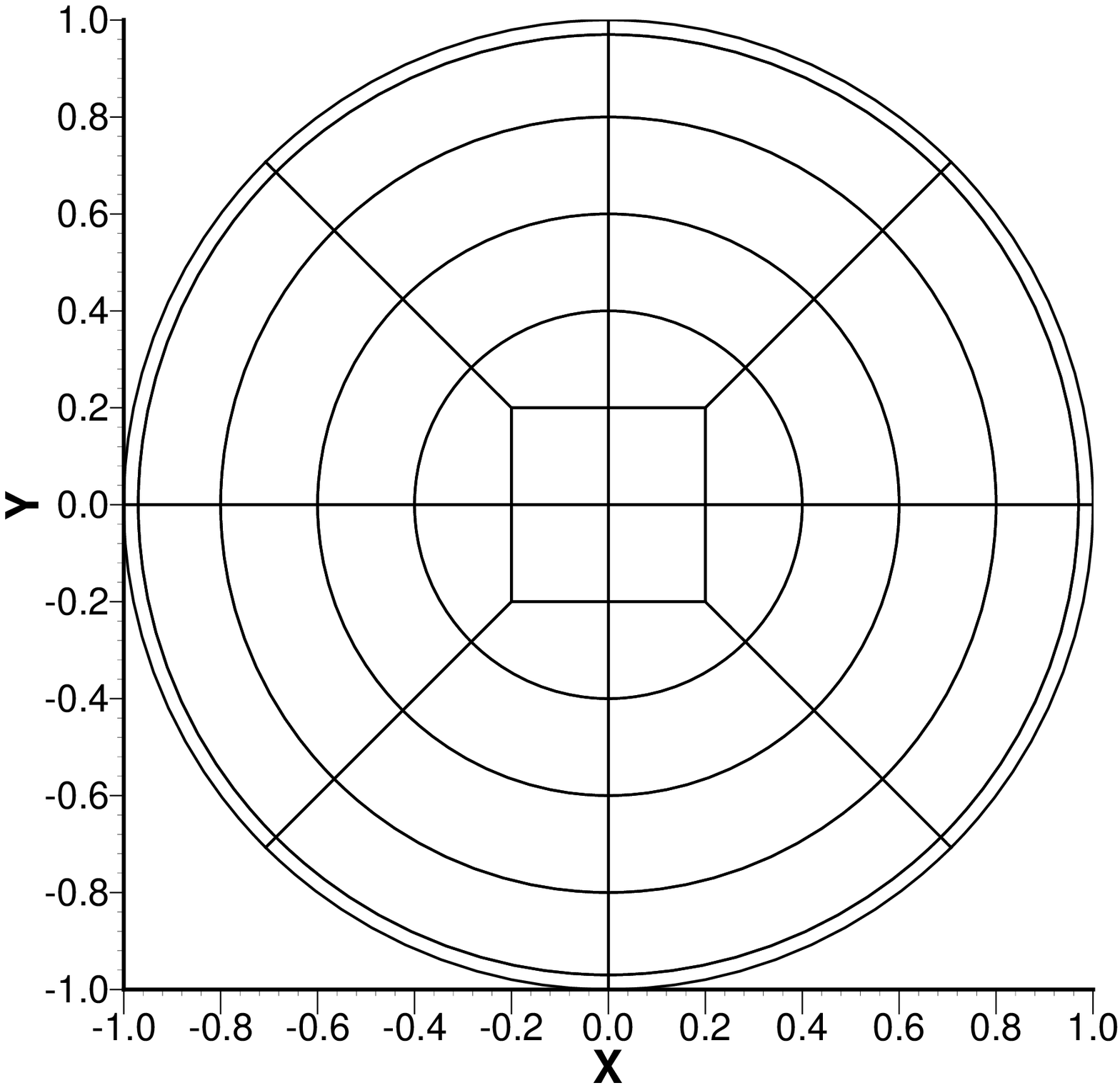}
\includegraphics[width=0.32\textwidth]{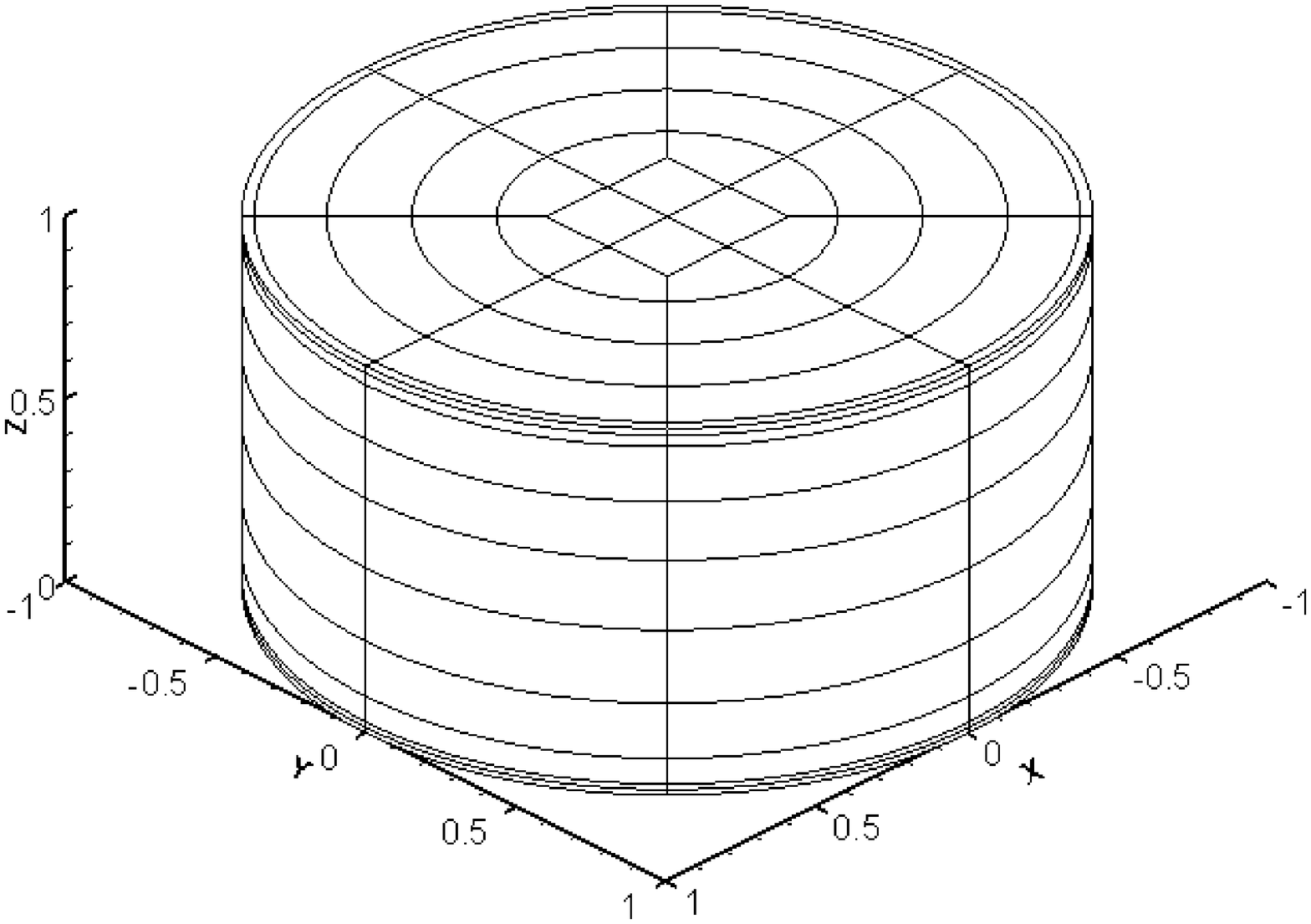}
\caption{Typical meshes view used throughout this study. Left:
  half-meridional grid. Center: spectral-element grid in any plane
  normal to the $z$-direction. Right: Three-dimensional grid
  comprising 10 cylindrical layers of nonuniform heights made of 44
  spectral elements each. Case $\Lambda=1$.}\label{fig:grid-cuts}
\end{figure}

The Navier--Stokes equations \eqref{eq:NS-1}--\eqref{eq:NS-2},
supplemented with the boundary conditions \eqref{eq:no-slip-tubular},
\eqref{eq:velocity-distribution-1}--\eqref{eq:velocity-distribution-3},
\eqref{eq:w}, \eqref{eq:stress-free-1}--\eqref{eq:stress-free-3},
constitute the set of governing equations for this free-surface
swirling flow to be discretized and ultimately solved. The numerical
method treats Eqs.~\eqref{eq:NS-1}--\eqref{eq:NS-2} within the weak
Galerkin formulation framework. The spatial discretization uses
Lagrange--Legendre polynomial interpolants. The reader is referred to
the monograph by Deville \etal\ \cite{deville02:_high} for full
details. The velocity and pressure are expressed in the
$\mathbb{P}_N-\mathbb{P}_{N-2}$ functional spaces where $\mathbb{P}_N$
is the set of polynomials of degree lower than $N$ in each space
direction. This spectral element method avoids the presence of
spurious pressure modes as it was proved by Maday \& Patera
\cite{maday89:_spect_navier_stokes} and Maday \etal\ \cite{maday93:_analy_stokes_applic}. The
quadrature rules are based on a Gauss--Lobatto--Legendre (GLL) grid
for the velocity nodes and a Gauss--Legendre grid (GL) for the
pressure nodes. The spectral element grid used for all simulations is
presented on Fig.~\ref{fig:grid-cuts}, in the particular case
$\Lambda=1$. This mesh comprises 440 spectral elements distributed
into 10 cylindrical layers of different heights, but all made of the
same distribution of 44 spectral elements, see
Fig.~\ref{fig:grid-cuts} (right). In order to resolve the boundary
layer along the tubular side-wall, the Ekman shear layer above the
rotating bottom end-wall and the surface shear layer below the free
surface, the spectral elements are unevenly distributed as can be seen
in Figs.~\ref{fig:grid-cuts}. The choice of polynomial degree in the
three space directions, defining the inner GLL and GL grid into each
spectral element is deferred to Sec.~\ref{sec:convergence}.
The essential Dirichlet boundary conditions---homogeneous for $\uu$ on
the tubular side-wall as expressed by Eq.~\eqref{eq:no-slip-tubular},
homogeneous for $w=u_z$ on the free surface as expressed by
Eq.~\eqref{eq:w}, and non-homogeneous for $\uu$ on the rotating bottom
end-wall as expressed by
Eqs.~\eqref{eq:velocity-distribution-1}--\eqref{eq:velocity-distribution-3}---are
embodied into the choice of test and trial functions chosen for the
velocity field.

The stress-free condition \eqref{eq:stress-free} at the free surface,
further expressed by
Eqs.~\eqref{eq:stress-free-1}--\eqref{eq:stress-free-3}, appears as a
homogeneous natural boundary condition in the weak Galerkin
framework. This central point is discussed in detail and in a more
general framework, in Bodard \etal\ \cite{bodard06:_solut}, where a
non-homogenous natural boundary condition is accounted for in the weak
formulation of the problem. Based on this previous analysis, the
treatment of the present stress-free condition at the free surface
appears straightforward and is automatically incorporated into the
weak formulation of the problem.

Borrowing the notation from Deville \etal\ \cite{deville02:_high}, the
semi-discrete Navier--Stokes equations resulting from space
discretization are
\begin{align}
  \bM \frac{\dd \underline{\uu}}{\dd t}+ \bC \underline{\uu}
  +\nu \AA \underline{\uu} -\bD^T \underline{p}&=0,\label{eq:odes}\\
  -\bD\underline{\uu} &=0\label{eq:constr}.
\end{align}
The diagonal mass matrix $\bM$ is composed of three blocks, namely the
mass matrices $M$. The global vector $\underline{\uu}$ contains all
the nodal velocity components while $\underline{p}$ is made of all
nodal pressures. The matrices $\AA$, $\bD^T$, $\bD$ are the discrete
Laplacian, gradient and divergence operators, respectively. The matrix
operator $\bC$ represents the action of the non-linear term written in
convective form $\underline{\uu}\cdot \bnabla$, on the velocity field
and depends on $\underline{\uu}$ itself. The semi-discrete equations
constitute a set of non-linear ordinary differential equations
\eqref{eq:odes} subject to the incompressibility condition
\eqref{eq:constr}.
%
%
\subsection{Time integration}
The time discretization of the semi-discrete set of governing
equations \eqref{eq:odes}--\eqref{eq:constr} is the same as the one
previously used in Bouffanais \etal\ \cite{bouffanais06:_large} and
Habisreutinger \etal\ \cite{habisreutinger06}. We only briefly recall
here the fundamentals of the method. The set of semi-discrete
equations \eqref{eq:odes}--\eqref{eq:constr} is discretized in time
using finite-difference schemes in a decoupled approach. The
computation of the linear Helmholtz problem---corresponding to the
stiffness matrix $\AA$---is integrated based on an implicit backward
differentiation formula of order 2, the nonlinear convective
term---corresponding to the operator $\CC$---is integrated based on a
relatively simple extrapolation method of order 2, introduced by
Karniadakis \etal\ \cite{karniadakis91:_high_navier}, see Bouffanais
\cite{bouffanais07:_simul} for full details.
%
%
\subsection{Convergence tests}\label{sec:convergence}
In order to demonstrate the spatial and temporal convergence of the
simulation method, time series data have been analyzed, while varying
separately the time-step $\Delta t$ and the polynomial degree $N$ of
the GLL basis in each space direction, at the upper bound in Reynolds
number $\Rey=6\,000$, and with $\Lambda=H/R=1$. As no
experimental, nor numerical reference results are available for the
present problem, three integral and one local quantities have been
computed and compared. These three integral quantities are the total
kinetic energy $Q$, enstrophy $E$ and helicity $\cH$ of the flow,
which definitions are recalled
\begin{align}
Q    & = \frac{1}{2} \int_\cV \uu \cdot \uu\, \dd \cV,\\
E    & = \frac{1}{2} \int_\cV \bomega \cdot \bomega\, \dd \cV,\\
\cH  & =             \int_\cV \uu \cdot \bomega\, \dd \cV.
\end{align}
The local quantity monitored is the axial velocity component $w=u_z$
at the point $\Pi_0$ of coordinates $(r/R=0,z/H=0.75)$, located along
the cylinder axis. The location choice of this monitoring point is
motivated by the study of Piva \& Meiburg \cite{piva05:_stead} for a
very similar configuration but at smaller Reynolds numbers. They show
that in the vicinity of this point, the axial velocity component
reaches a local maximum. Given the relatively high Reynolds number of
our benchmark simulation, a quite long transient---approximately 500
time units in $\Omega_0^{-1}$ units---is observed. Performing
convergence tests involving a simulation time of the order of this
transient time would simply be prohibitive. Consequently, it was
chosen to assess the convergence after only 50 time units of
simulations, which corresponds to the appearance of the first
vortex-breakdown recirculation bubble in the fluid domain.
First, the spatial convergence is assessed by varying the polynomial
degree in the range $6\leq N \leq 10$, while keeping the time-step
values constant $\Delta t= 0.0025$. Results are reported in
Table~\ref{tab:spatial-convergence}, and suggest that the spatial
convergence is achieved using a polynomial degree $N=8$ in all three
space directions. This value is used for all the direct numerical
simulations presented in the sequel, except for one single case
corresponding to the steady laminar case $(\Lambda = 1,
\Rey=900)$, for which $N=7$ is chosen.
\begin{table}[htbp]
\begin{center}
  \begin{tabular}{lcccc}
    $N$ & $Q$ & $E$ & $\cH$ & $w(\Pi_0)$ \\
    \hline\hline
    6  & \texttt{2.02745e-02} & \texttt{2.11900e+01} & \texttt{1.29876e-01} & \texttt{3.28923e-03}\\
    7  & \texttt{2.08244e-02} & \texttt{2.18923e+01} & \texttt{1.33612e-01} & \texttt{3.40034e-03}\\
    8  & \texttt{2.19036e-02} & \texttt{2.37953e+01} & \texttt{1.66448e-01} & \texttt{3.81373e-03}\\
    9  & \texttt{2.19034e-02} & \texttt{2.37957e+01} & \texttt{1.66450e-01} & \texttt{3.81376e-03}\\
    10 & \texttt{2.19035e-02} & \texttt{2.37955e+01} & \texttt{1.66447e-01} & \texttt{3.81375e-03}\\
    \hline
  \end{tabular}
  \caption{Spatial convergence analysis for the case
    $(\Rey=6\,000,\Lambda=1)$ with $\Delta t=
    0.0025\,\Omega_0^{-1}$. $Q$ in $R^3(R\Omega_0)^2$ units, $E$ in
    $R^3\Omega_0^2$ units, $\cH$ in $R^4\Omega_0^2$ units, and $w$ in
    $R\Omega_0$ units. Instant
    $t=50\Omega_0^{-1}$.}\label{tab:spatial-convergence}
\end{center}
\end{table}

Finally, the temporal convergence is assessed by varying the
time-step, while keeping the polynomial degree constant at the value
$N=8$, in agreement with the earlier spatial convergence
analysis. Results are reported in Table~\ref{tab:temporal-convergence}
below, and suggest that the temporal convergence is achieved using a
time-step $\Delta t =0.0025$. At a reduced Reynolds number compared to
the one employed for this convergence analysis $\Rey=6\,000$,
greater values of the time-step have been chosen in relation with the
more laminar nature of the flow without affecting the convergence of
the simulations.

\begin{table}[htbp]
\begin{center}
  \begin{tabular}{lcccc}
    $\Delta t$ & $Q$ & $E$ & $\cH$ & $w(\Pi_0)$ \\
    \hline\hline
    0.0050  & \texttt{2.08574e-02} & \texttt{2.14302e+01} & \texttt{1.36342e-01} & \texttt{3.49221e-03}\\
    0.0035  & \texttt{2.11896e-02} & \texttt{2.25311e+01} & \texttt{1.48303e-01} & \texttt{3.61923e-03}\\
    0.0025  & \texttt{2.19036e-02} & \texttt{2.37953e+01} & \texttt{1.66448e-01} & \texttt{3.81373e-03}\\
    0.0010  & \texttt{2.19034e-02} & \texttt{2.37960e+01} & \texttt{1.66446e-01} & \texttt{3.81379e-03}\\
    \hline
  \end{tabular}
  \caption{Temporal convergence analysis for the case
    $(\Rey=6\,000,\Lambda=1)$ with $N=8$. $\Delta t$ in
    $\Omega_0^{-1}$ units, $Q$ in $R^3(R\Omega_0)^2$ units, $E$ in
    $R^3\Omega_0^2$ units, $\cH$ in $R^4\Omega_0^2$ units, and $w$ in
    $R\Omega_0$ units. Instant
    $t=50\Omega_0^{-1}$.}\label{tab:temporal-convergence}
\end{center}
\end{table}
%
\section{Numerical simulations and results} \label{sec:numerical}
\subsection{General physical characteristics of the flow}\label{sec:general}
The central characteristics of the flow in a closed cylindrical
container with a bottom rotating end-wall is a large recirculation of
the fluid. The features of the intense shear layer induced by the
rotation of the bottom wall can be obtained from the analogy with the
analysis by von K\'arm\'an for the flow generated by a spinning plate
in an unbounded fluid domain, see the review by Zandbergen \& Dijkstra
\cite{zandbergen87:_vonkar} for full details. The rotation of the
bottom wall has a suction effect on the fluid in the near-axis region
and a pumping effect, while accelerating the fluid radially outwards
in an Ekman shear layer of thickness $O(\Rey^{-1/2})$. In the
framework of our problem, this Ekman layer is bounded by the tubular
cylinder side-wall, which forces the recirculation of the fluid in the
upward direction along the side-wall, and towards the top wall. As the
fluid approaches the curved corner, the radial velocity contribution
to the kinetic energy is progressively transformed into an axial
velocity contribution to the kinetic energy. It has been observed that
the fluid turns and subsequently spirals upward along the tubular
side-wall.

As mentioned in Sec.~\ref{sec:open-swirling-flow}, replacing the fixed
top solid wall with a free surface changes significantly the physics
of the flow and the recirculation mechanisms. In absence of tangential
stresses at the free surface, the boundary layer is replaced by a
surface layer in the sense of Shen \etal\
\cite{shen99,shen00:_turbul}. In addition, the inward spiraling fluid
elements conserve their angular momentum at the free surface. When the
related centrifugal force is large enough to balance the radial
pressure gradient, the flow separates from the free surface and leads
to the generation of a vortex breakdown bubble. The most striking
difference between the flow patterns observed in the present open
cylinder case and compared to the close cylinder one are the possible
appearances of recirculation bubbles, which are generally attached to
the free surface. Such flow patterns are simply impossible in presence
of the no-slip condition imposed on the top wall in the closed
cylinder case. More precisely, Iwatsu \cite{iwatsu05:_numer}
determined 24 different flow states in the steady regime according to
the meridional streamline patterns observed. Spohn \etal\
\cite{spohn98:_exper} summarized those flow states in a simplified
bifurcation diagram in the $(\Lambda, \Rey )$ space. Based on his
extensive and comprehensive study, Iwatsu \cite{iwatsu05:_numer} came
out with a more detailed and complex bifurcation diagram.

In subsequent studies, the stability of those steady axisymmetric
flows were investigated. Young \etal\ \cite{young95:_period},
Miraghaie \etal\ \cite{miraghaie03:_flow}, Lopez \etal\
\cite{lopez04:_symmet}, and Lopez \& Marques \cite{lopez04:_mode}
observed unstable azimuthal modes which are triggered at different
values of the Reynolds number depending on the nature ``shallow''
($\Lambda < 1$) or ``deep'' ($\Lambda >1$) of the system. Valentine \&
Jahnke \cite{valentine94:_flows}, Lopez \cite{lopez95:_unstead}, and
Br{\o}ns \etal\ \cite{brons01:_topol} associated the axisymmetry
breakage to instability modes. These modes appear following a Hopf
bifurcation which generally occurs at relatively high Reynolds number.

\subsection{Cases studied}\label{sec:cases}
As mentioned in the previous sections, the physics of these
free-surface swirling flows depends critically on the Reynolds
number. Nevertheless, the height-to-radius aspect ratio $\Lambda$ also
has considerable impact on the observed nature of the flow. Very
often, situations corresponding to extreme values of $\Lambda$ have
been studied, as they generally lead to simplified flow
mechanisms. For instance, shallow systems ($\Lambda<1$) are often
referred to as ``rotor-stator'' configurations, in which the fluid is
almost in a state of solid-body rotation. On the other hand, deep
systems associated with large values of $\Lambda$, generate
recirculation bubbles away from the free surface and generally on the
cylinder axis. Consequently, systems corresponding to values of
$\Lambda$ close to the unity are intermediate in the sense that the
physics of the flow observed is a complex combination of the general
features reported for the shallow and deep systems.

\begin{table}[htbp]
  \centering
  \begin{tabular}{llccll}
    Case  & \Rey\     & $\Lambda=H/R$ & Time-step $\Delta t$ & Time evolution & Vortex breakdown\\
    \hline \hline
    $(a)$ & $900$      & $1.0$           & 0.0050   & steady      & one attached bubble\\
    $(b)$ & $1\,500$   & $1.0$           & 0.0050   & steady      & one attached toroidal bubble\\
    $(c)$ & $6\,000$   & $1.0$           & 0.0025  & unsteady    & complex dynamics\\
    $(d)$ & $2\,000$   & $1/3$           & 0.0040   & steady      & two long attached bubbles\\
    $(e)$ & $2\,000$   & $3.0$           & 0.0040   & steady      & one short detached bubble\\
    \hline
  \end{tabular}
  \caption{Parameters and characteristics of the cases considered. The time-step $\Delta t$ is expressed in $\Omega_0^{-1}$ units.}\label{tab:parameters}
\end{table}
\begin{figure}[htbp]
  \centering
  \includegraphics[width=0.6\textwidth]{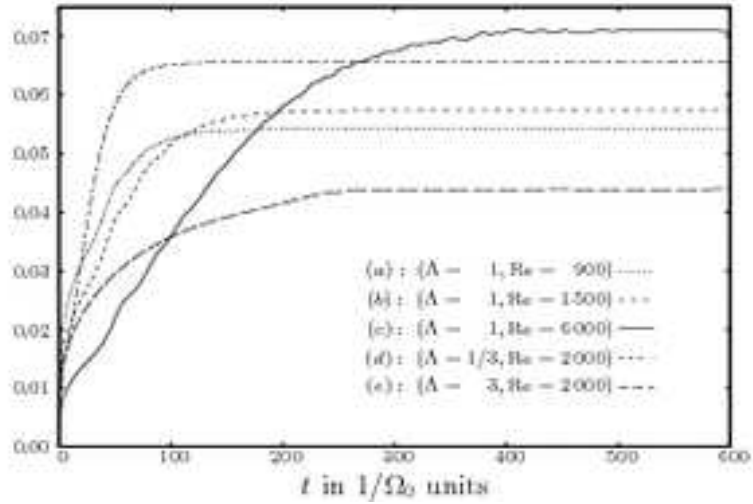}
  \caption{Time history of the volume integral of the kinetic energy
    $Q$ of the flow, in $R^5\Omega_0^2$ units for cases
    $(a)$--$(e)$.}\label{fig:T-K-O-H}
\end{figure}
The details related to the five cases considered in this article are
summarized in Table~\ref{tab:parameters}. The primary focus is on the
flow defined by $\Rey=6\,000$ and $\Lambda=1$, and corresponding to
case $(c)$. The value of the Reynolds number is intentionally set to a
high value compared to previous studies---the highest to our
knowledge, in order to obtain fields of a relative intensity at the free surface. 
The choice of the value of $\Lambda$ follows the earlier comment on
systems being intermediate between shallow and deep. This central case
$(\Lambda=1,\Rey=6\,000)$ is supplemented with four secondary cases
described in Table~\ref{tab:parameters}. The study of those secondary
flows is of prime importance for the understanding of the complex
dynamics of the primary case $(\Lambda=1,\Rey=6\,000)$.

In terms of initial conditions, the steady rotation is impulsively
started from a quiescent fluid state for all cases presented in the
sequel. At this point, it is worth noting the timescales of the
evolution of these flows. Figure~\ref{fig:T-K-O-H} displays the time
history of the volume integral of the kinetic energy of the flow
$Q$. For all cases except case $(c)$, the flow reaches a steady state
after a given time scale, which is, as expected, shorter for shallow
systems. Case $(c)$ leads to an unsteady flow which does not display
any oscillatory evolution. The value of the Reynolds number for this
case is large enough to produce a non-trivial evolution of the
recirculation zones as will be seen in the sequel.

The time histories of the volume integral of the kinetic energy $Q$
for the five cases $(a)-(e)$ can be compared to the ones, reported in
Bouffanais \cite{bouffanais07:_simul} for the closed swirling flow
problem with $\Lambda=2.5$, see Fig~B.4. It should be noted that for a
fixed value of $\Lambda$, the total kinetic energy $Q$ of the flow
decreases with the Reynolds number for the closed cylinder case, while
it increases in the open cylinder case. This decreasing trend for $Q$
in the closed cylinder case can easily be resolved by transposing the
analysis given by Leriche \& Gavrilakis \cite{leriche00:_direc} in
their study of the closed lid-driven cubical cavity flow. Leriche \&
Gavrilakis argue that the most significant part of the kinetic energy
of the flow is contained in the viscous layer developing on the
driving wall. Consequently, the total energy varies like the energy
contained in this viscous layer which can approximately be expressed
as $U_0^2\cV \Rey^{-1/2}$, where $U_0$ is the characteristic velocity
of the driving wall and $\cV$ the volume of the cavity. Such argument
and estimate can easily be transposed for the closed swirling flows,
and explains the decreasing trend for $Q$ with respect to
\Rey. Furthermore this argument is confirmed by the measurements of
the kinetic energy $Q(L_1)$ of the cylindrical layer of fluid $L_1$
located right above the spinning disk and of height $0.015H$, reported
in Table~\ref{tab:Q-E}. This thin layer of fluid which only represents
$1.5\%$ of the total volume of fluid, contributes for approximately
$10\%$ to the total kinetic energy of the flow. Concurrently, its
contribution to the total kinetic energy of the flow decreases with
the Reynolds number.

\begin{table}[htbp]
\begin{center}
  \begin{tabular}{lccc}
    \Rey\ & $Q(L_1)$ & $E(L_1)$ & $E_z(L_1)$ \\
    \hline\hline
    $900$     & \texttt{7.42243e-03} & \texttt{3.04128e+00} & \texttt{3.99817e-01}\\
    $1\,500$  & \texttt{7.05011e-03} & \texttt{4.46885e+00} & \texttt{4.03147e-01}\\
    $6\,000$  & \texttt{5.71943e-03} & \texttt{1.18305e+01} & \texttt{4.25622e-01}\\
    \\
    \Rey\ & $Q(L_{10})$ & $E(L_{10})$ & $E_z(L_{10})$ \\
    \hline\hline
    $900$     & \texttt{3.88037e-04} & \texttt{6.14708e-03} & \texttt{5.99848e-03}\\
    $1\,500$  & \texttt{5.47285e-04} & \texttt{9.04519e-03} & \texttt{8.25633e-03}\\
    $6\,000$  & \texttt{1.21966e-03} & \texttt{6.43344e-02} & \texttt{2.00737e-02}\\
    \hline
  \end{tabular}
  \caption{Measurements of the kinetic energy $Q$, enstrophy $E$, and
    enstrophy associated solely with the axial vorticity component
    $E_z$ for the cylindrical layer $L_1$ comprised between $z=0$ and
    $z=0.015H$, and for the cylindrical layer $L_{10}$ comprised
    between $z=0.98H$ and $z=H$. $Q$ in $R^3(R\Omega_0)^2$ units, $E$
    and $E_z$ in $R^3\Omega_0^2$ units. Instant
    $t=600\Omega_0^{-1}$.}\label{tab:Q-E}
\end{center}
\end{table}

On the contrary, a reverse trend is observed for the variations of $Q$
with respect to \Rey\ in the open swirling flow. It therefore requires
another physical justification. Nevertheless, the previous energetic
argument associated with the viscous layer still holds for the viscous
layer above the spinning disk and near the tubular side-wall in the
open cylinder swirling flow. Below the surface at $z=H$, the viscous
layer in the closed cylinder case is replaced by an intense shear
layer. As mentioned in Sec.~\ref{sec:free-surface}, in the present
flat-free-surface problem, the axial vorticity $\omega_z$ is the only
component of the vorticity field which is not vanishing at the free
surface and it provides a measure of the internal shear at the free
surface. The part $E_z$ of the enstrophy associated with the axial
vorticity component is measured in the cylindrical layer of fluid
$L_{10}$, of height $0.02H$ and located below the free
surface. Results are reported in Table~\ref{tab:Q-E} and clearly show
a significant increase of $E_z$ with respect to \Rey\ in the layer
$L_{10}$, while it is almost constant in $L_1$. These results allow us
to infer that the shear layer below the free surface becomes more and
more intense and energetic---see $Q(L_{10})$--- when increasing the
Reynolds number. But this observed energetic trend of the free-surface
shear layer is not the only factor responsible for counterbalancing
the decreasing trend of the viscous layers. The internal structure of
the free-surface layer is itself physically different as reported by
Shen \etal\ \cite{shen99,shen00:_turbul}. The so-called surface layer
corresponds to a thin region adjacent to the free surface
characterized by fast variations of the tangential vorticity
components. This surface layer is caused by the dynamic zero-stress
boundary conditions \eqref{eq:stress-free} at the free surface and
lies inside a thicker blockage (or ``source'') layer, which is due to
the kinematic boundary condition \eqref{eq:w} at the free surface. The
importance of the outer blockage layer is manifested mainly in the
redistribution of the kinetic energy, \ie\ in the increase of the
kinetic energy of the tangential velocity components at the expense of
the kinetic energy of the axial velocity component. 
This point is to be further discussed in Sec.~\ref{sec:unsteady}, where a comprehensive comparison of the flow below the surface $z=H$ for the case $(c)$ with a free surface and the equivalent closed case is given.

\subsection{Physical description of flow states}\label{sec:physical}
\subsubsection{Steady flows}\label{sec:steady}
\begin{figure}[htbp]
  \centering
  \includegraphics[width=0.39\textwidth]{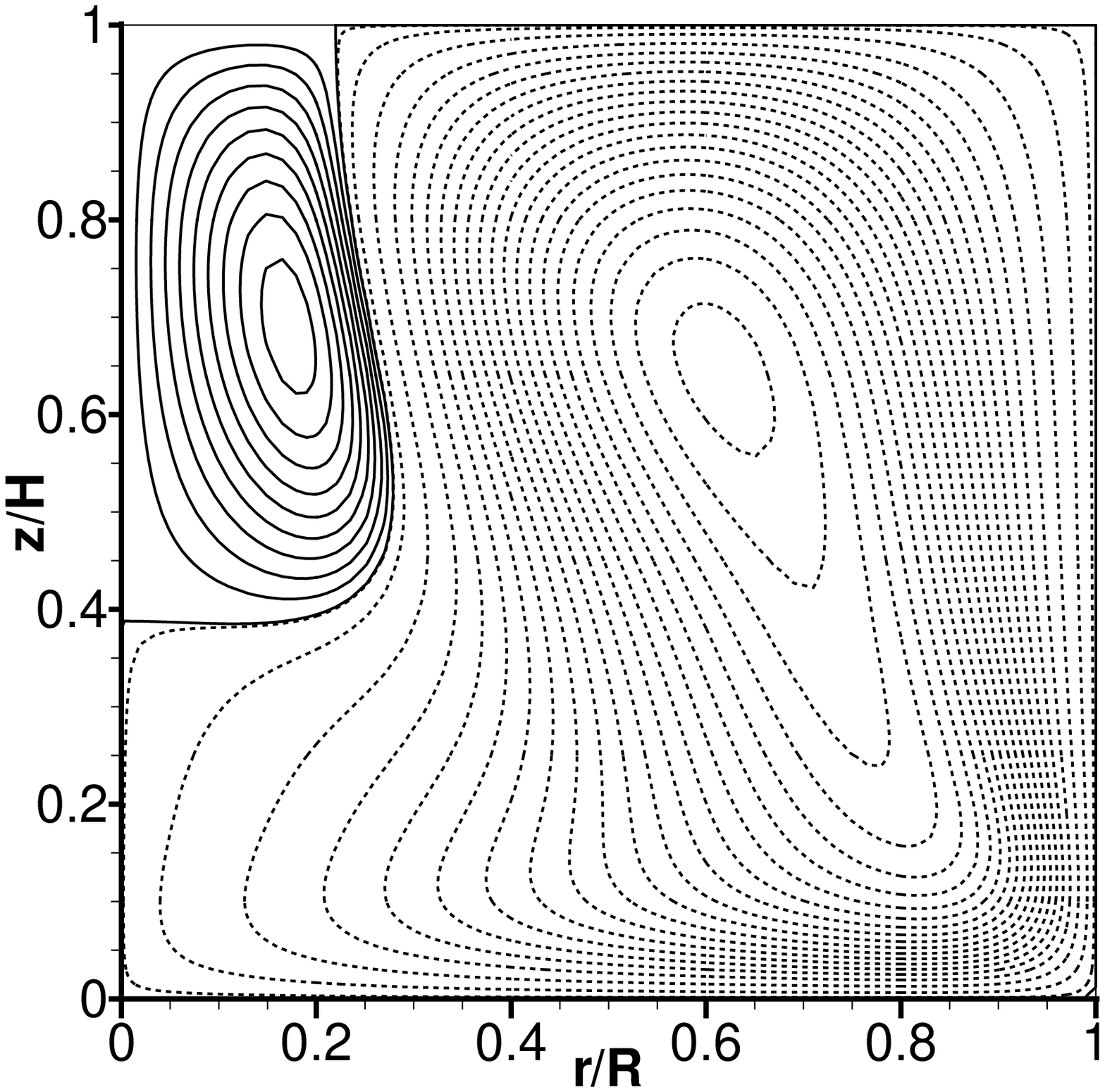}
  \includegraphics[width=0.39\textwidth]{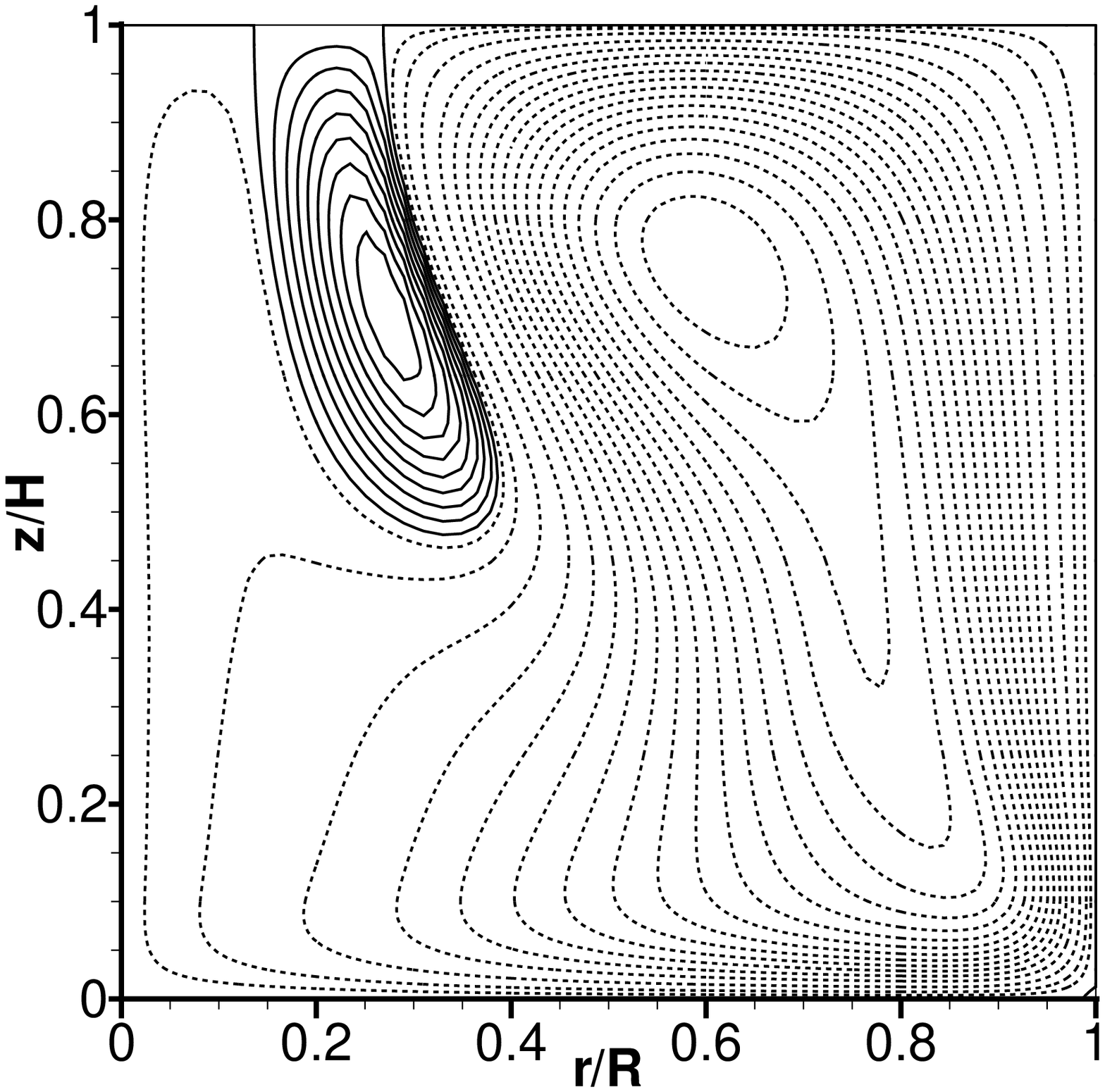}\\
  \caption{Contours of streamlines in a meridional plane, case
    $\Lambda=1$. Left: case $\Rey = 900$; Right: case
    $\Rey=1\,500$. The 30 contours are non-uniformly spaced for
    visualization purposes, 20 equally-spaced negative contours and 10
    equally-spaced positive contours for $\Rey=900$ and
    $\Rey=1\,500$.}\label{fig:streamlines-900-1500}
\end{figure}

As a first step, we present the two steady flows for $\Lambda=1$ at
$\Rey=900$ and $\ 1\,500$ corresponding to cases $(a)$ and $(b)$
respectively. Figure~\ref{fig:streamlines-900-1500} displays the
streamlines of these flows into any meridional plane. Both of these
flows present a large axisymmetric vortex breakdown bubble attached to
the free surface in agreement with the experimental results from Spohn
\etal\ \cite{spohn98:_exper} summarized in their bifurcation diagram. These recirculation zones are characteristic of
these swirling flows due to the conjugate effects of the centrifugal
force and the overturning flow induced by the presence of the tubular
side-wall. The central difference between the low-Reynolds-number
cases $(a)$ and $(b)$ is the shape of the recirculation, which becomes
toroidal after leaving the axis when the Reynolds number is increased
from $900$ up to $1\,500$. These results can be further validated by
comparing them to the experimental results (dye visualizations)
obtained by Piva for $\Rey = 1\,120$. Finally, case
$(a)=(\Lambda=1,\Rey=900)$ has also been computed by Piva \& Meiburg
\cite{piva05:_stead}.

\begin{landscape}
  \begin{figure}[ht!]
    \centering
    \includegraphics[width=0.34\textwidth]{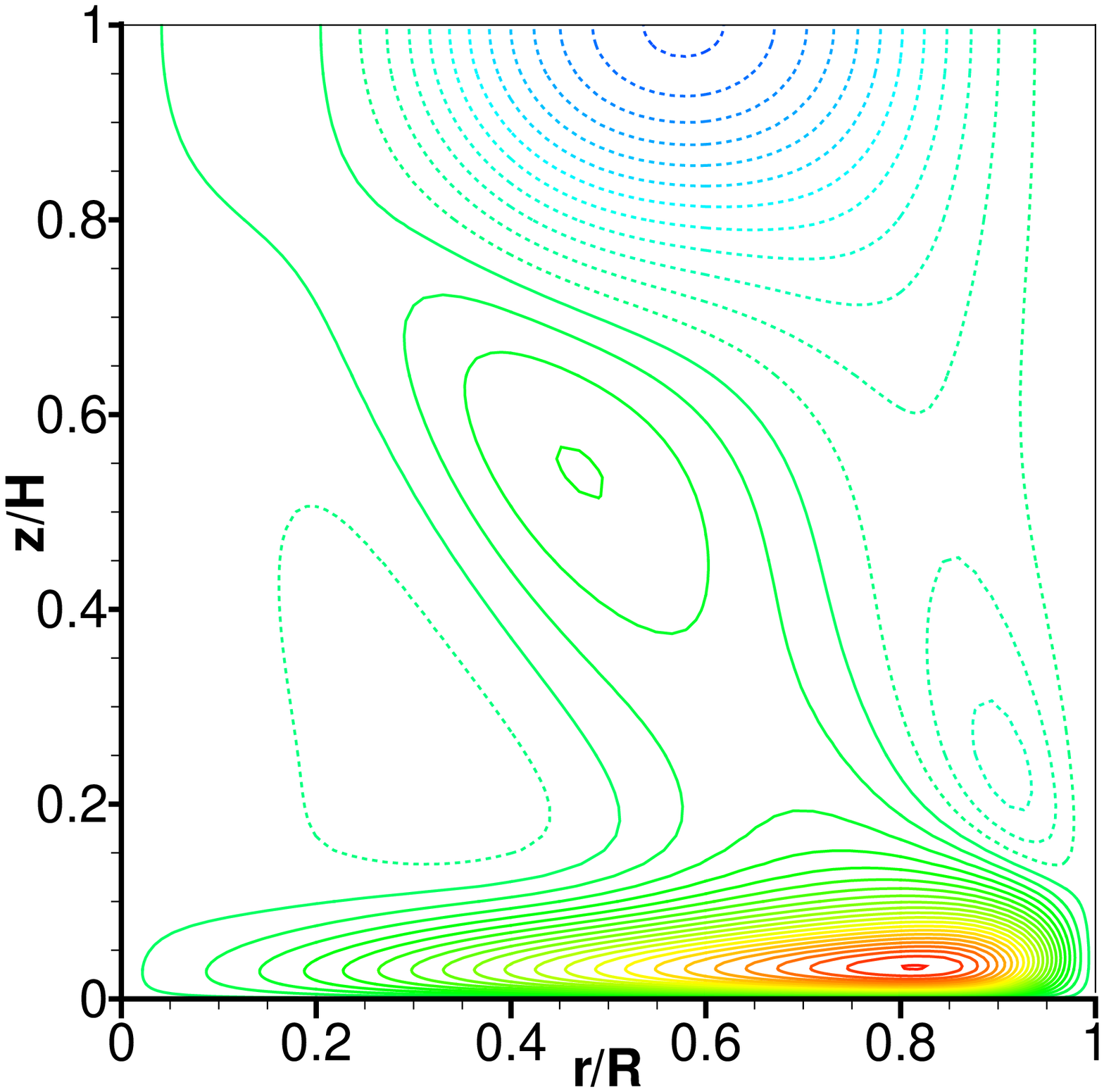}
    \includegraphics[width=0.34\textwidth]{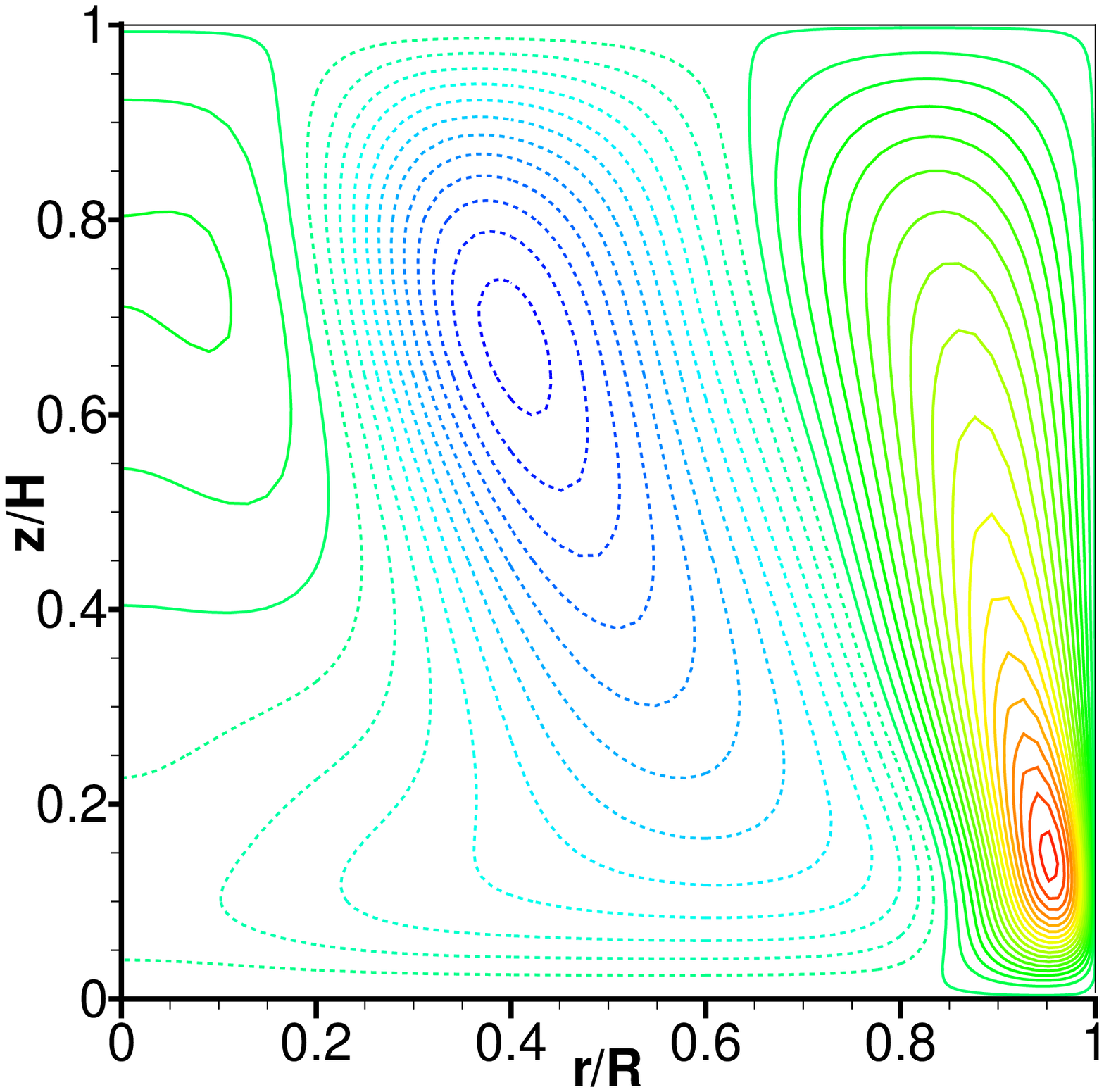}
    \includegraphics[width=0.34\textwidth]{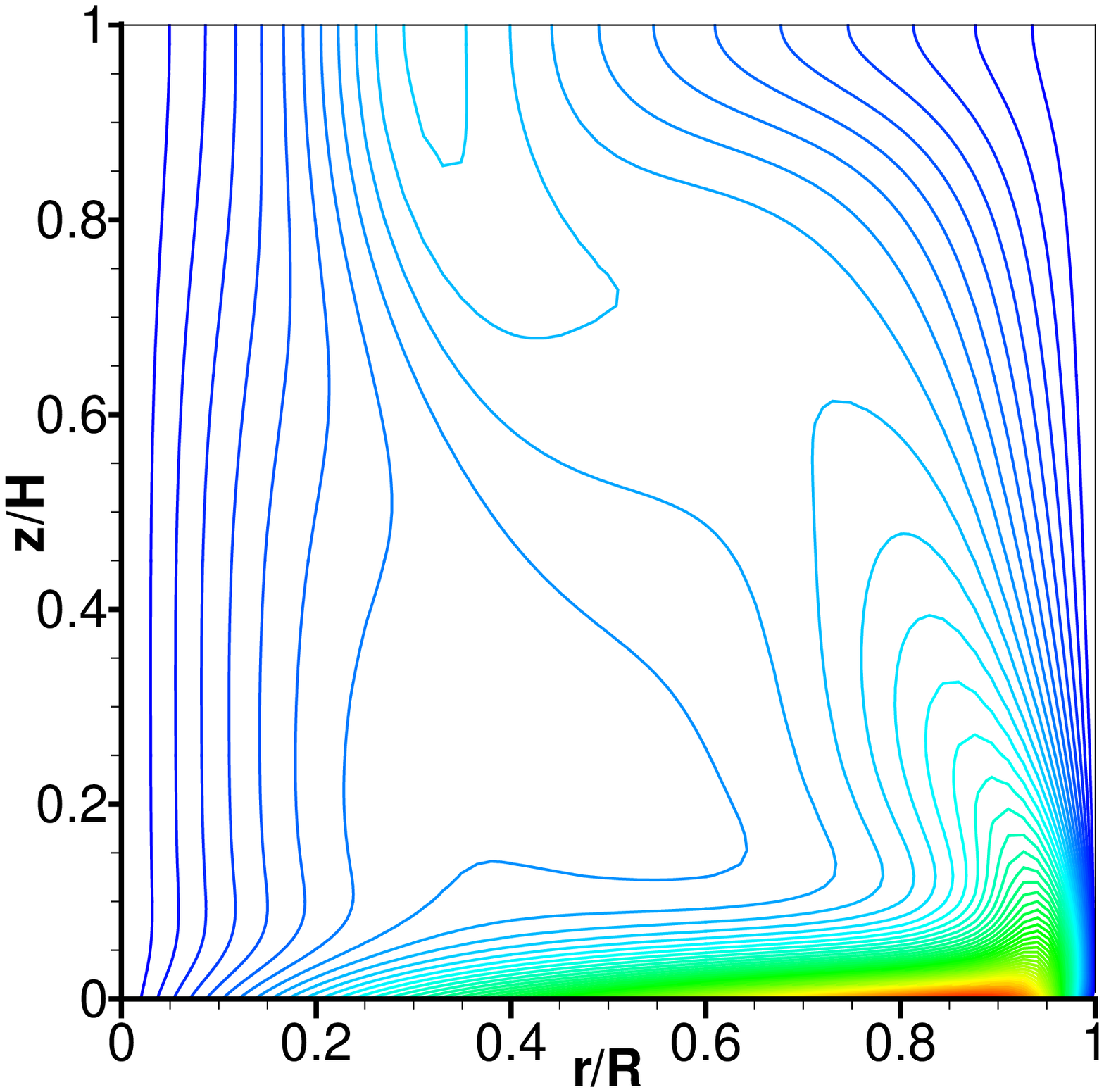}
    \includegraphics[width=0.34\textwidth]{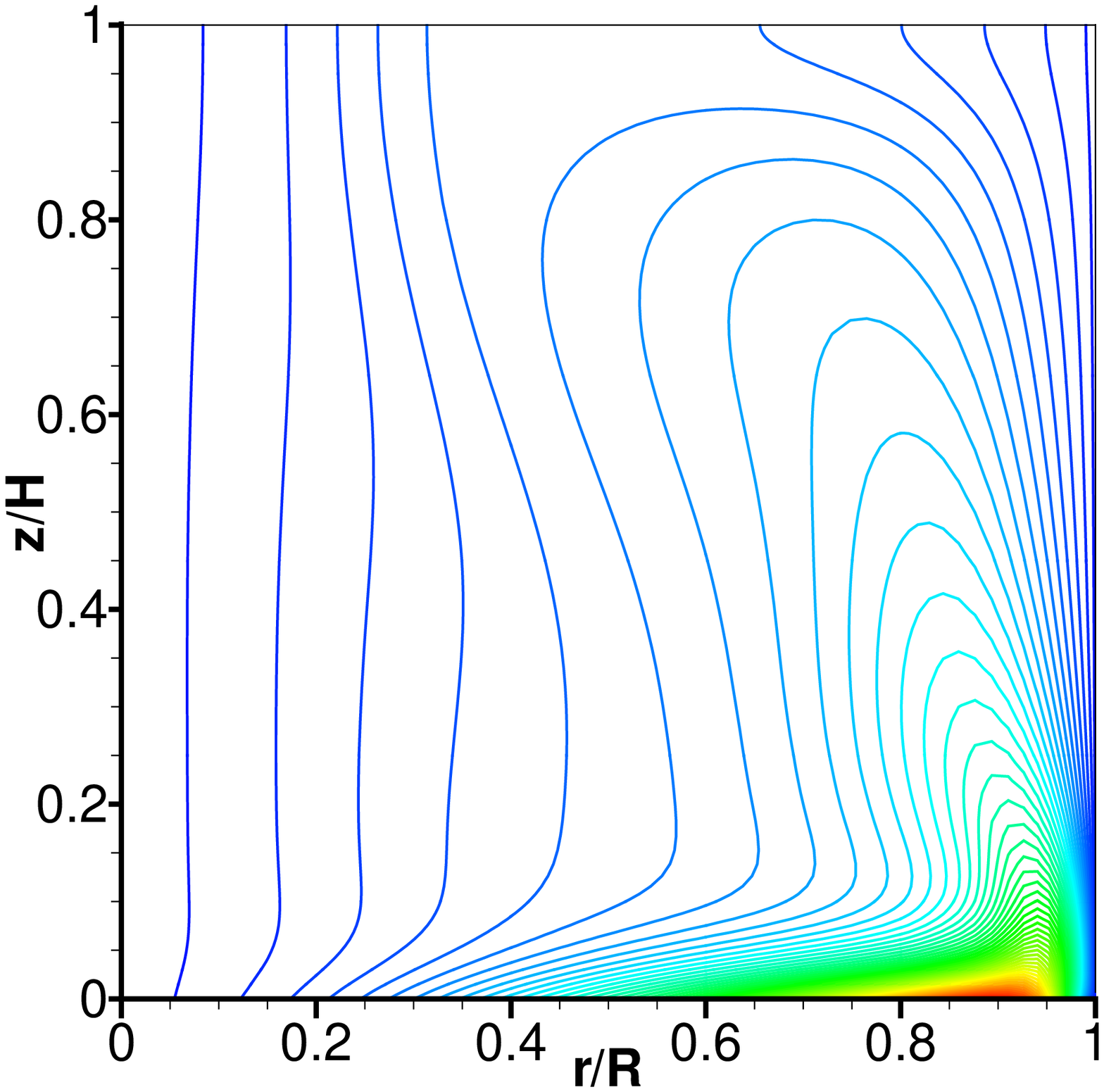}\\
    \includegraphics[width=0.34\textwidth]{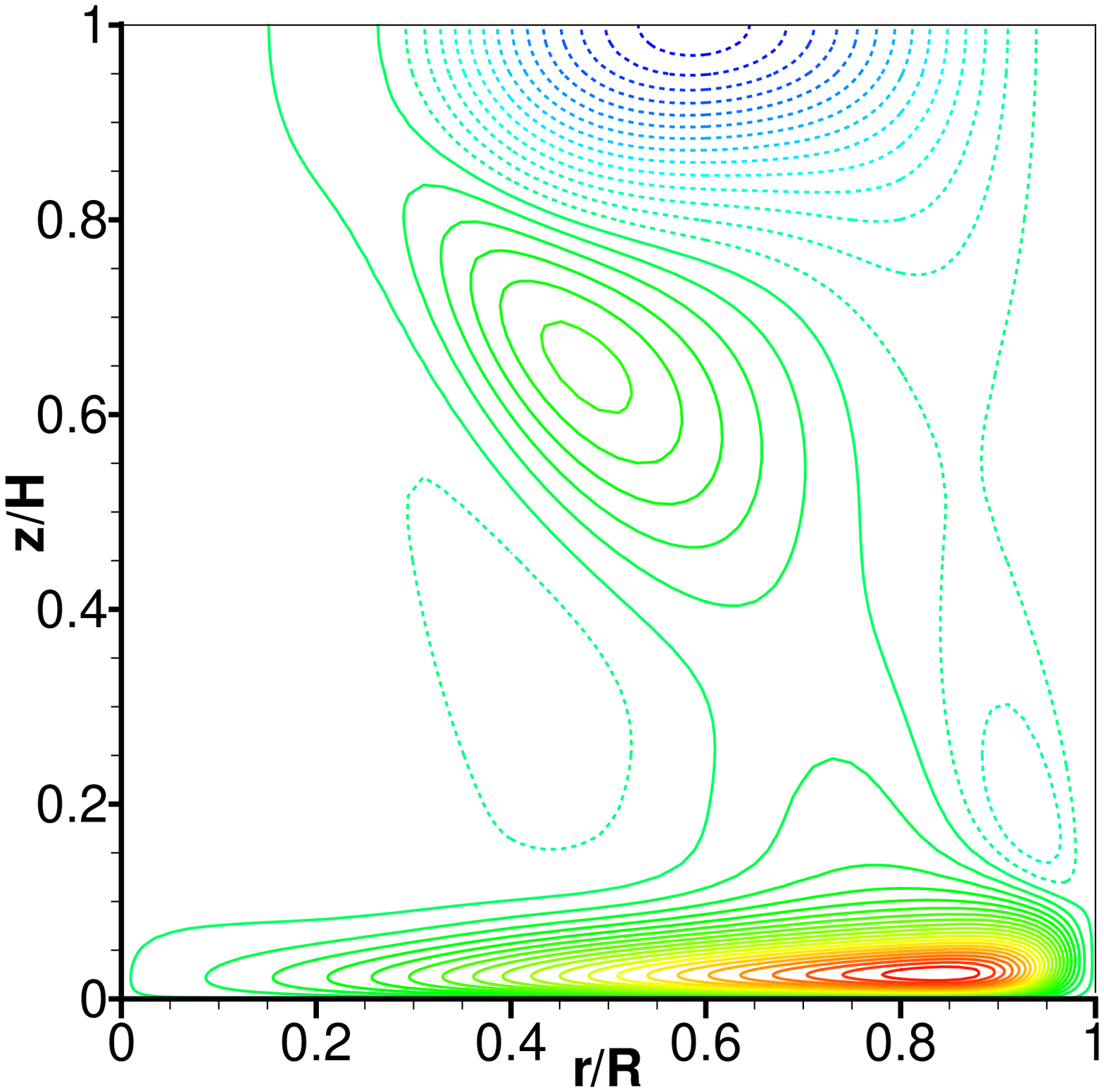}
    \includegraphics[width=0.34\textwidth]{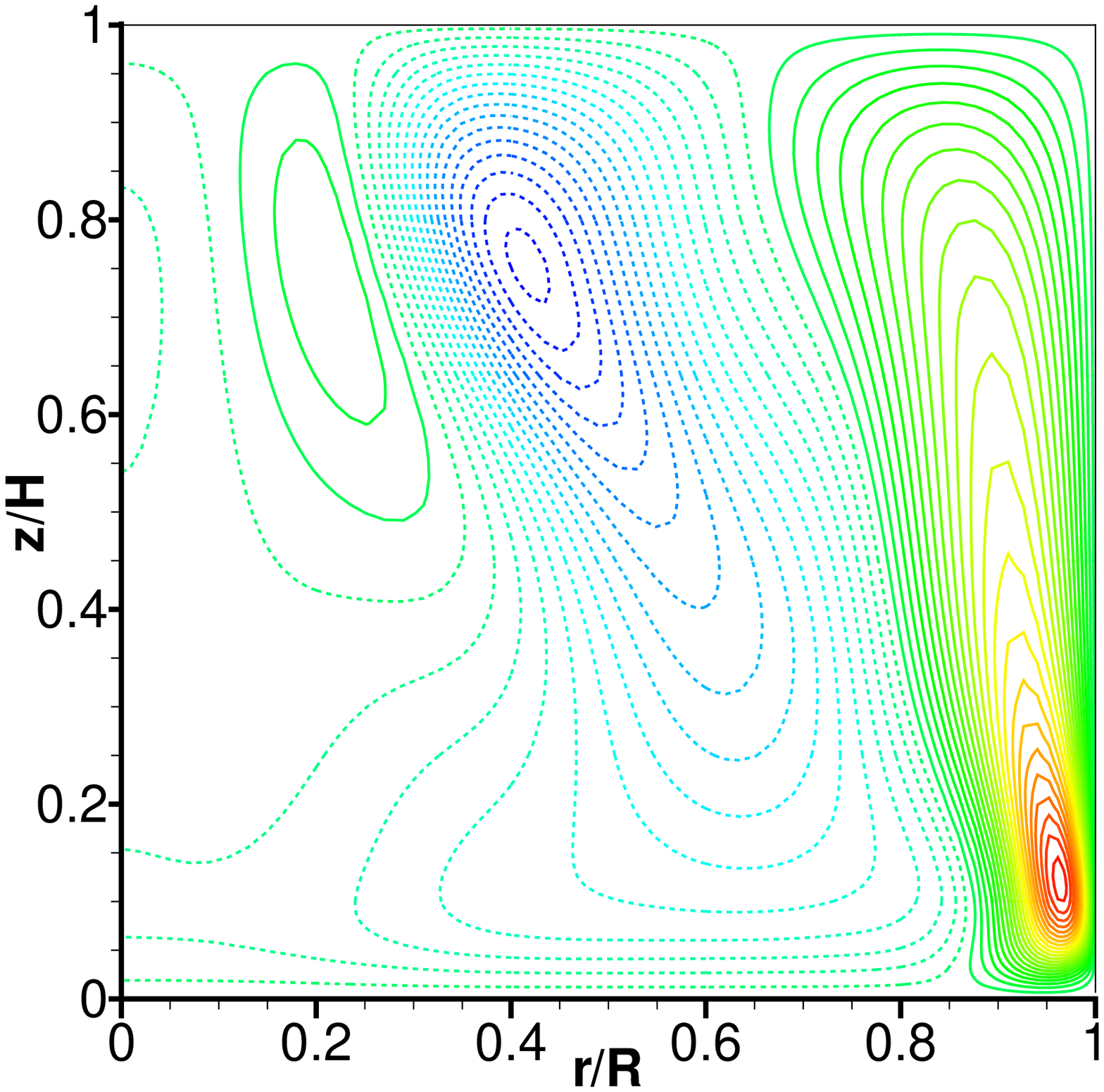}
    \includegraphics[width=0.34\textwidth]{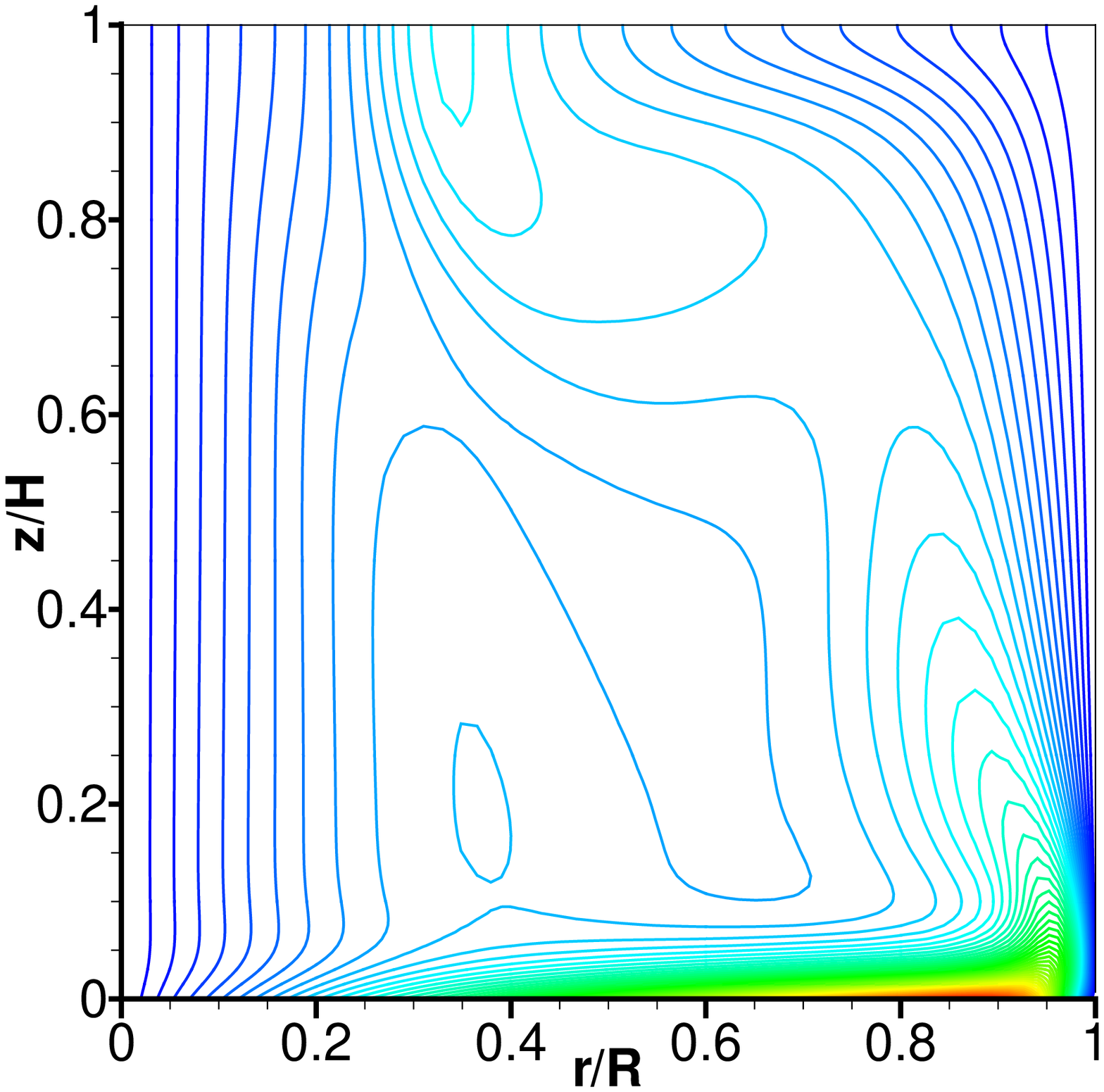}
    \includegraphics[width=0.34\textwidth]{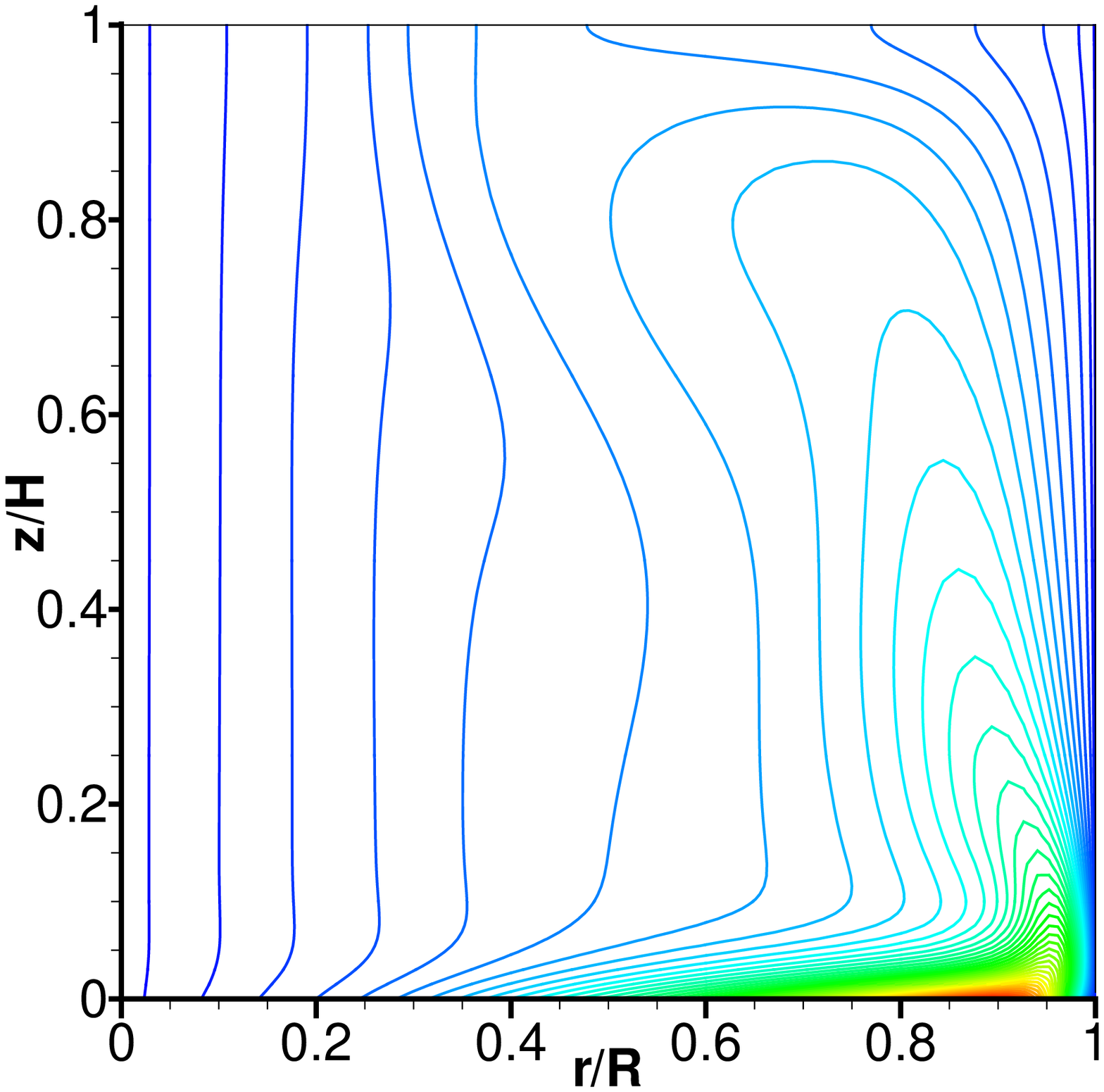}
    \caption{Contours in a meridional plane for the case
      $\Lambda=1$. Top row: case $\Rey = 900$; Bottom row: case $\Rey
      = 1\,500$. From left column to right column: radial velocity
      component $u_r$; axial velocity component $w=u_z$; azimuthal
      velocity component $u_\theta$; axial angular momentum component
      $\Gamma = r u_\theta$. The 35 contours are uniformly spaced,
      between $-0.1$ and $0.145$ for $u_r$; and between $-0.08$ and
      $0.115$ for $u_z$; The 50 contours are uniformly spaced, between
      $0$ and $1$ for $u_\theta$ and
      $\Gamma$.}\label{fig:vr-vz-vt-gamma-900-1500}
  \end{figure}
\end{landscape}

As a second step, the contours of the radial, azimuthal and axial
velocity components in any meridional plane are given in
Fig.~\ref{fig:vr-vz-vt-gamma-900-1500}. These data are supplemented
with the contours of the axial component of the angular momentum
$\Gamma=r u_\theta$ still in Fig.~\ref{fig:vr-vz-vt-gamma-900-1500},
extreme right column. The interest for $\Gamma$ lies in the fact that
it plays the role of a streamfunction for the part of the velocity
field comprised in any meridional plane, see Bragg \& Hawthorne
\cite{bragg50:_some} and Keller \cite{keller96} for full
details. Therefore, the contours of $\Gamma$ deliver the intersection
of vortex surfaces with the corresponding meridional plane where they
are drawn, and as such provide us with the local direction of the
meridional vorticity field. One can notice from the velocity
components and axial angular momentum component that the meridional
structure of these flows is far from being trivial. It consists of an
intense boundary layer above the spinning bottom end-wall that is
turned into the interior by the presence of the tubular side-wall,
forming a shear layer having a jet-like velocity profile in the
azimuthal direction. The contour lines of the axial component of the
angular momentum shown in Fig.~\ref{fig:vr-vz-vt-gamma-900-1500}
(extreme right column) simply represent the vortex lines, which all
emanate from the rotating disk; the structure of the shear layer is
apparent. It is worth noting here that the vortex lines distribution
at their origin varies like $r^2$. 
As a consequence of the regularized
profile of angular velocity of the rotating disk---see
Sec.~\ref{sec:lid}, this distribution in $r^2$ is slightly affected in
the vicinity of $r=R$. This regularization of this profile has the
advantage of preventing the appearance of vortex lines terminating at
the circular corner $(r=R,z=0)$. The overturning nature of these flows is also apparent in the vicinity
of the tubular side-wall, which is the vortex surface corresponding to
$\Gamma=0$, together with the cylinder axis. As non-zero azimuthal
velocities are possible at the free surface, vortex lines emanating
from the rotating end-wall have the option of terminating orthogonally
to the free surface. 
This observation is one of the major difference
with the closed cylinder swirling flow where all vortex lines have to
terminate in the corner. Furthermore, the termination of vortex lines
at the free surface is responsible for the possibility of having
vortex breakdown bubbles being attached to the free surface as
observed in Fig.~\ref{fig:streamlines-900-1500}. A careful analysis of
Fig.~\ref{fig:vr-vz-vt-gamma-900-1500} (extreme right column) reveals
that one vortex line marks the limit between an inner region
comprising only vortex lines terminating at the free surface, and an
outer region, where they terminate near the circular corner, like in
the closed cylinder case. In summary, it appears that the main effect
of this overturning flow is to bring high-angular-momentum fluid
towards the cylinder axis.

The results for case $(a)=(\Lambda=1, \Rey=900)$ presented in
Fig.~\ref{fig:streamlines-900-1500} (left) and
Fig.~\ref{fig:vr-vz-vt-gamma-900-1500} (top row) show a qualitative
good agreement with the numerical results of Piva \& Meiburg
\cite{piva05:_stead}. The differences related to the features of the
recirculation bubbles for cases $(a)$ and $(b)$ have been discussed
earlier. One can notice in Fig.~\ref{fig:vr-vz-vt-gamma-900-1500} (two first
left columns), that the thicknesses of the intense radial velocity layer as well as
the axial wall jet are reduced when the Reynolds number is increased
from $900$ to $1\,500$, as expected. The contours of the axial
velocity component reveal that the downward-directed flow induced by
the suction effect of the Ekman layer, is more intense at higher
Reynolds number. In addition, the region of the flow where $w=u_z$ has
a negative extremum tends to move closer to the free surface when
increasing \Rey. Regarding the vortex lines shown in the extreme right
column, their bending towards the cylinder axis is more pronounced at
the higher Reynolds number of $1\,500$. In relation with the previous
analysis, this latter observation highlights the fact that more
high-angular-momentum fluid is brought towards the axis when
increasing \Rey.

As mentioned in Sec.~\ref{sec:cases}, our primary interest lies in
case $(c)=(\Lambda=1,\Rey = 6\,000 )$, thereby justifying the study of
cases $(a)$ and $(b)$, having the same aspect ratio $\Lambda$ but
corresponding to laminar cases. Nevertheless, the study of cases $(d)$
and $(e)$, which both correspond to ``extreme'' cases in terms of
height-to-radius aspect ratio, illustrate some essential features of
the open swirling flow. In case $(c)$ some of these features may
prevail only in specific regions of the flow as it corresponds to an
intermediate case between a shallow system characterized by case $(d)$
and a deep system characterized by case $(e)$. These features are as
follows:
\begin{itemize}
\item[--] solid-body rotation of the inner core region, predominantly
  for small $\Lambda$;
\item[--] radial jet of angular momentum at the free surface;
\item[--] lateral jet-like shear layer along the tubular side-wall;
\end{itemize}

\begin{figure}[htbp]
  \centering
  \includegraphics[width=0.39\textwidth]{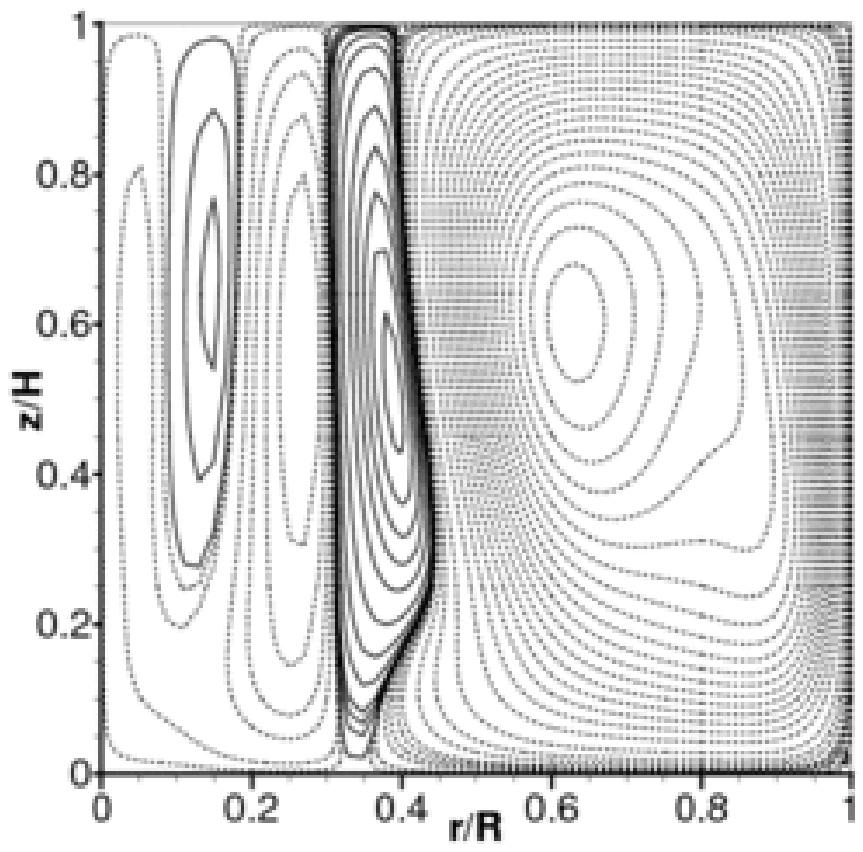}
  \includegraphics[width=0.39\textwidth]{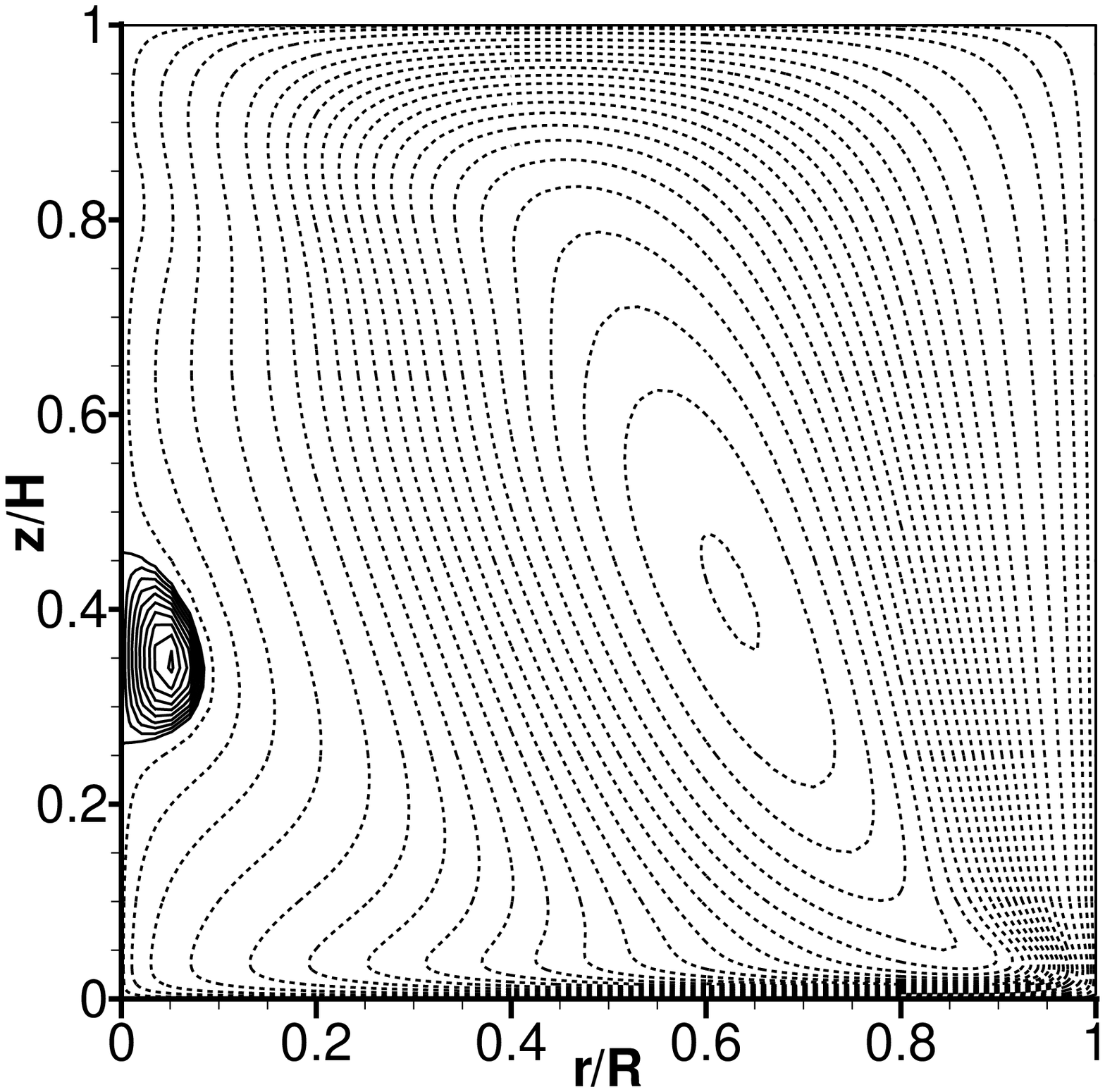}
  \caption{Contours of streamlines in a meridional plane. Left: case
    $(d):\ (\Lambda=1/3,\Rey = 2\,000)$; Right: case $(e):\
    (\Lambda=3,\Rey = 2\,000)$. The 30 contours are non-uniformly
    spaced for visualization purposes, 20 equally-spaced negative
    contours and 10 equally-spaced positive contours for $(d)$ and
    $(e)$.}\label{fig:streamlines-H13-H3}
\end{figure}

The very different flow patterns developed by both cases $(d)$ and
$(e)$ are highlighted by the very distinctive streamlines shown in
Fig.~\ref{fig:streamlines-H13-H3}. The shallow case $(d)$ yields two
vortex breakdown bubbles, which are off the cylinder axis while
remaining attached to the free surface. The recirculation is more
intense in the largest bubble, which is elongated enough to produce a
recirculation of the fluid from the free surface all the way down to
the rotating disk, and so forth. Consequently, this elongated bubble
completely separate the outer region of the flow ($r/R\geq 0.5$) from
the inner core, where the second less intense, vortex breakdown
occurs. Comparatively, the recirculation in case $(e)$ is fairly
limited. A vortex breakdown still occurs in this case, leading to the
formation of a small on-axis bubble, which is detached from the free
surface. One can notice from the bending of the streamlines near the
axis at the height $z/H=0.8$, that another vortex breakdowns is in
preparation---compare this effect to the similar effect on the
streamlines prior the vortex breakdown in the closed swirling flow at
$\Rey=1\,900$, see Bouffanais \cite{bouffanais07:_simul}.

\begin{landscape}
  \begin{figure}[ht!]
    \centering
    \includegraphics[width=0.34\textwidth]{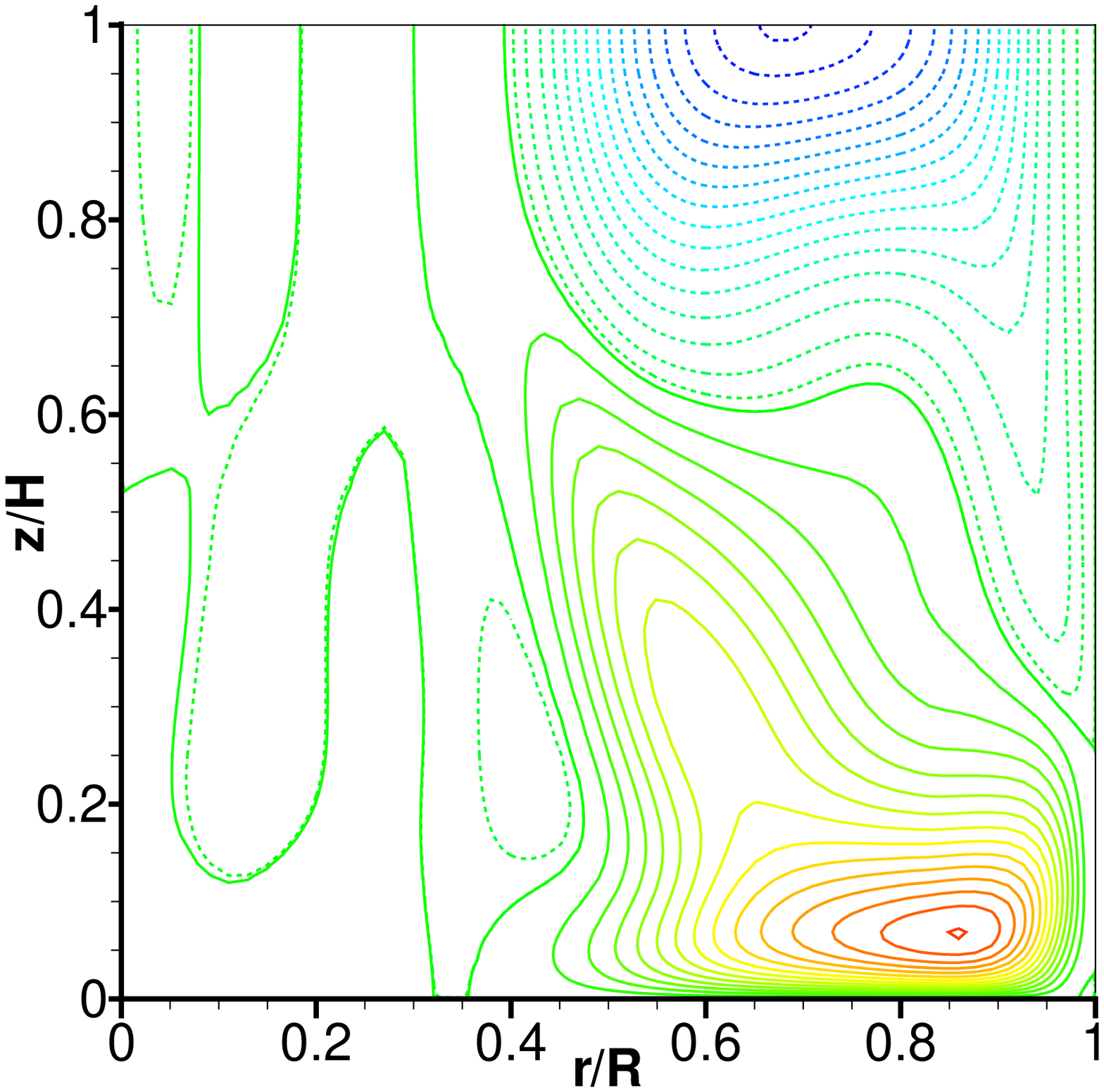}
    \includegraphics[width=0.34\textwidth]{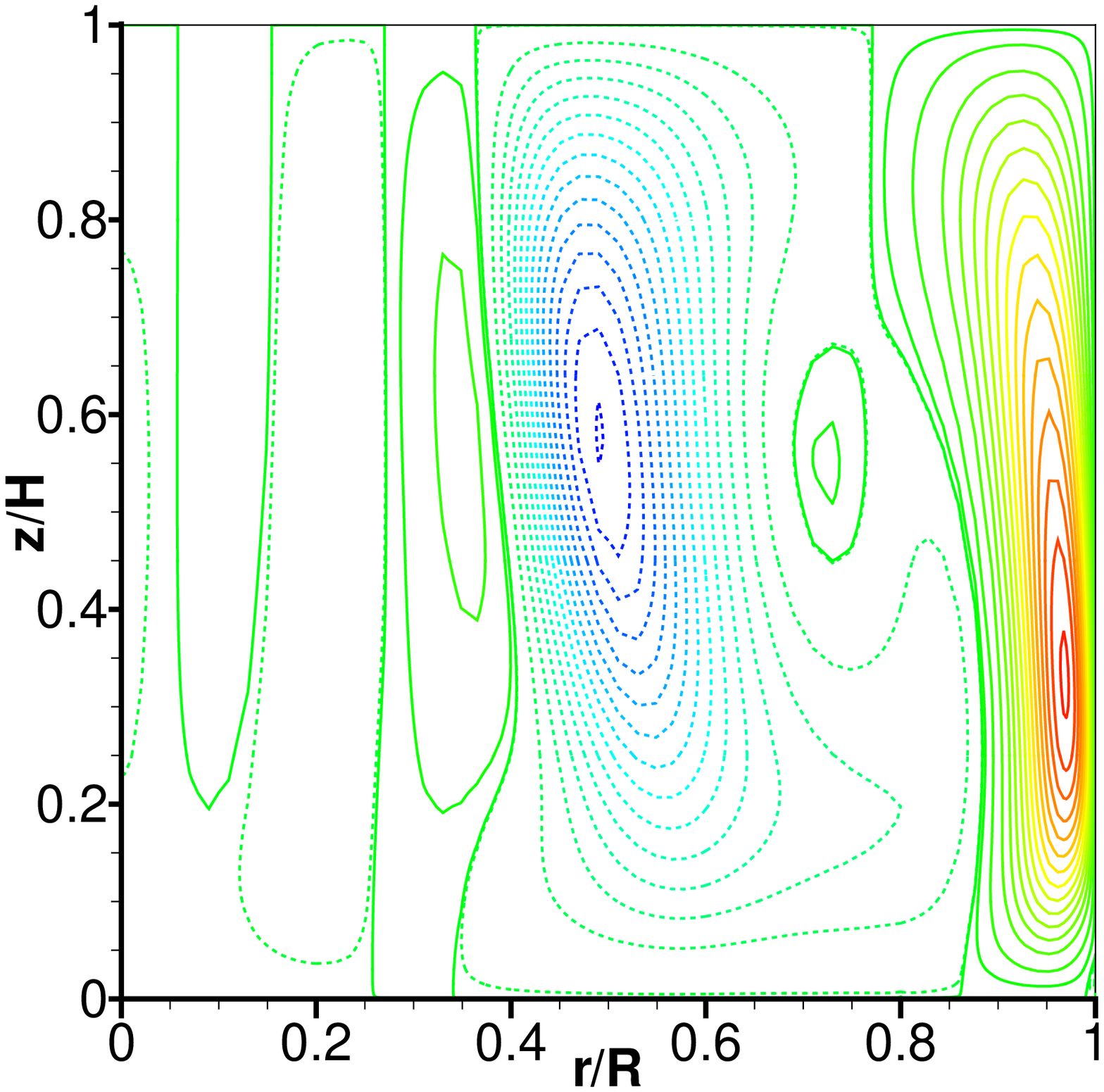}
    \includegraphics[width=0.34\textwidth]{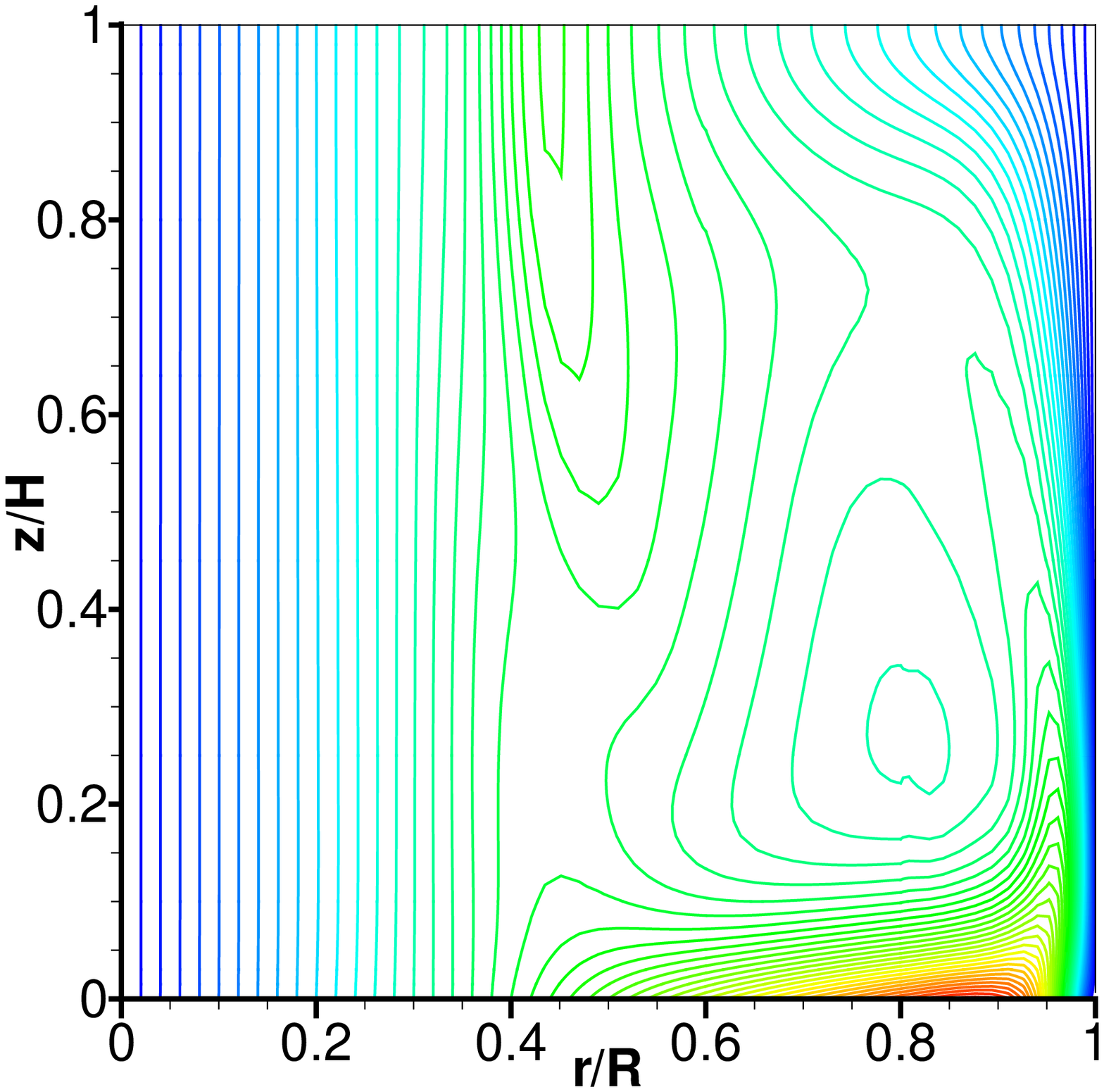}
    \includegraphics[width=0.34\textwidth]{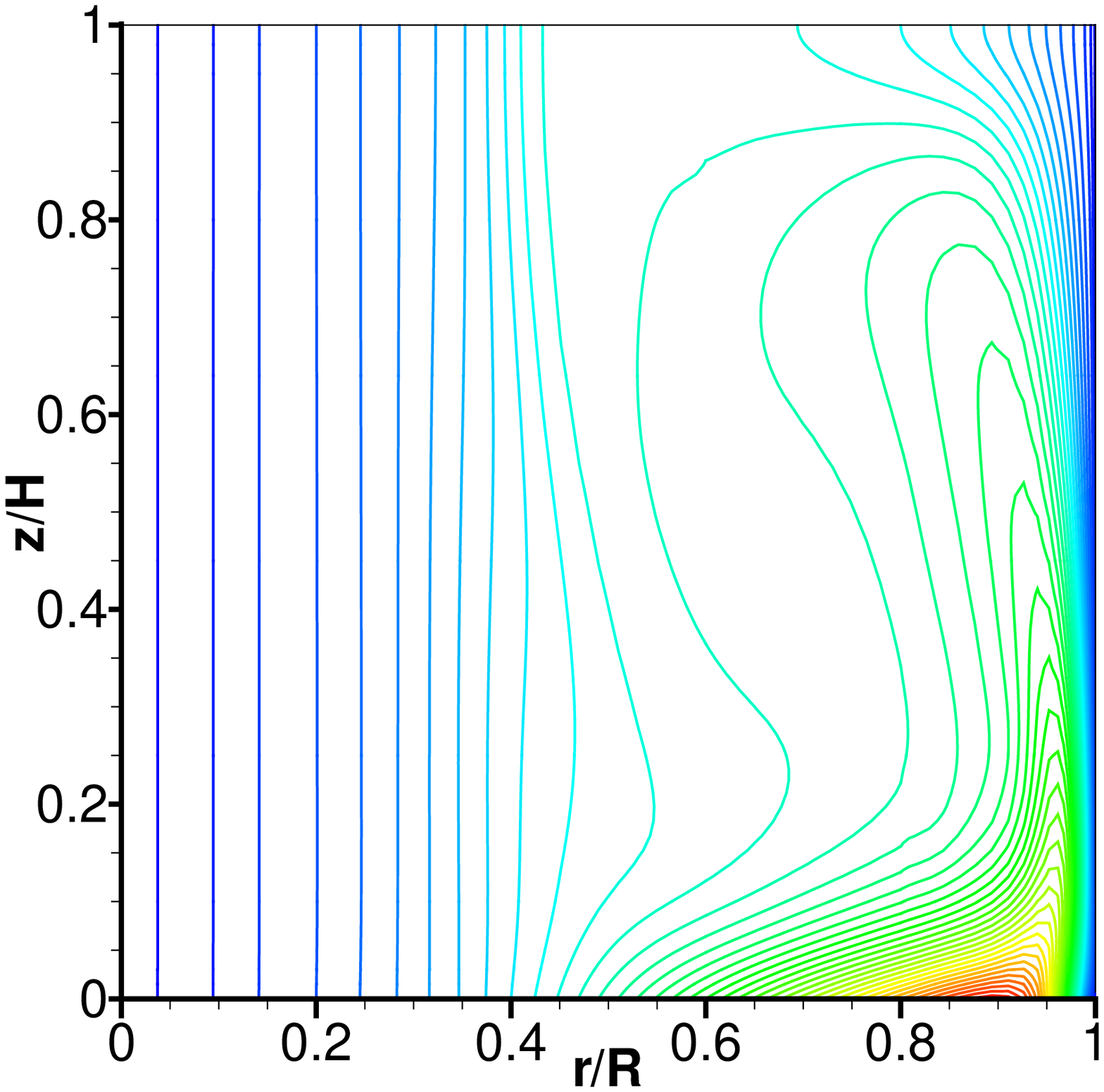}\\
    \includegraphics[width=0.34\textwidth]{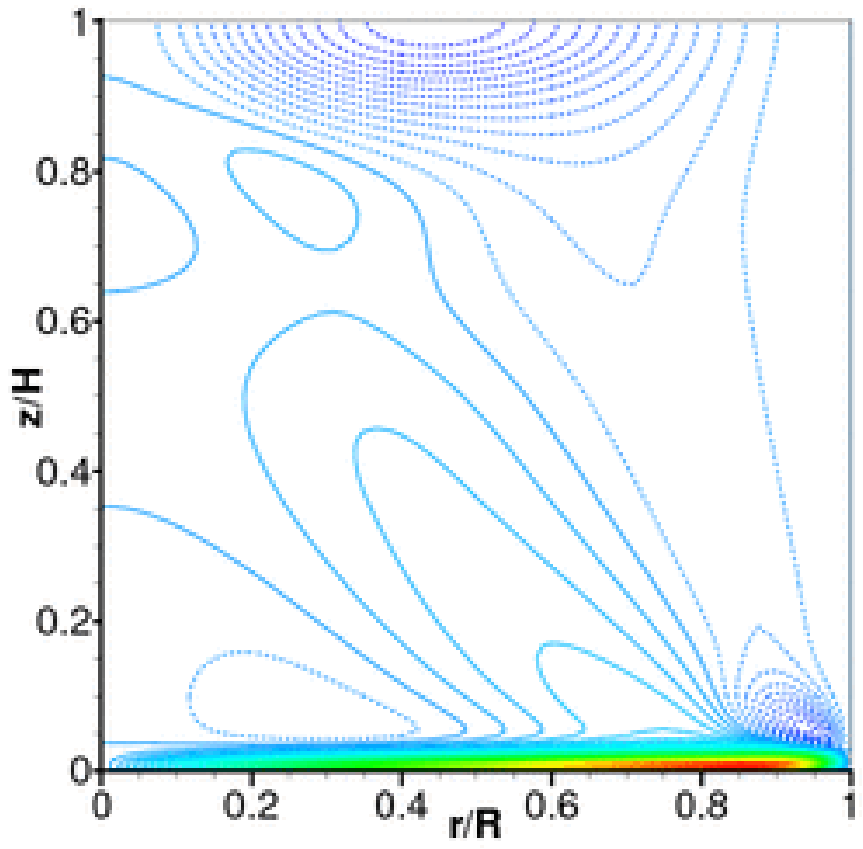}
    \includegraphics[width=0.34\textwidth]{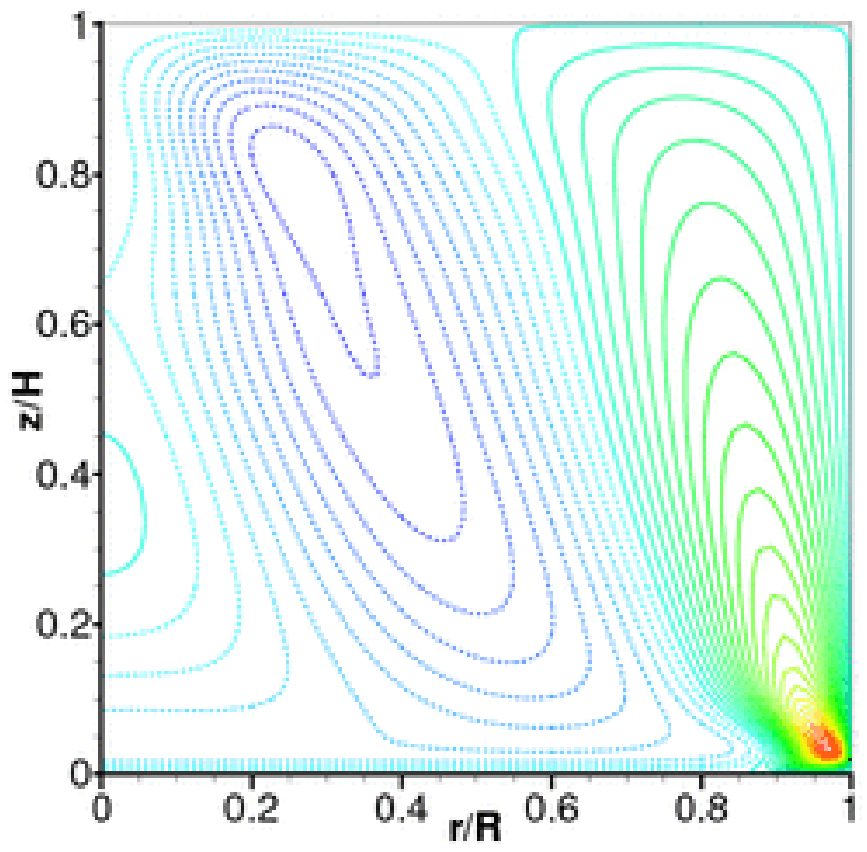}
    \includegraphics[width=0.34\textwidth]{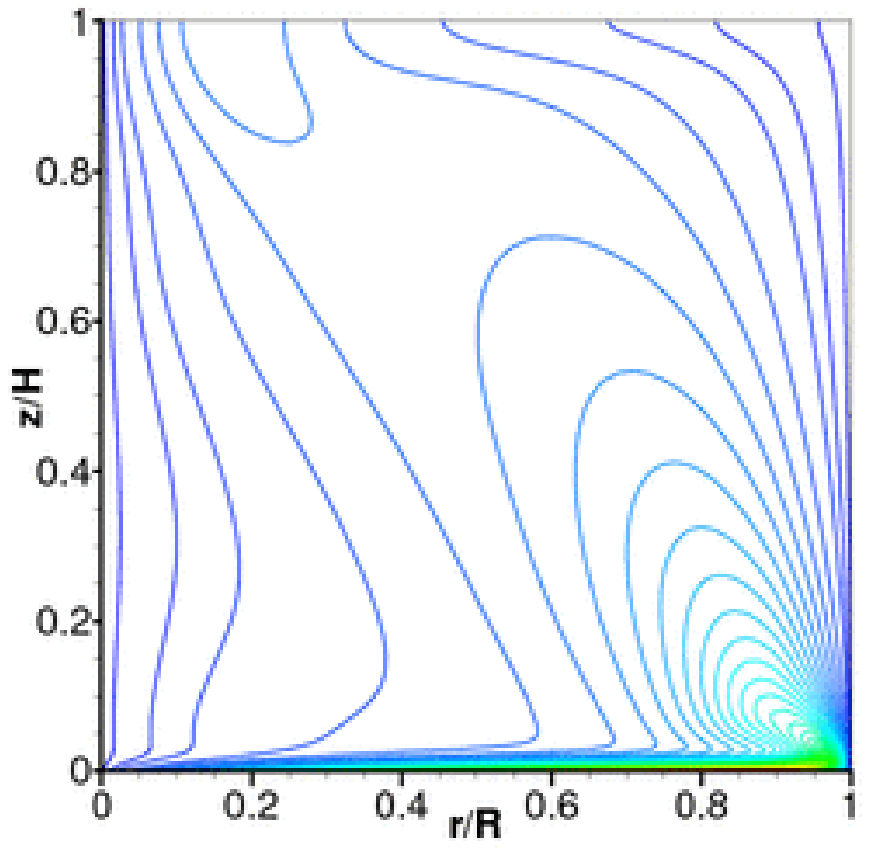}
    \includegraphics[width=0.34\textwidth]{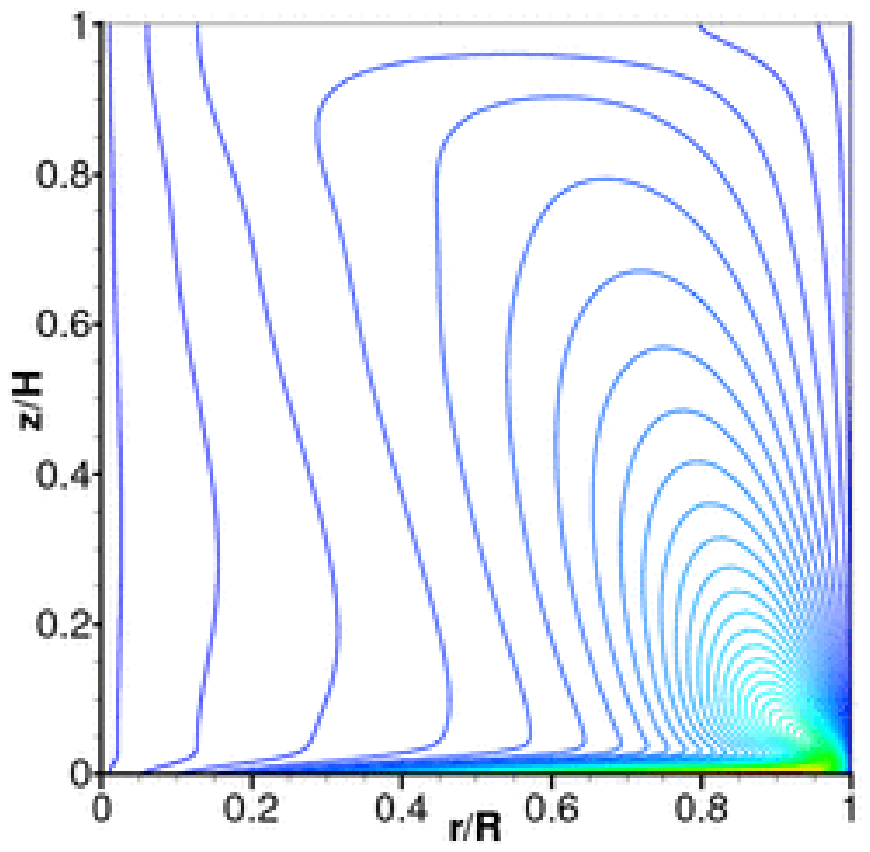}
    \caption{Contours in a meridional plane for the case $(d):\
      (\Lambda=1/3,\Rey = 2\,000)$ (top row) and case $(e):\
      (\Lambda=3,\Rey = 2\,000)$ (bottom row). From left column to
      right column: radial velocity component $u_r$; axial velocity
      component $w=u_z$; azimuthal velocity component $u_\theta$;
      axial angular momentum component $\Gamma = r u_\theta$. All
      contours are uniformly spaced; 35 contours between $-0.17$ and
      $0.15$ for $u_r$ case $(d)$; 90 contours between $-0.03$ and
      $0.15$ for $u_r$ case $(e)$; 35 contours between $-0.1$ and
      $0.12$ for $u_z$ case $(d)$; 35 contours between $-0.05$ and
      $0.12$ for $u_z$ case $(e)$; 50 contours between $0$ and $1$ for
      $u_\theta$ and $\Gamma$, cases $(d)$ and
      $(e)$.}\label{fig:all-v-H13-H3}
  \end{figure}
\end{landscape}

The shallow system $(d)=(\Lambda=1/3,\Rey=2\,000)$ possesses some very
distinctive features as can be seen in Fig.~\ref{fig:all-v-H13-H3}
(top row). The vortex lines for $r/R<0.4$ being aligned with the
rotation axis, one can easily conclude that the flow is essentially in
solid-body rotation in this inner core region of the cavity. The
meridional flow in this inner part of the cavity has a very weak
intensity as attested by the values close to zero of the contours of
the axial and radial velocity components---solid contour lines are
positive and negative contour lines being dashed. In contrast, for
$r>0.4$ the primary recirculation of the flow is intense and
predominates. The vortex lines bending is limited to this region and
again is at the origin of the vortex breakdown appearing near
$r/R=0.4$. The boundary layer on the rotating disk is limited to the
region $r/R>0.5$ and the internal jet-like shear layer close the
tubular side-wall has a structure quite different from the cases with
$\Lambda=1$. Indeed in this shear layer, the axial velocity is
relatively intense all along the tubular side-wall, unlike cases
$(a)$, $(b)$ and $(e)$, where the axial velocity $u_z$ decreases
rapidly with $z/H$. This observation is easily explained by the
shorter height in the case studied, but has several considerable
consequences on the flow itself. A more intense wall-jet implies a
more intense angular momentum jet at the free surface, which
facilitates the vortex breakdown phenomena. The intense angular
momentum free-surface jet produces an elongated recirculation bubble
located as seen earlier, around $r/R=0.4$. In this elongated bubble,
the axial velocity field is globally positive, thereby producing an
effect similar to the jet-like shear layer near the tubular
side-wall. In turn it generates a secondary angular momentum
free-surface wall jet responsible for the second vortex breakdown.

Increasing the height-to-radius aspect ratio to $\Lambda=3$, modifies
considerably the flow dynamics as can be seen in
Fig.\ref{fig:all-v-H13-H3} (bottom row). It seems clear from the
previous analysis for the shallow case $(d)$, that the influence of
the free surface on the flow is more important when $\Lambda$ is
small. The proximity between the driving disk, which generates the
primary flow and the free surface with its specific boundary
conditions, leads to the complex flow dynamics earlier
explained. Conversely, for large values of $\Lambda$ the important
distance between the spinning disk and the free surface is so
important that it significantly reduces the effect of the presence of
the free surface. The flow pattern presents \textit{in fine} a
structure very similar to the flow pattern observed in the closed
cylinder swirling flow, except very close to the free surface. As
mentioned earlier the recirculation bubble itself is fairly small and
located on the cylinder axis likewise in the closed cylinder
case. Finally, it is worth adding that the region of solid-body
rotation is almost completely eliminated. Even the closest-to-the-axis
vortex lines present some bending.

As a brief conclusion of the previous study of the steady and laminar
free surface swirling flows, it appears that the choice $\Lambda=1$
for the height-to-radius aspect ratio of the cavity in presence of a
free surface, ensures us to deal with a complex flow
dynamics. Different mechanisms are in competition in different regions
of the cavity, and in the end make the cases with $\Lambda=1$
physically more challenging and more interesting. This
conclusion---valid in presence of a free surface---stops being valid
for the closed cylinder swirling flow, and thereby explains the focus
in the literature on cases with $\Lambda\geq 2$.
\subsubsection{Unsteady flow}\label{sec:unsteady}
For sufficiently small Reynolds number and irrespective of $\Lambda$,
the basic flow state is stable. As noted by Lopez \etal\
\cite{lopez04:_symmet}, when \Rey\ is increased, the basic flow state
loses stability via a variety of Hopf bifurcations. It is worth noting
that when \Rey\ tends to infinity, the stream surfaces and vortex
surfaces---giving the streamlines and vortex lines by intersection
with a meridional plane---must coincide. At this point, the presence
of a flat free surface poses problem because of the constraint of
having orthogonal streamlines and vortex lines on it. This apparent
paradox is unraveled by simply letting the free surface
move. Nevertheless, we know from the experiments carried out by Spohn
\etal\ \cite{spohn98:_exper,spohn93:_obser}, that even at a
$\Rey=6\,000$ the tangential flow is extremely intense compared to the
normal one, leading to small free-surface deformations. It is very
likely that these small amplitude deformations are not sufficient to
solve our apparent paradox. At low Reynolds number, like those of
cases $(a)$, $(b)$, $(d)$, and $(e)$, the viscosity acts on the
velocity field to allow the latter condition of orthogonality to be
fulfilled. But when the Reynolds number is increased, the action of
viscosity and the limited deformation of the free surface are not
sufficient to bring back the orthogonality of the two sets of
lines. Therefore, the flow must either lose its axisymmetry or become
unsteady in order to allow to drop the orthogonality condition. The
experiments by Spohn \etal\ \cite{spohn98:_exper} suggest that the
open swirling flow first go through the unsteady path.

\begin{figure}[htbp]
  \centering
  \includegraphics[width=0.75\textwidth]{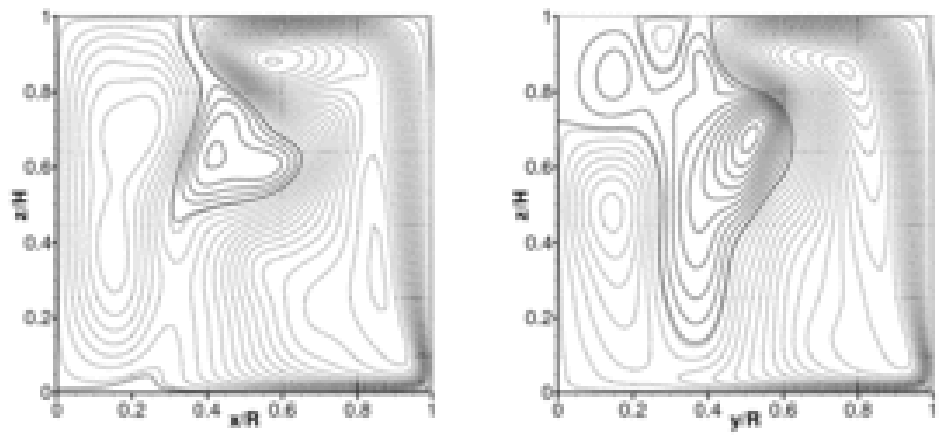}
  \includegraphics[width=0.75\textwidth]{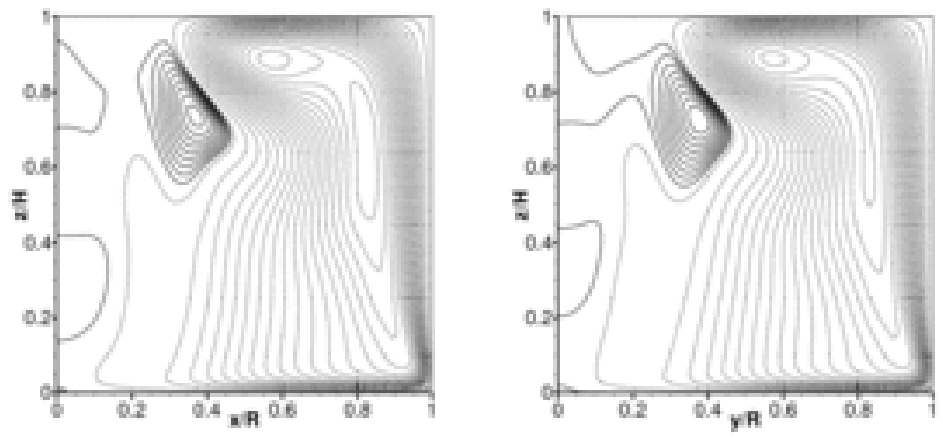}
  \caption{Contours of streamlines in two orthogonal meridional
    planes, case $\Lambda=1$ and $\Rey=6\,000$. Top row: instantaneous
    flow; Bottom row: mean flow. Left column: meridional plane
    $y/R=0$; Right column: meridional plane $x/R=0$.  The 30 contours
    are non-uniformly spaced for visualization purposes, 20
    equally-spaced negative contours and 10 equally-spaced positive
    contours.}\label{fig:orthogonal-meridian}
\end{figure}

In this section, the study is focused on the unsteady swirling flow
corresponding to case $(c)=(\Lambda=1,\Rey = 6\,000)$. To our
knowledge, such transitional regime at this relatively high \Rey\ has
never been investigated nor reported in the literature. At this
Reynolds number the loss of axisymmetry in this flat-free-surface case
is evident from the observation of the contours of streamlines in
Fig.~\ref{fig:orthogonal-meridian} (top row). As the flow is unsteady,
these recirculation bubbles are instantaneous and correspond to a flow
sample taken in the statistically-steady regime for $t>600$ in
$\Omega_0^{-1}$ units, see Fig.~\ref{fig:T-K-O-H}. The streamlines of
this flow sample are represented in two orthogonal meridian planes
corresponding to $y/R=0$ and $x/R=0$, in
Fig.~\ref{fig:orthogonal-meridian} (top row). Once again, the loss of
axisymmetry appears clearly from the complex and nonaxisymmetric
structure of the recirculation bubbles. Compared to the laminar and
steady cases $(a)$ and $(b)$, the recirculation bubbles have their own
dynamics and evolution. In a common approach to such unsteady
problems, this complex dynamics is analyzed by the means of an
averaging process, which is supplemented with an analysis of
instantaneous flow samples equally-spaced in time. The mean flow is
obtained by averaging 500 flow samples corresponding to successive
flow states extracted every 0.25 times units (or equivalently every
100 iterations). Subsequently, the root-mean-square (rms) fluctuations
of flow fields are calculated using the same extracted flow samples
and the mean flow field obtained earlier.

\begin{landscape}
\begin{figure}[htbp]
  \centering
  \includegraphics[width=0.34\textwidth]{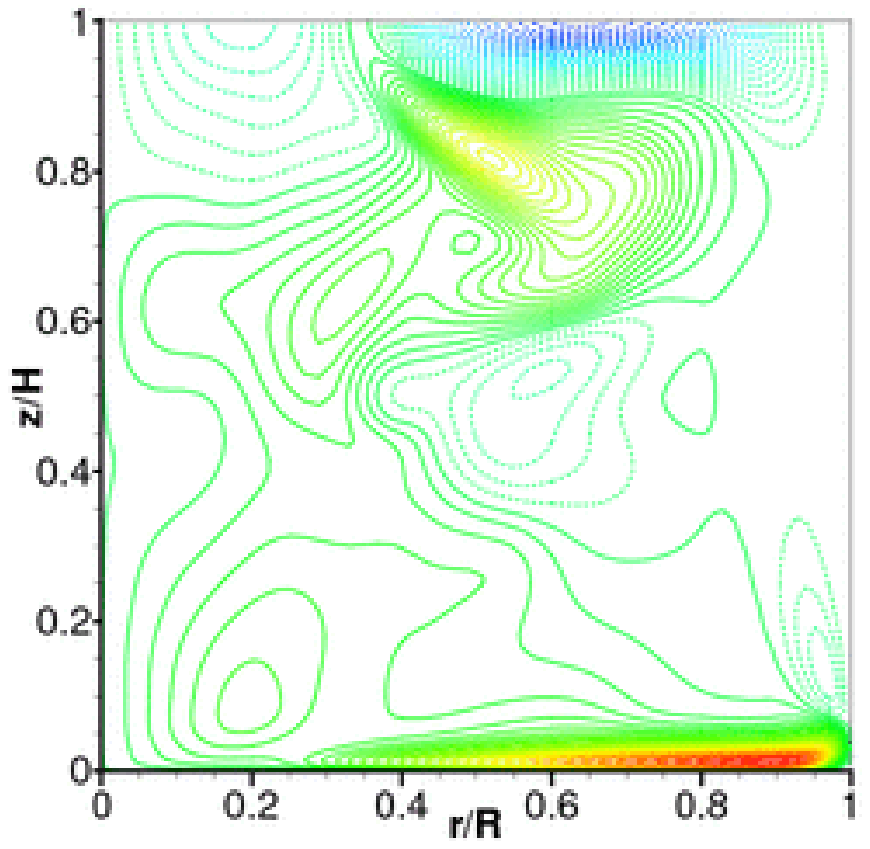}
  \includegraphics[width=0.34\textwidth]{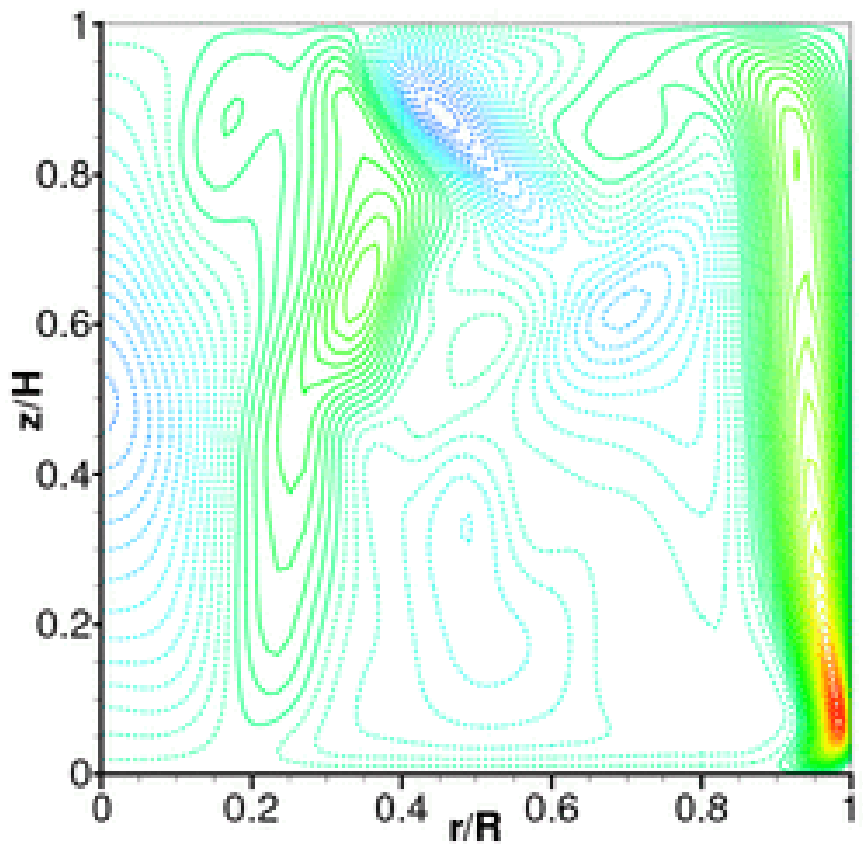}
  \includegraphics[width=0.34\textwidth]{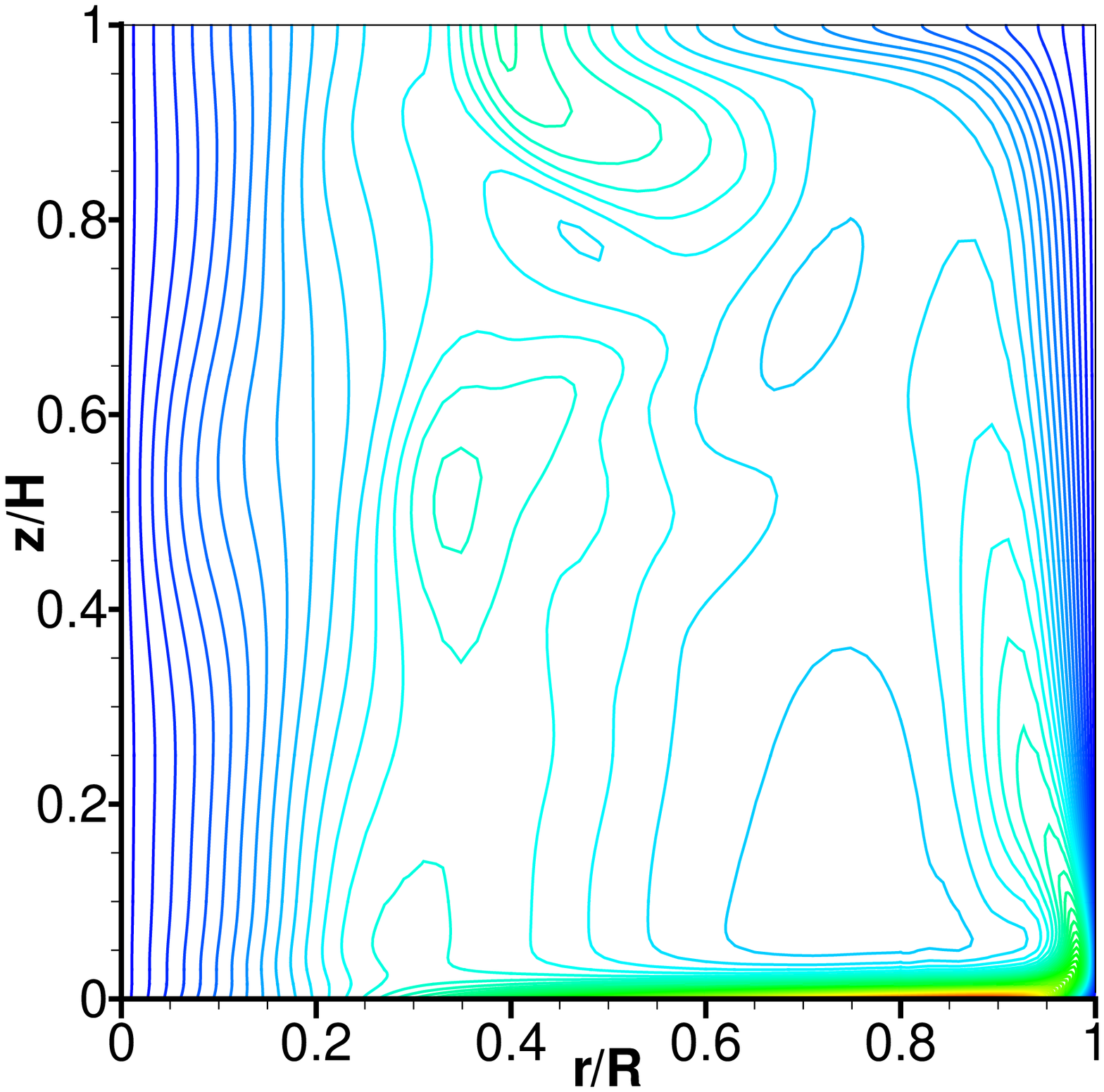}
  \includegraphics[width=0.34\textwidth]{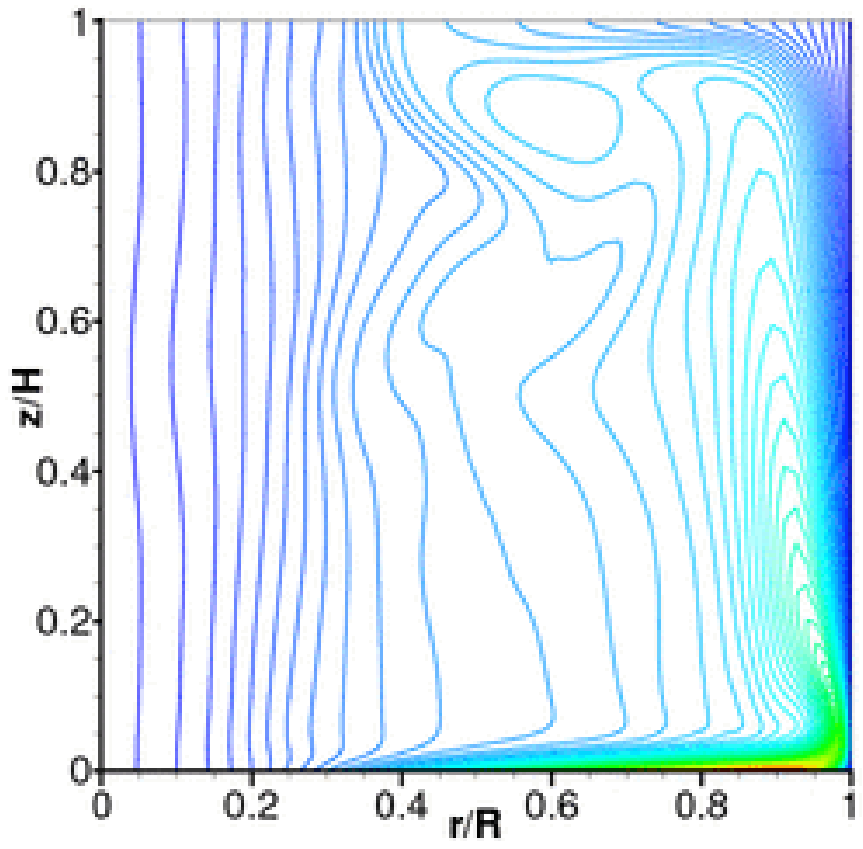}\\
  \includegraphics[width=0.34\textwidth]{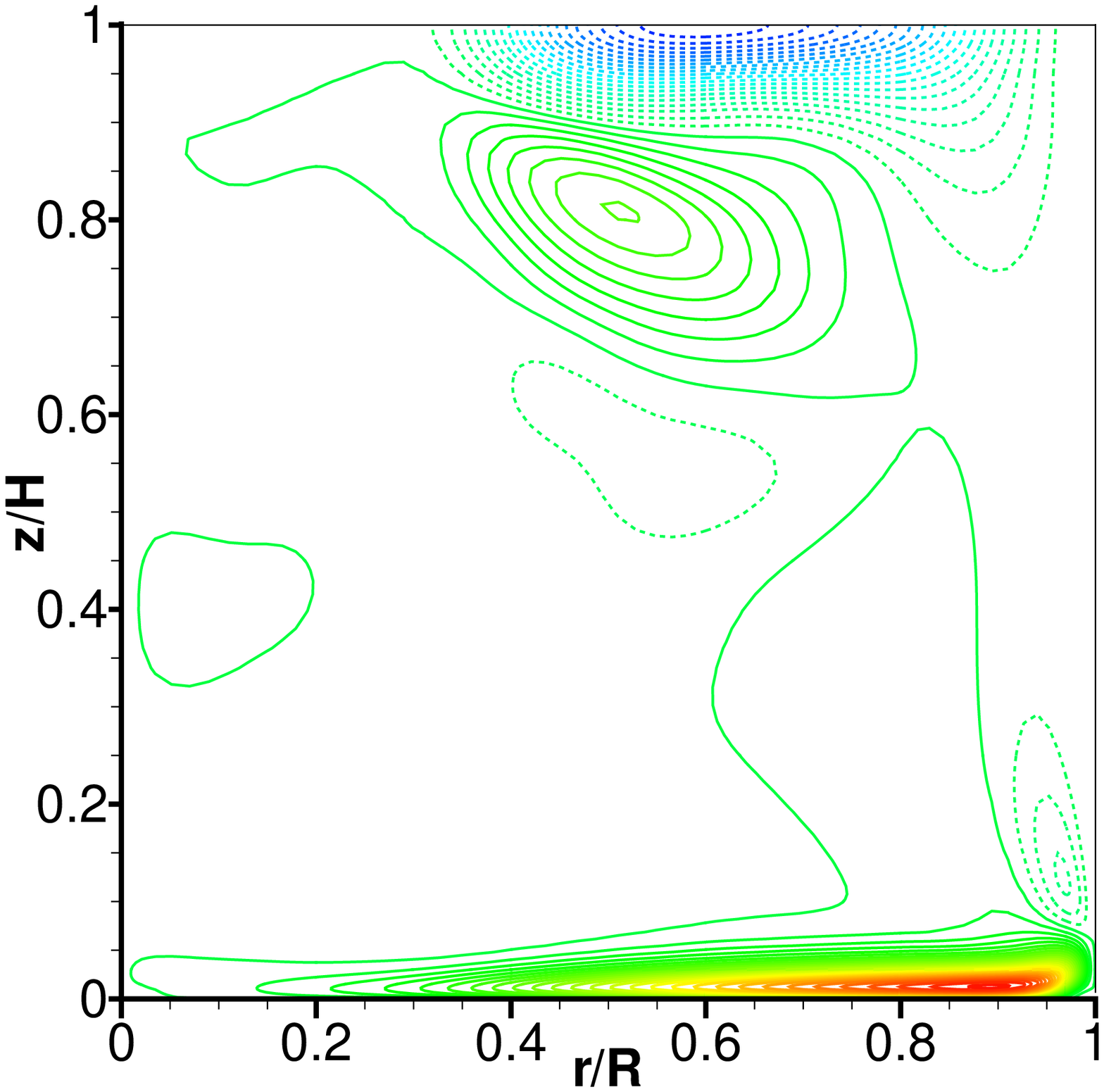}
  \includegraphics[width=0.34\textwidth]{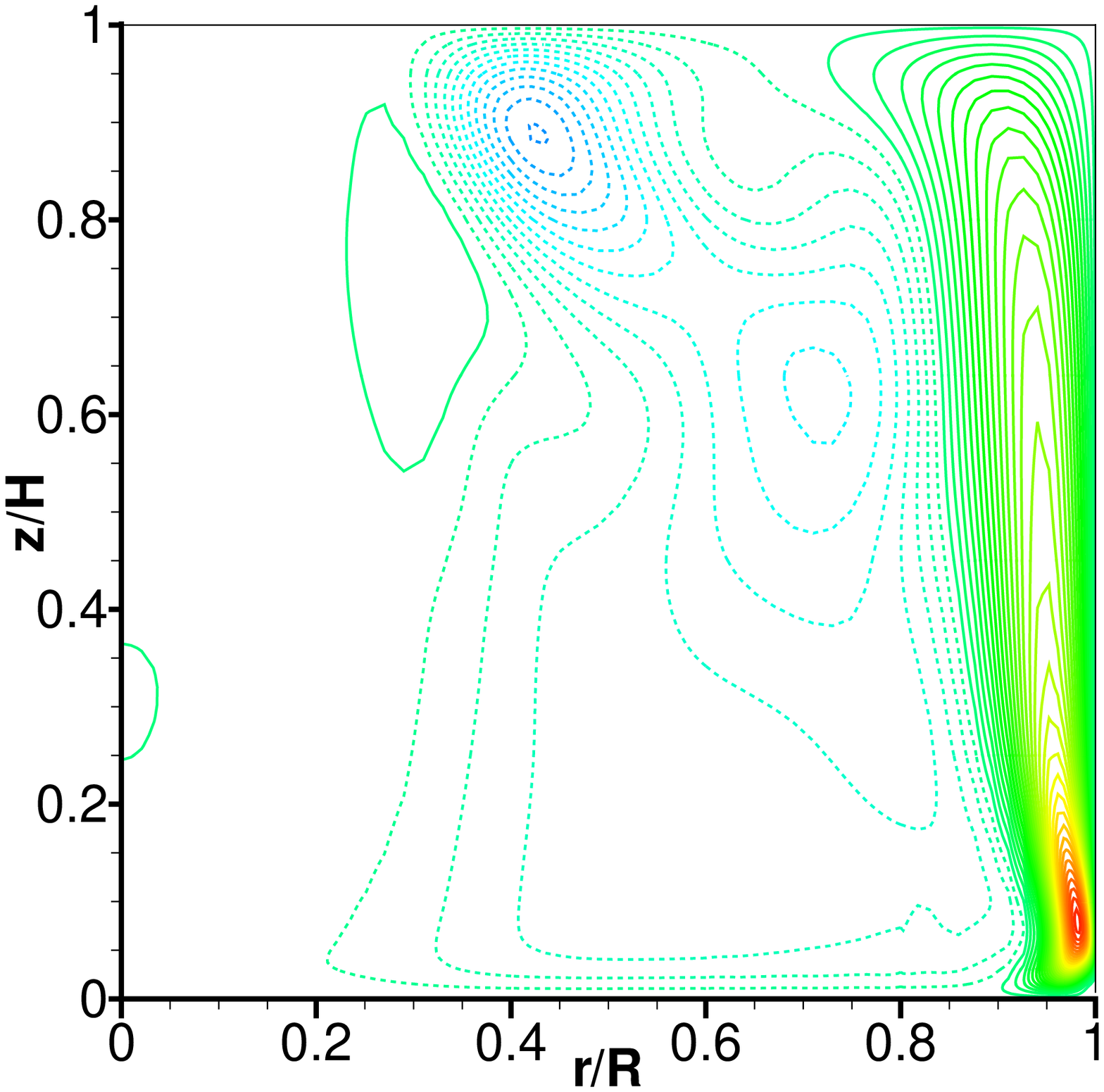}
  \includegraphics[width=0.34\textwidth]{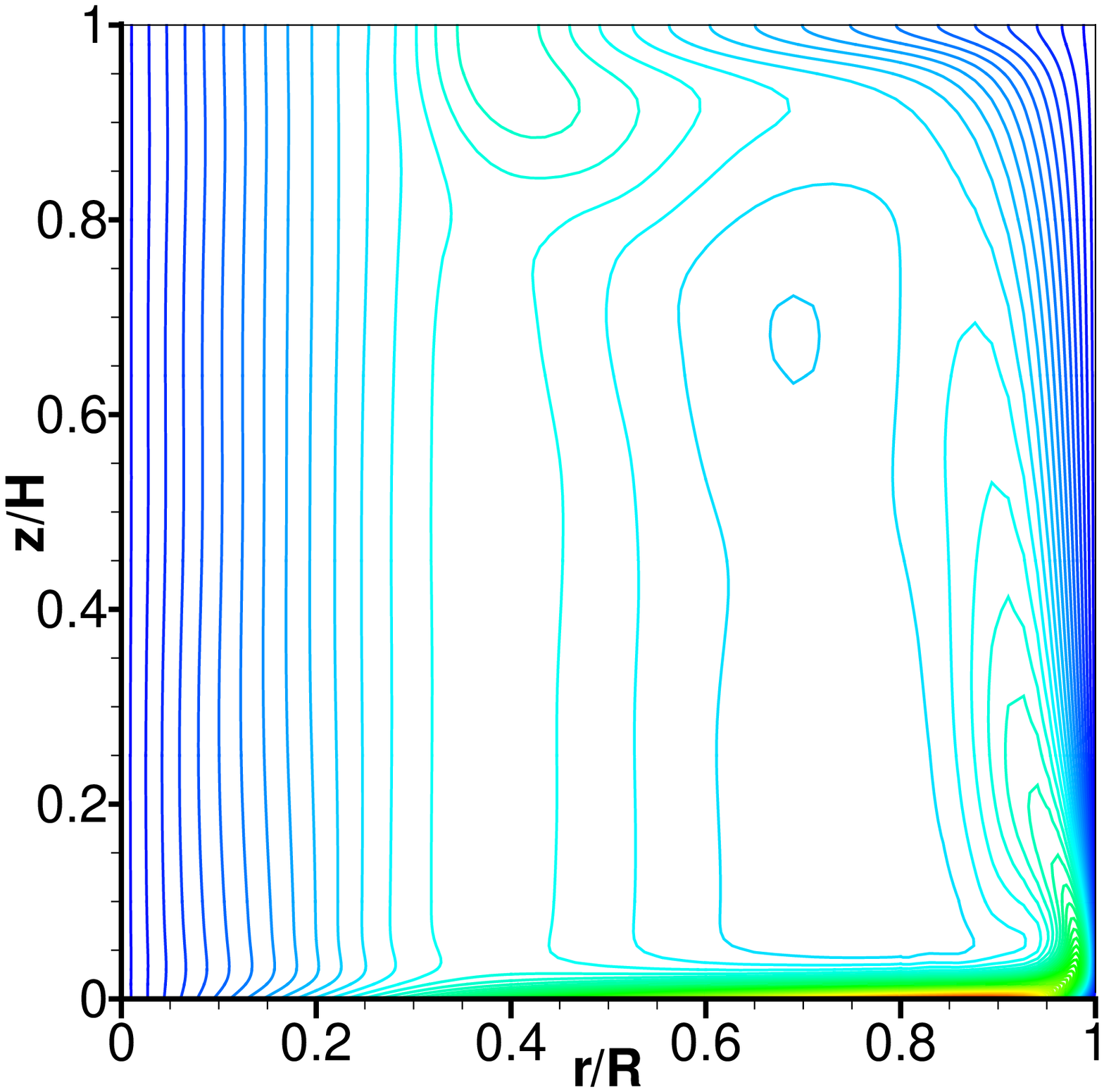}
  \includegraphics[width=0.34\textwidth]{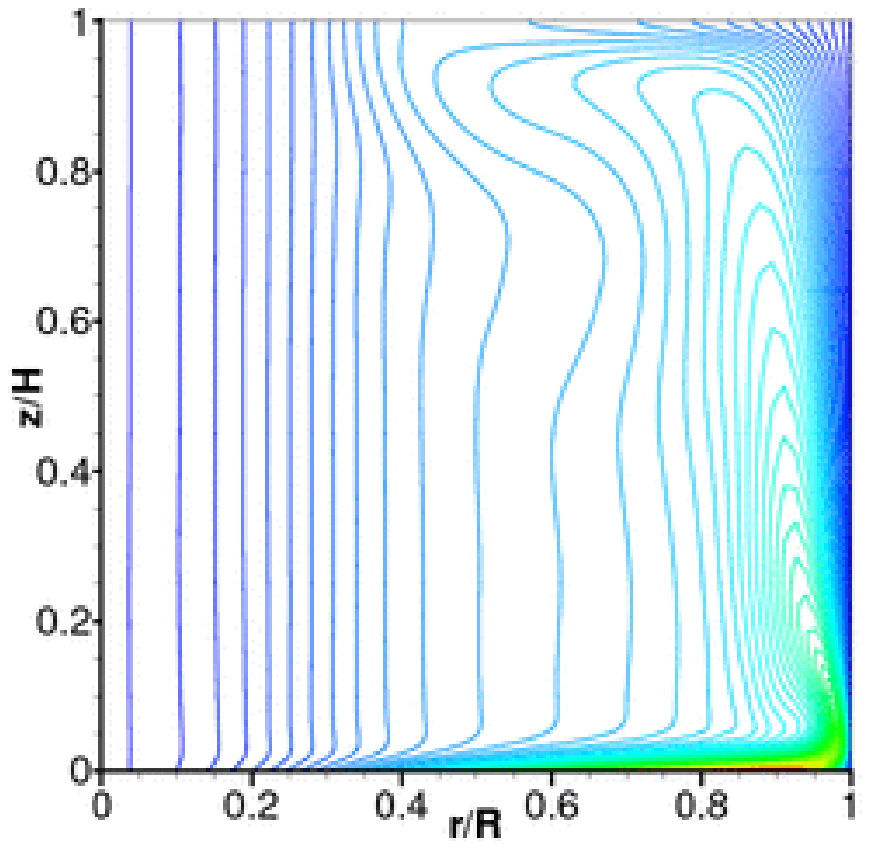}
  \caption{Contours in a meridional plane for the case $\Lambda=1$ and
    $\Rey = 6\,000$. Top row: instantaneous flow; Bottom row: mean
    flow. From left column to right column: radial velocity component
    $u_r$; axial velocity component $u_z$; azimuthal velocity
    component $u_\theta$; axial angular momentum
    $\Gamma=ru_\theta$. The 50 contours are uniformly spaced, between
    $-0.13$ and $0.16$ for $u_r$; between $-0.09$ and $0.14$ for
    $u_z$; and between $0$ and $1$ for $u_\theta$. The 100 contours
    are uniformly spaced between $0$ and $1$ for
    $\Gamma$.}\label{fig:ur-ut-uz-6000}
\end{figure}
\end{landscape}

The streamlines associated with the mean flow are shown in
Fig.~\ref{fig:orthogonal-meridian} (bottom row). The streamlines of
the mean flow reveal the existence of a toroidal recirculation bubble,
located off the cylinder axis and more surprisingly detached from the
free surface. The toroidal shape and off-axis location of the mean
recirculation bubble is in agreement with the increased-\Rey\ trend
observed with cases $(a)$ and $(b)$ in
Sec.~\ref{sec:steady}. Regarding the detachment from the free surface
of the mean bubble, it is more relevant here to notice that the
instantaneous bubbles are still attached to the free surface. More
precisely, one may notice two points:
\begin{itemize}
\item[--] in the meridional plane $y/R=0$, a small recirculation zone
  appears attached to the rotating disk for $r/R\simeq 0.2$;
\item[--] in the meridional plane $x/R=0$, the recirculation bubble is
  stretched from the free surface $z/H=1$ down to $z/H=0.15$, in a
  radial position $r/R\simeq 0.4$.
\end{itemize}
These two observations remind the streamline patterns described in the
case $(d)=(\Lambda =1/3 , \Rey = 2\,000)$, with a long bubble
stretching from the free surface down to the driving disk. The
previous analysis is further confirmed by the contours of the three
velocity components and of the axial angular momentum for both an
instantaneous flow sample and the mean flow, presented in
Fig.~\ref{fig:ur-ut-uz-6000}, in the meridional plane $x/R=0$. A
careful analysis of the vortex lines for the instantaneous flow sample
shows a bending in the whole meridian plane. This bending is very
significant in the region $0.3 \leq r/R \leq 0.8$ and $0.4 \leq z/H
\leq 1$, which corresponds to the limit between the primary
recirculation of the flow and the secondary recirculation bubble. On
the contrary, the vortex lines structure of the mean flow is as
expected much more regular. The inner core region of the flow $r/R
\leq 0.4$ displays a state of solid-body rotation. For both the
instantaneous and mean flow, the jet-like shear layer along the
tubular side-wall is turned into the interior of the flow by the free
surface. Compared to the previous cases $(a)$, $(b)$, $(d)$, $(e)$,
and also the closed swirling flow $(\Lambda=1, \Rey=6\,000)$ in
Bouffanais \cite{bouffanais07:_simul}, the structure of this shear
layer at $\Rey =6\,000$ reveals the presence of an intense radial jet
of angular momentum at the free surface.

The fluctuations of the flow with respect to its mean state have been
calculated with the same flow samples as before. It should be noted
that the fluctuation level corresponds to less than $5\%$ of the
maximal intensity of the respective mean flow fields. Despite the
relatively low level of fluctuation encountered, these fluctuations
are very localized in space as can be seen in
Fig.~\ref{fig:rms-ur-uz-ut-gamma-6000}. Similarly to the mean flow
fields, the rms fluctuations of the velocity field and of $\Gamma$
appear to be slightly nonaxisymmetric. All the three velocity
components present a noticeable level of fluctuation near the free
surface for radii close to $0.4$. In this region, the free-surface
radial jet of angular momentum reaches the inner flow, which is
solid-body rotation. These fluctuations are therefore located in the
vicinity of the stagnation point where the vortex breakdown is
initiated. The rms-fluctuations of the vortex lines, \ie\ $\Gamma$,
are the highest in the corner region between the free surface and the
tubular side-wall. It is in this corner, where the shear layer is
turned into the interior by the presence of the free surface.

\subsection{One-dimensional momentum budgets}\label{sec:one}
This section is devoted to the careful analysis of the momentum
balance for the radial, azimuthal and axial components. This study is
performed along different radial and axial lines within the cavity. It
is of interest to determine the predominant physical terms, which are
responsible for the complex flow dynamics depicted in the previous
sections. As noticed in these previous sections, the structure of the
flow in the inner core region is far different from the one close to
the tubular side-wall. Similarly, the flow above the rotating driving
disk has properties, which are not comparable to the ones below the
flat free surface. For the sake of conciseness, this momentum balance
analysis is limited to cases $(a)-(c)$ for which $\Lambda=1$.

\subsubsection{General considerations}
The numerical integration of the Navier--Stokes equations using the
spectral element method as described in Sec.~\ref{sec:computational},
is performed in Cartesian coordinates $(x,y,z)$ for the velocity
components $(u,v,w)$. Nevertheless, the axisymmetric nature of the
container and of the boundary conditions imposed to the flow suggests
the use of cylindrical coordinates. Indeed, the different physical
terms involved in the momentum equation represented here by the
Navier--Stokes equations---nonlinear advective term, viscous strain,
pressure gradient, etc.---are better apprehended when expressed in
cylindrical coordinates. Accordingly, all vectors and physical terms
are recast as functions of $(r,\theta,z )$, and for instance the
velocity components are $(u_r,u_\theta,u_z)$.

The complete expression of the momentum equations in cylindrical
coordinates reads
\begin{align}
  \ddp{u_r}{t} + u_r \ddp{u_r}{r} + \frac{u_\theta}{r} \ddp{u_r}{\theta} -\frac{u_\theta^2}{r}+ u_z \ddp{u_r}{z} & = -\ddp{p}{r} +\frac{1}{\Rey} \left[\frac{\partial}{\partial r}\left(\frac{1}{r}\ddp{(ru_r)}{r}  \right)+\frac{1}{r^2} \frac{\partial^2 u_r}{\partial \theta^2} -\frac{2}{r^2}\ddp{u_\theta}{\theta} + \frac{\partial^2 u_r}{\partial z^2}    \right],\label{eq:momentum-1}\\
  \ddp{u_\theta}{t} + u_r \ddp{u_\theta}{r} + \frac{u_\theta}{r} \ddp{u_\theta}{\theta} +\frac{u_r u_\theta}{r}+ u_z \ddp{u_\theta}{z} & = -\frac{1}{r}\ddp{p}{\theta}+\frac{1}{\Rey} \left[\frac{\partial}{\partial r}\left(\frac{1}{r}\ddp{(ru_\theta)}{r}  \right) +\frac{1}{r^2} \frac{\partial^2 u_\theta}{\partial \theta^2}+\frac{2}{r^2}\ddp{u_r}{\theta} + \frac{\partial^2 u_\theta}{\partial z^2}    \right],\label{eq:momentum-2}\\
  \ddp{u_z}{t} + u_r \ddp{u_z}{r} + \frac{u_\theta}{r}
  \ddp{u_z}{\theta} + u_z \ddp{u_z}{z} & = -\ddp{p}{z}+\frac{1}{\Rey}
  \left[\frac{1}{r}\frac{\partial}{\partial r}\left(r\ddp{u_z}{r}
    \right) +\frac{1}{r^2} \frac{\partial^2 u_z}{\partial \theta^2} +
    \frac{\partial^2 u_z}{\partial z^2} \right],\label{eq:momentum-3}
\end{align}
where successively appears, the velocity time derivative, the
nonlinear advective term, the pressure gradient and the viscous
strain. The central objective of this study is to compare the relative
importance of some of these terms along different lines. Equation
\eqref{eq:momentum-1} (resp. \eqref{eq:momentum-2}) represents the
momentum balance in the radial (resp. azimuthal) direction, and is
analyzed along four radial lines at four different heights $z/H=0.03,\
0.64,\ 0.95,\ 1$, ranging from right above the rotating disk up to the
free surface. Equation \eqref{eq:momentum-3} represents the momentum
balance in the axial direction, and is analyzed along three different
axial vertical lines at three radial positions $r/R=0.08,\ 0.48,\
0.98$, ranging from near the cylinder axis to near the tubular
side-wall.

For the sake of simplicity, some of the terms appearing in
\eqref{eq:momentum-1}--\eqref{eq:momentum-3} are identified and
denoted specifically in Table~\ref{tab:name-terms}. In the sequel, the
various graphs reporting the variations of these terms will use this
nomenclature.

\begin{table}[htbp]
\begin{center}
  \begin{tabular}{cc|cc|cc}
    Name & Expression & Name & Expression & Name & Expression\\
    \hline
    \hline
           & & & & & \\[-1ex]
    NL$r1$ & $\displaystyle{- u_r \ddp{u_r}{r}}$                                                                                                                        & NL$t1$ & $\displaystyle{-u_r \ddp{u_\theta}{r}}$ & NL$a1$ & $\displaystyle{-u_r \ddp{u_z}{r}}$\\[2ex]
    NL$r2$ & $\displaystyle{+\frac{u_\theta^2}{r}}$                                                                                                                    & NL$t2$ & $\displaystyle{-\frac{u_r u_\theta}{r}}$ & NL$a2$ & $\displaystyle{-u_z \ddp{u_z}{z}}$ \\[2ex]
    NL$r3$ & $\displaystyle{-u_z \ddp{u_r}{z}}$                                                                                                                         & NL$t3$ & $\displaystyle{- u_z \ddp{u_\theta}{z}}$ & PG$a$ & $\displaystyle{-\ddp{p}{z}}$ \\[2ex]
    PG$r$  & $\displaystyle{-\ddp{p}{r}}$                                                                                                                               & VS$t1$ & $\displaystyle{\frac{1}{\Rey}\left[\frac{\partial}{\partial r}\left(\frac{1}{r}\ddp{(ru_\theta)}{r}\right)\right]}$ & VS$a1$ & $\displaystyle{\frac{1}{\Rey}\left[\frac{1}{r}\frac{\partial}{\partial r}\left(r\ddp{u_z}{r}\right)\right]}$ \\[2ex]
    VS$r$  & $\displaystyle{\frac{1}{\Rey} \left[\frac{\partial}{\partial r}\left(\frac{1}{r}\ddp{(ru_r)}{r}\right)+\frac{\partial^2 u_r}{\partial z^2}\right]}$ & VS$t2$ & $\displaystyle{\frac{1}{\Rey}\left(\frac{\partial^2 u_\theta}{\partial z^2}\right)}$ & VS$a2$ & $\displaystyle{\frac{1}{\Rey}\left(\frac{\partial^2 u_z}{\partial z^2}\right)}$ \\[2ex]
    \hline
  \end{tabular}
 \end{center}
 \caption{Name definitions of different terms appearing in the momentum budget equations \eqref{eq:momentum-1}--\eqref{eq:momentum-3}.}\label{tab:name-terms}
\end{table}

\subsubsection{Steady flows}\label{sec:momentum-steady}
As discussed in Sec.~\ref{sec:steady}, the steady swirling flows are
also fully axisymmetric. The first-order time derivative of the
velocity fields is identically zero in
Eqs.~\eqref{eq:momentum-1}--\eqref{eq:momentum-3}. The axisymmetric
property makes the velocity field independent of the azimuthal angle
$\theta$, and consequently all partial derivatives with respect to
this variable vanish. Therefore, the balance in the momentum equation
solely involves the terms described in Table~\ref{tab:name-terms}.

\begin{figure}[ht!]
  \centering
  \includegraphics[width=0.6\textwidth]{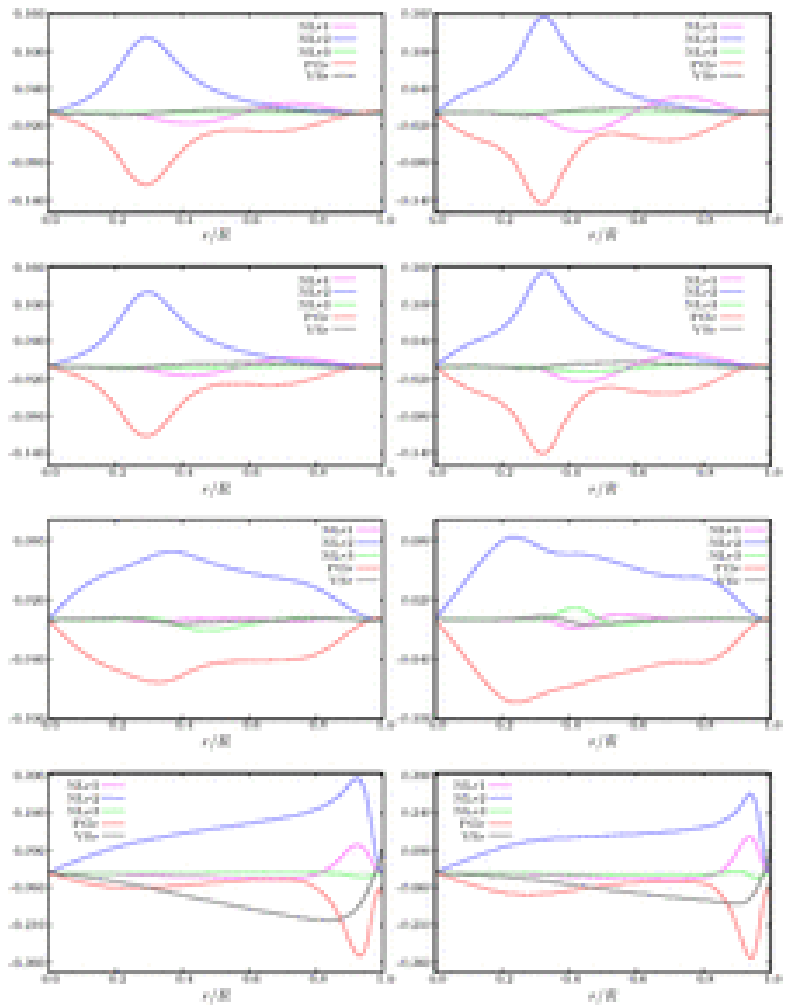}  
  \caption{Momentum balance in radial direction plotted along
    horizontal radial lines at four different vertical
    positions. First row: $z/H=1$; second row: $z/H=0.95$; third row:
    $z/H=0.64$; fourth row: $z/H=0.03$. Left column: case $\Rey=900$;
    Right column: case $\Rey=1\,500$. Case $\Lambda=1$. The
    terminology refers to
    Tab.~\ref{tab:name-terms}.}\label{fig:radial-900-1500}
\end{figure}

As a first step, the momentum balance in the radial direction is
presented in Fig.~\ref{fig:radial-900-1500} for case $(a)$ (left
column) and case $(b)$ (right column). At the free surface $z/H=1$
(top row), the viscous terms are insignificant, and the flow is driven
by the radial pressure deceleration, which is mainly counterbalanced
by the centrifugal acceleration NL$r2=u_\theta^2/r$ and to a certain
extent by NL$r$1. This analysis at the free surface still holds below
the free surface at $z/H=0.94$ and at $z/H=0.64$. At this latter
height, both the radial pressure deceleration and the centrifugal
acceleration NL$r2$ have a lower magnitude than at the free surface
but their magnitude is less localized than at the free surface. As
expected, above the disk, at $z/H=0.03$, all the terms have a higher
magnitude and the momentum balance is more complex as only one single
term NL$r3$ does not really contribute to the balance. The centrifugal
acceleration NL$r2$ keeps its predominant position, but its maximum is
now shifted towards the outer radial region, which corresponds to the
region of highest angular momentum $0.8\leq r/R \leq 1$. The other
acceleration term NL$r1$ becomes relatively important. Both of these
acceleration terms are counterbalanced by the radial pressure gradient
and now also by the viscous strain. The importance of the viscous
strain at this height $z/H=0.03$ can easily be understood, as we are
located in the viscous layer generated by the motion of the disk. One
may add that in the inner core region of the flow, say $r/R<0.15$, the
linear trend observed for both the centrifugal acceleration and the
pressure deceleration are well-know features of a flow in solid-body
rotation, as observed with the vertical vortex lines in
Fig.~\ref{fig:vr-vz-vt-gamma-900-1500} (extreme right column).

The evolution of those momentum balances along radial lines at
different heights does not really change when increasing \Rey\ from
900 up to $1\,500$. But some noticeable trends are observable. For
instance, even if the viscous strain does not play a central role at
the free surface, it is worth noting that its effect is increased with
\Rey\, while conversely it is decreased when getting closer to the
rotating bottom end-wall.

\begin{figure}[htbp]
  \centering
  \includegraphics[width=0.6\textwidth]{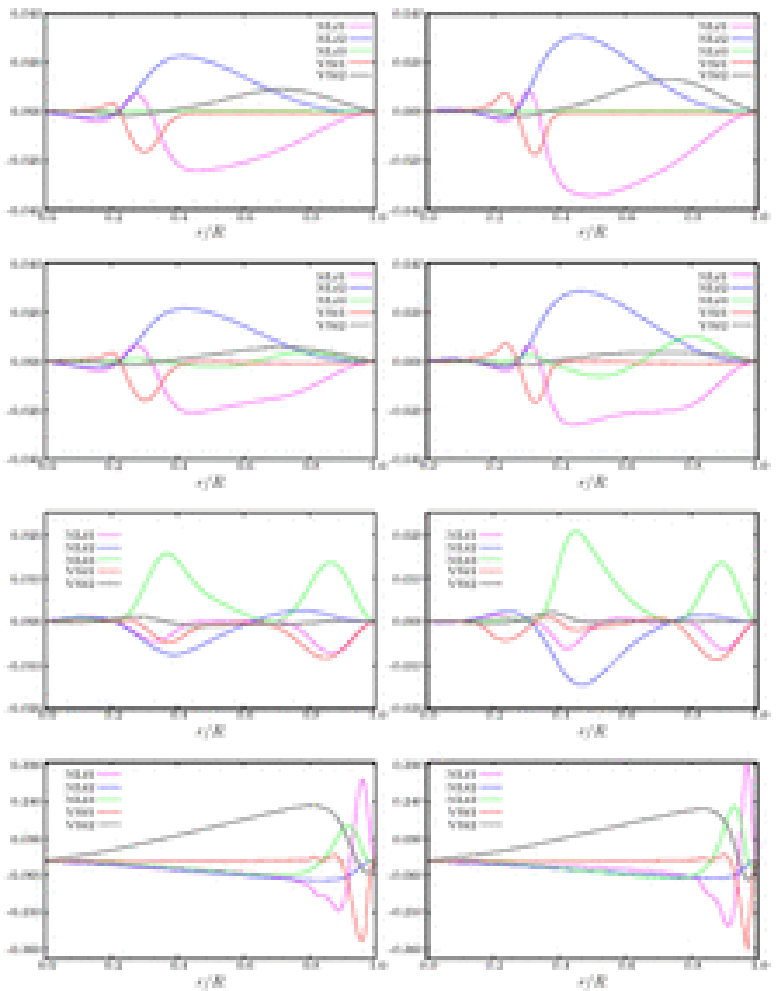}
  \caption{Momentum balance in azimuthal direction plotted along
    horizontal radial lines at four different vertical
    positions. First row: $z/H=1$; second row: $z/H=0.95$; third row:
    $z/H=0.64$; fourth row: $z/H=0.03$. Left column: case $\Rey=900$;
    Right column: case $\Rey=1\,500$. Case $\Lambda=1$. The
    terminology refers to
    Tab.~\ref{tab:name-terms}.}\label{fig:azimuthal-900-1500}
\end{figure}

As a second step, we aim at analyzing the momentum balance in the
azimuthal direction along radial lines at the same different heights
as before. The results are presented in
Fig.~\ref{fig:azimuthal-900-1500} for case $(a)$ (left column) and
case $(b)$ (right column). It is important to note at this point that
the two terms NL$t1$ and VS$t1$ involve partial derivatives with
respect to the radial variable $r$. Given the fact that our solution
is continuous and first-order differentiable within a spectral element
and only continuous at the element edges, one expects some slight
unphysical deformations of the plots associated with these two
terms. Along the radial lines of interest, the spectral element edges
are located at $r/R=0.2,\ 0.4,\ 0.6,\ 0.8,\ 0.97$. As a consequence,
some rapid variations of the terms VS$t1$ and NL$t1$ are going to be
simply disregarded in the coming discussions.

A rapid glance at all the plots in Fig.~\ref{fig:azimuthal-900-1500}
allows to conclude that the importance of the azimuthal momentum
transfers resides in the near bottom end-wall region. The magnitude of
all terms is over ten times smaller at $z/H=0.64,\ 0.95,\ 1$, compared
to $z/H=0.03$. Given the solid-body rotation in the inner core region
of the flow $r/R<0.15$, most of the terms are vanishing
small---excluding the unphysical values of VS$t1$.

At $z/H=0.03$, one can notice the vigorous action of the viscous
strain term VS$t2$ which literally drives the fluid in the viscous
Ekman layer. This driving viscous term is being compensated by the
convective terms NL$t1$ and NL$t3$, and by the Coriolis term
NL$t2$. As one gets closer to the corner between the rotating disk and
the tubular side-wall, say $r/R\geq 0.8$, the interplay between the
various terms is being reversed. The term VS$t2$, which is driving the
fluid in the inner region of the cavity is now a dissipative term in
the jet-like shear layer. Conversely, the convective term NL$t1$
becomes large and is driving the fluid in the shear layer. Very close
to the tubular side-wall, this term starts being counterbalanced by
the second viscous term VS$t1$.

The viscous driving effect of the term VS$t2$ becomes insignificant at
$z/H=0.64$, but when $z/H$ is increased, VS$t2$ starts growing again
to reach a local maximum value at the free surface, but with a
magnitude slightly smaller than the two other nonlinear convective
terms NL$t1$ and NL$t2$. Close to the free surface and at the free
surface, the flow is primarily driven by the Coriolis term NL$t2$
together with the viscous term VS$t2$. Their global action is
counterbalanced by the nonlinear convective term NL$t1$. This momentum
balance at the free surface in the outer region $r/R\geq 0.4$ reflects
the central effect of the free-surface jet of angular momentum. Again
the viscous effects are more intense at the free surface when \Rey\ is
increased from $900$ to $1\,500$, despite the presence of the
kinematic viscosity term $1/\Rey$ in their definitions.

\begin{figure}[htbp]
  \centering
  \includegraphics[width=0.6\textwidth]{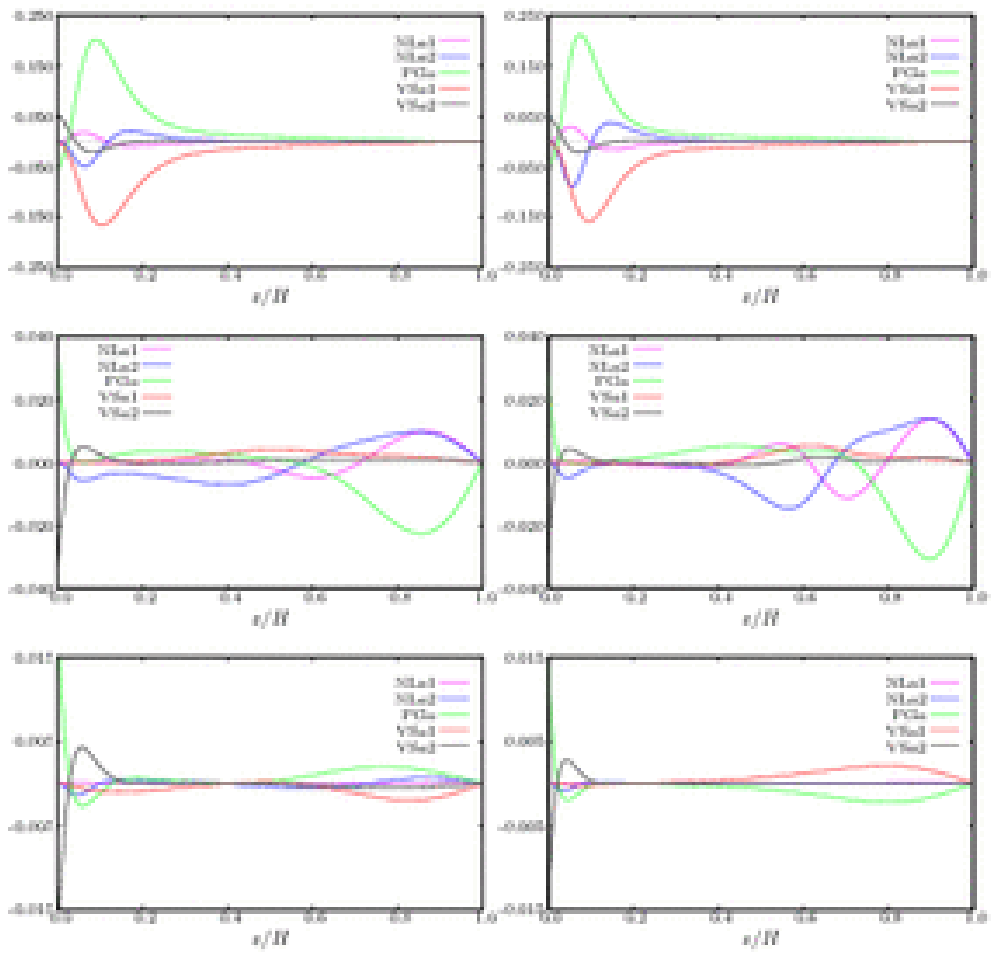}
  \caption{Momentum balance in axial direction plotted along vertical
    lines at three different radial positions. First row: $r/R=0.98$;
    second row: $r/R=0.48$; third row: $r/R=0.08$. Left column: case
    $\Rey=900$; Right column: case $\Rey=1\,500$. Case
    $\Lambda=1$. The terminology refers to
    Tab.~\ref{tab:name-terms}.}\label{fig:axial-900-1500}
\end{figure}

As a last step for these two steady flows $(a)$ and $(b)$, we
investigate the momentum transfer in the axial direction, but now
along three different vertical lines corresponding to three different
radii $r/R=0.08,\ 0.48,\ 0.98$. The graphs for the five different
terms involved are reported in Fig.~\ref{fig:axial-900-1500}. We start
from the top row, which is associated with the results for $r/R=0.98$
in the jet-like shear layer and which present the highest magnitudes
of all radii considered. The lack of axial momentum transfers is
clearly visible in the region $z/H\geq 0.4$ and even inexistent at the
free surface. Conversely, in the corner between the rotating bottom
end-wall and the tubular side-wall the flow is driven by the axial
pressure gradient, and to some extent by the two nonlinear terms
NL$a1$ and NL$a2$ independently. The viscous term VS$a1$ is primarily
responsible for counterbalancing the driving pressure effects. The
second viscous term VS$a2$ has a less important effect in terms of
magnitude but is driving very close to the disk (viscous Ekman layer),
when $z/H$ is increased it becomes dissipative as we are out of the
Ekman layer but within the wall-jet shear layer.

Near the medium radial position $r/R=0.48$, the axial momentum
transfers near the free surface show the relative importance of the
two nonlinear terms NL$a1$ and NL$a2$, which are driving the fluid
against the axial pressure gradient, which is negative as a
consequence of the kinematic boundary condition imposing a vanishing
axial velocity component at the free surface.

Near the cylinder axis, for $r/R=0.08$, the axial momentum transfers
are limited and relatively simples. Nonlinear convective effects seem
almost insignificant and the axial component of the flow is obtained
from the balance between the axial pressure gradient and the two
viscous terms: VS$a2$ in the bottom of the cavity and VS$a1$ in the
top of the cavity.

\subsubsection{Unsteady flow}
The momentum balance analysis developed in the previous section for
the two steady flows $(a)$ and $(b)$ cannot be directly transposed to
the unsteady case $(c)$. The two reasons for that are first the
unsteady character requiring to account for the first-order time
derivative $\partial \uu /\partial t$. The second reason is the loss
of axisymmetry of case $(c)$ imposing to account for all the terms
involving a partial derivative with respect to $\theta$ in
Eqs.~\eqref{eq:momentum-1}--\eqref{eq:momentum-3}. One can overcome
the issue associated with $\partial \uu /\partial t$ by performing the
analysis on the mean flow, which is obtained from the statistically
steady regime. On the other hand, the momentum equations for the mean
flow involve the Reynolds stress terms, expressing the influence of
the fluctuating velocity field onto the dynamics of the mean
field. Despite all these considerations and issues, we have
deliberately omitted the terms involving derivatives with respect to
the time and to the azimuthal coordinate, and we have calculated the
values of the terms in Table~\ref{tab:name-terms} on the same lines as
in Sec.~\ref{sec:momentum-steady}. By doing so, the objective is not
to reproduce a similar analysis as with the steady cases, but more to
investigate the evolution of the different terms for the instantaneous
flow and the mean flow as compared to the laminar cases $(a)$ and
$(b)$.

\begin{figure}[htbp]
  \centering
  \includegraphics[width=0.6\textwidth]{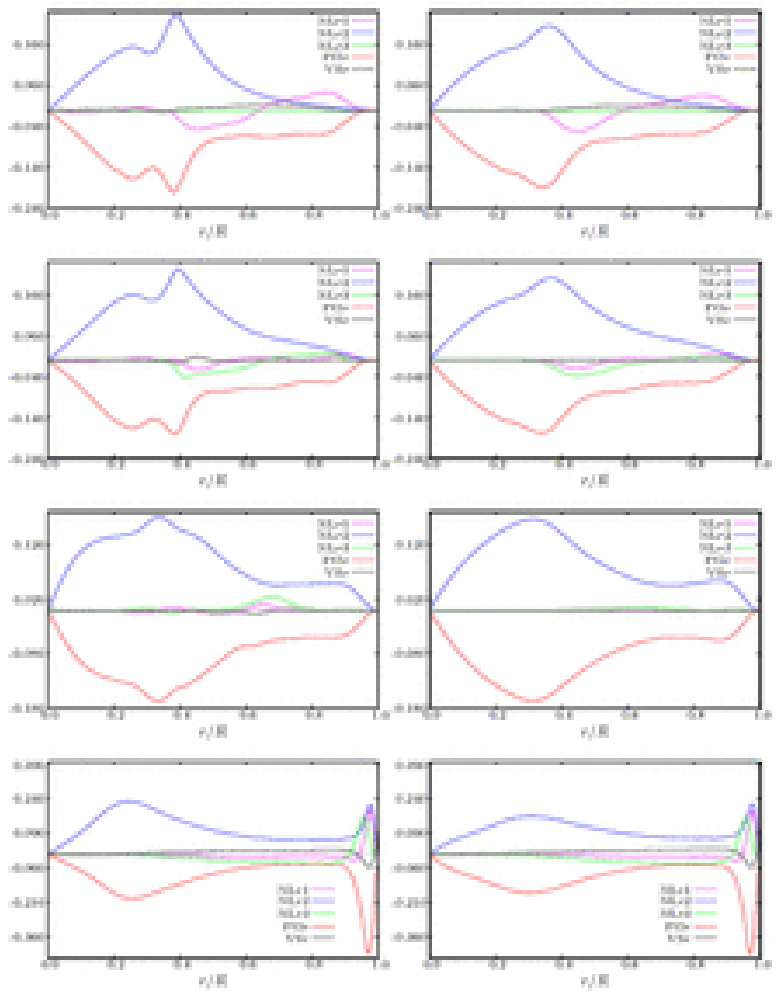}
  \caption{Momentum balance in radial direction plotted along
    horizontal radial lines at four different vertical
    positions. First row: $z/H=1$; second row: $z/H=0.95$; third row:
    $z/H=0.64$; fourth row: $z/H=0.03$. Left column: instantaneous
    flow; Right column: mean flow. Case $\Lambda=1$ and
    $\Rey=6\,000$. The terminology refers to
    Tab.~\ref{tab:name-terms}.}\label{fig:radial-6000}
\end{figure}

We compare the radial terms for the instantaneous and mean flows
presented in Fig.~\ref{fig:radial-6000}, to their laminar and steady
counterparts in Fig.~\ref{fig:radial-900-1500}. In terms of magnitude,
the leading terms have a slightly higher magnitude at
$\Rey=6\,000$. The general observations given in
Sec.~\ref{sec:momentum-steady} remain valid here for $z/H=0.64,\
0.95,\ 1$. However, the influence of the intense recirculation bubble
modifies locally and significantly the terms in the region $0.2 \leq
r/R\leq 0.4$, for the instantaneous flow. Indeed, in this interval
around the radial position $r/R=0.3$ and close to the free surface,
the centrifugal acceleration NL$r2$ presents a local minimum, while
the decelerating radial pressure gradient presents a local
maximum. These localized effects are directly related to the presence
of the recirculation bubble as can be seen in
Fig.~\ref{fig:orthogonal-meridian}. A similar observation can be done
for the mean flow but the effect is much less visible.

Close to the rotating disk, at $z/H=0.03$, the variations of the five
axial terms are notably different from their laminar counterparts, but
extremely similar for the instantaneous and mean flows. More
precisely, the viscous term VS$r$ has mainly a dissipative action in
the laminar regime, while it is slightly driving the flow at
$\Rey=6\,000$, except very close to the tubular side-wall where it
gets back its dissipative action in the jet-like shear
layer. Moreover, the two leading terms, namely the centrifugal
acceleration NL$r2$ and the radial pressure gradient, both presents a
local maximum in the region $0.2\leq r/R \leq 0.3$, for the case
$(c)$, while it keeps growing in the laminar regime. This particular
observation is again related to the presence of the recirculation
bubbles in this region, which locally strongly modifies the momentum
transfers. Finally, the third radial convective term NL$r3$, which is
very small in the laminar regime, acquires a magnitude as important as
the two other convective terms for $r/R\geq 0.9$.

\begin{figure}[htbp]
  \centering
  \includegraphics[width=0.6\textwidth]{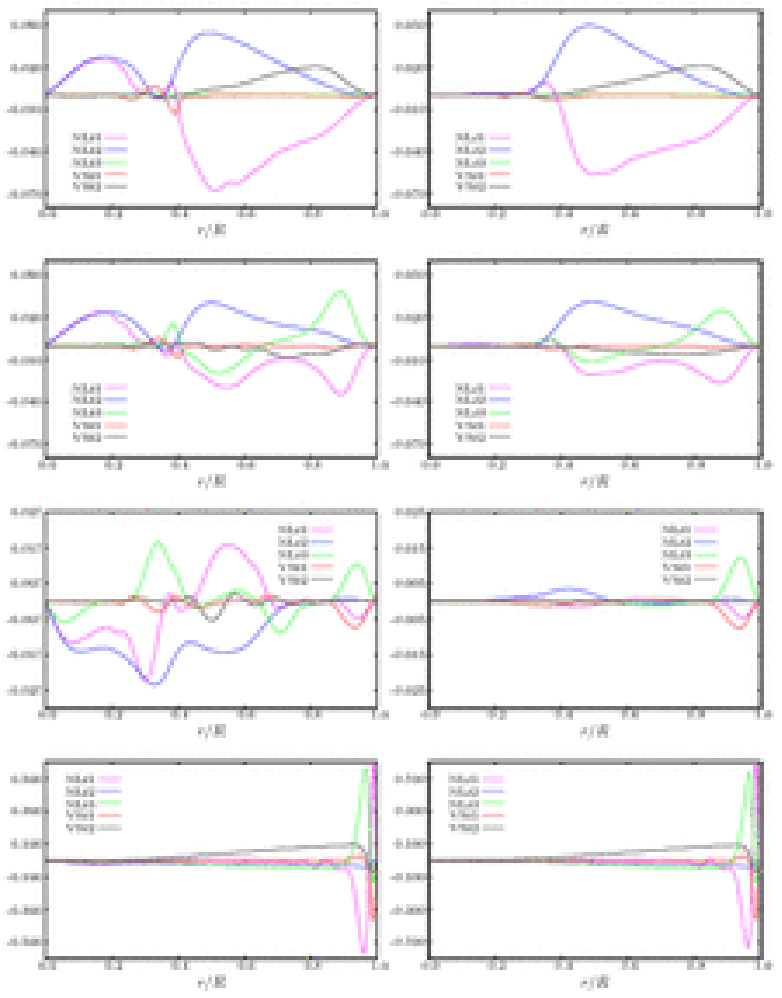}
  \caption{Momentum balance in azimuthal direction plotted along
    horizontal radial lines at four different vertical
    positions. First row: $z/H=1$; second row: $z/H=0.95$; third row:
    $z/H=0.64$; fourth row: $z/H=0.03$. Left column: instantaneous
    flow; Right column: mean flow. Case $\Lambda=1$ and
    $\Rey=6\,000$. The terminology refers to
    Tab.~\ref{tab:name-terms}.}\label{fig:azimuthal-6000}
\end{figure}

Let us consider now the radial variations of the five azimuthal terms
as shown in Fig.~\ref{fig:azimuthal-6000}. A rapid overlook of all
variations for the instantaneous flow sample (left column) allows to
conclude to a general agreement with the results obtained in the
laminar cases $(a)$ and $(b)$. The variations of the different terms
are similar for the instantaneous and mean flows in the outer radial
region, which implies again a relative steadiness of those terms for
$r/R\geq 0.7$. On the other hand, the mean flow yields vanishingly
small terms in the inner core region $r/R\leq 0.4$, where the
instantaneous flow have the three nonlinear convective terms
NL$t1$--NL$t3$ with a relatively high magnitude. The unsteady activity
of those three nonlinear terms and the intense fluctuating activity
generated by them is further discussed in Sec.~\ref{sec:modes}.

\begin{figure}[htbp]
  \centering
  \includegraphics[width=0.6\textwidth]{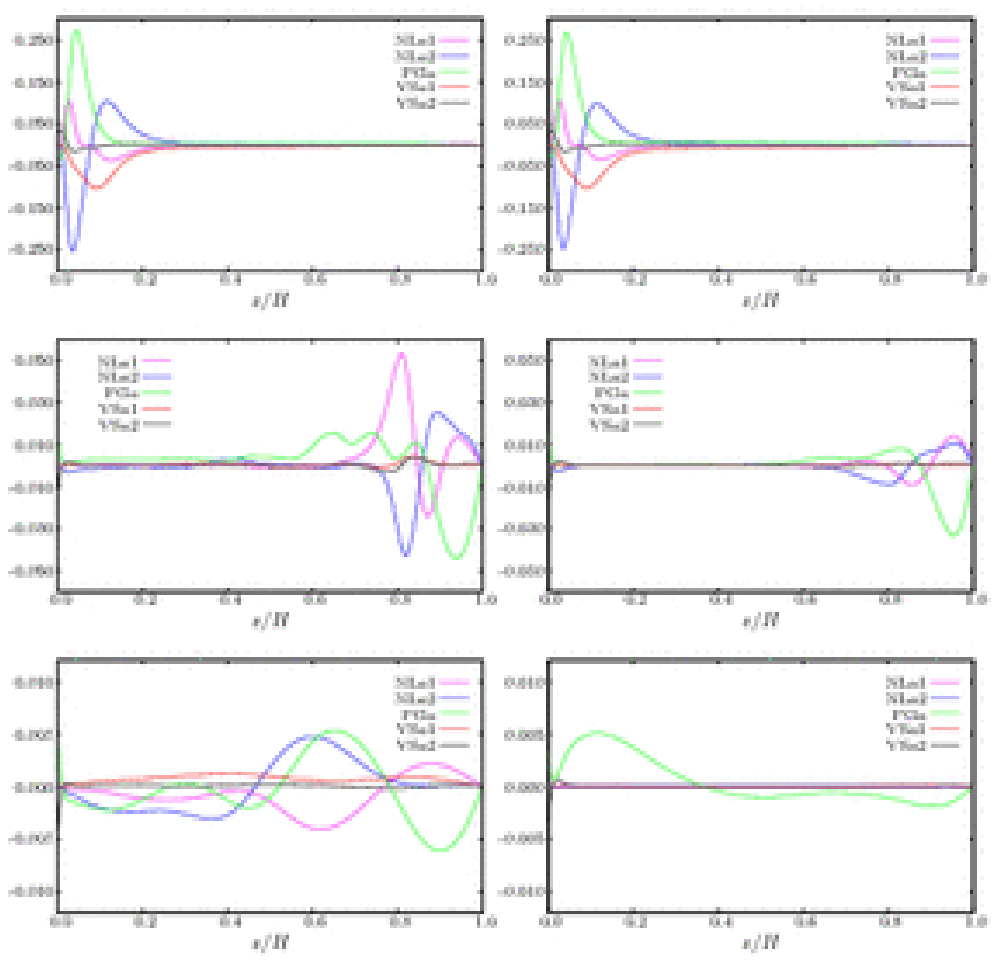}
  \caption{Momentum balance in axial direction plotted along vertical
    lines at three different radial positions. First row: $r/R=0.98$;
    second row: $r/R=0.48$; third row: $r/R=0.08$. Left column:
    instantaneous flow; Right column: mean flow. Case $\Lambda=1$ and
    $\Rey=6\,000$. The terminology refers to
    Tab.~\ref{tab:name-terms}.}\label{fig:axial-6000}
\end{figure}

As a last step, we compare the axial terms for the instantaneous and
mean flows presented in Fig.~\ref{fig:axial-6000} to their laminar and
steady counterparts in Fig.~\ref{fig:axial-900-1500}. We start from
the outer radial line $r/R=0.98$, where the results for the
instantaneous flow and the mean flow are extremely close, revealing an
almost steady behavior of the jet-like wall shear layer surrounding
the tubular side-wall. The comparison of these results with those of
cases $(a)$ and $(b)$ leads to several comments. First, the variations
of all the terms are limited to a smaller zone above the disk at
$\Rey=6\,000$. The magnitude of the axial pressure gradient is
increased with \Rey. More surprisingly the axial pressure gradient is
no longer counterbalanced by the convective term NL$a1$, but is now
counterbalanced by NL$a2$. By extension, one can infer that the
jet-like shear layer is dominated by axial effects at high Reynolds
number.

\begin{landscape}
\begin{figure}[htbp]
  \centering
  \includegraphics[width=0.34\textwidth]{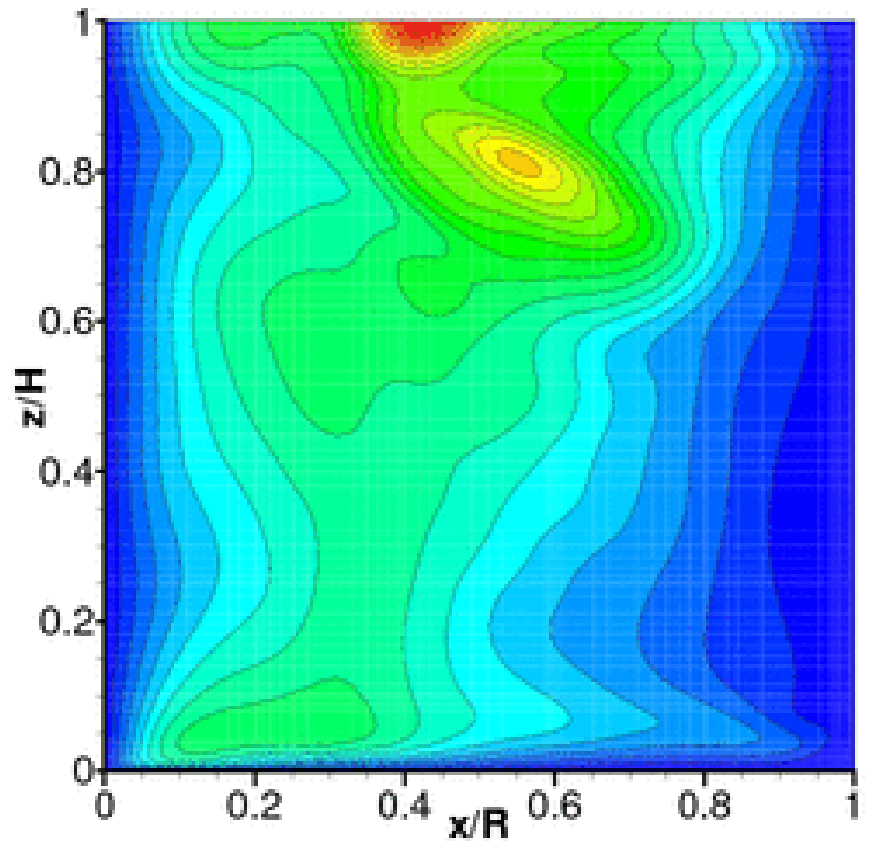}
  \includegraphics[width=0.34\textwidth]{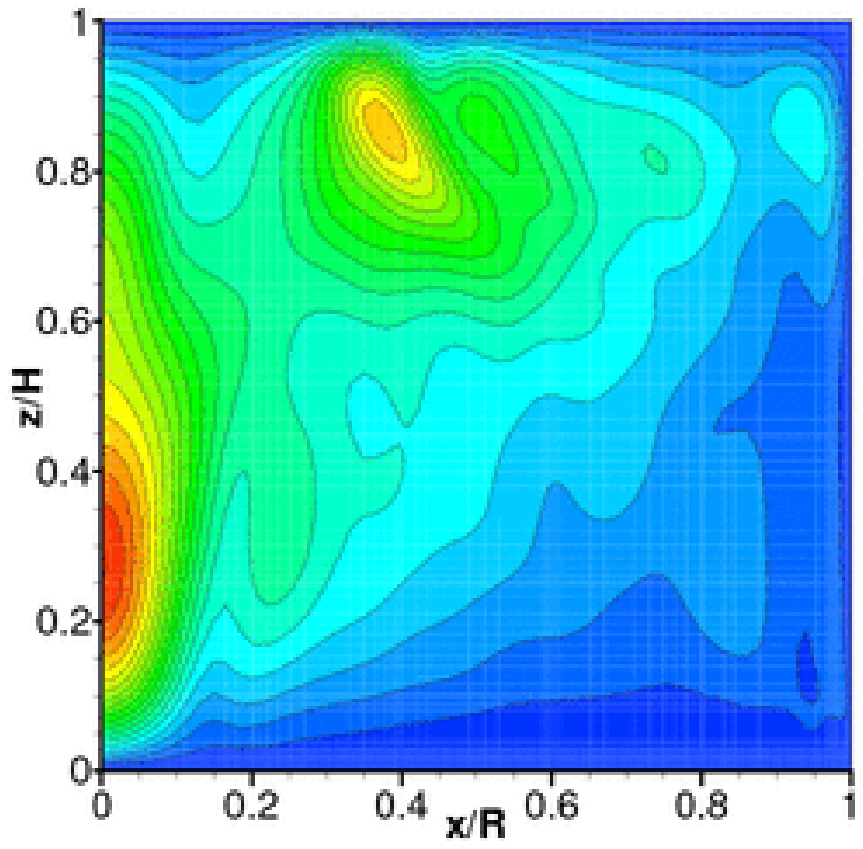}
  \includegraphics[width=0.34\textwidth]{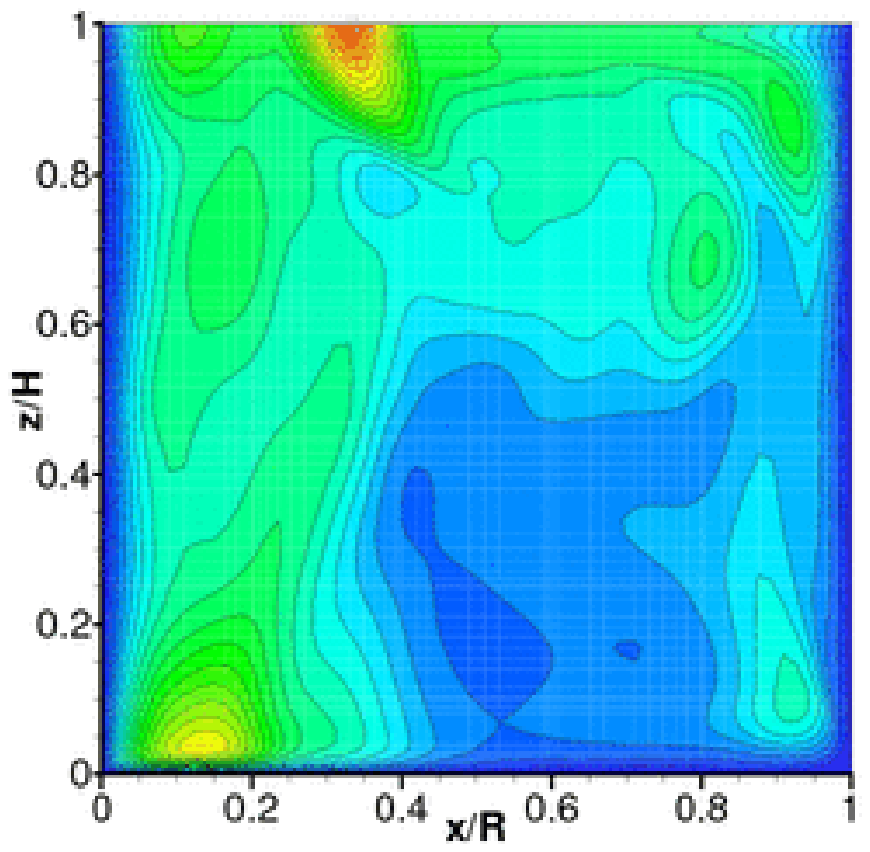}
  \includegraphics[width=0.34\textwidth]{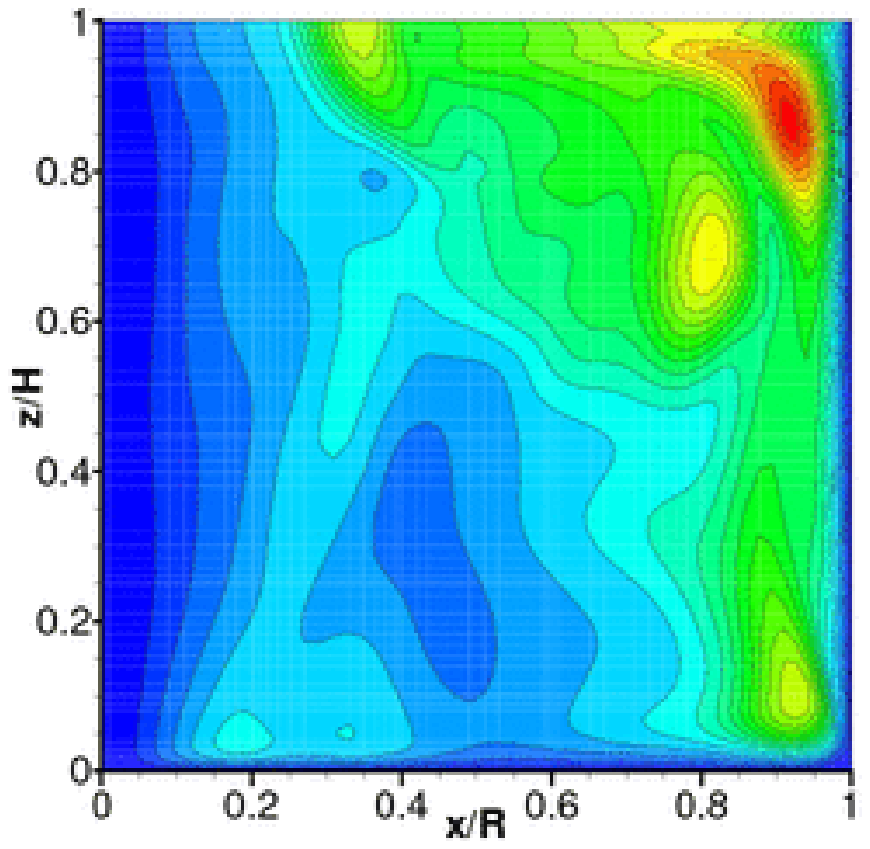}\\
  \includegraphics[width=0.34\textwidth]{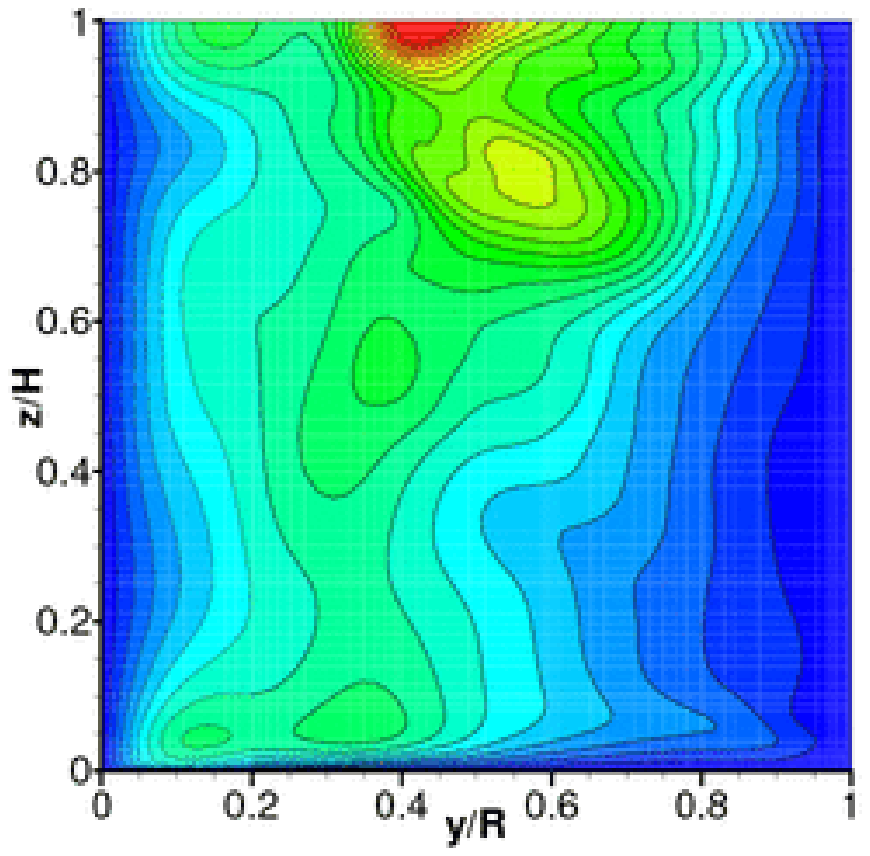}
  \includegraphics[width=0.34\textwidth]{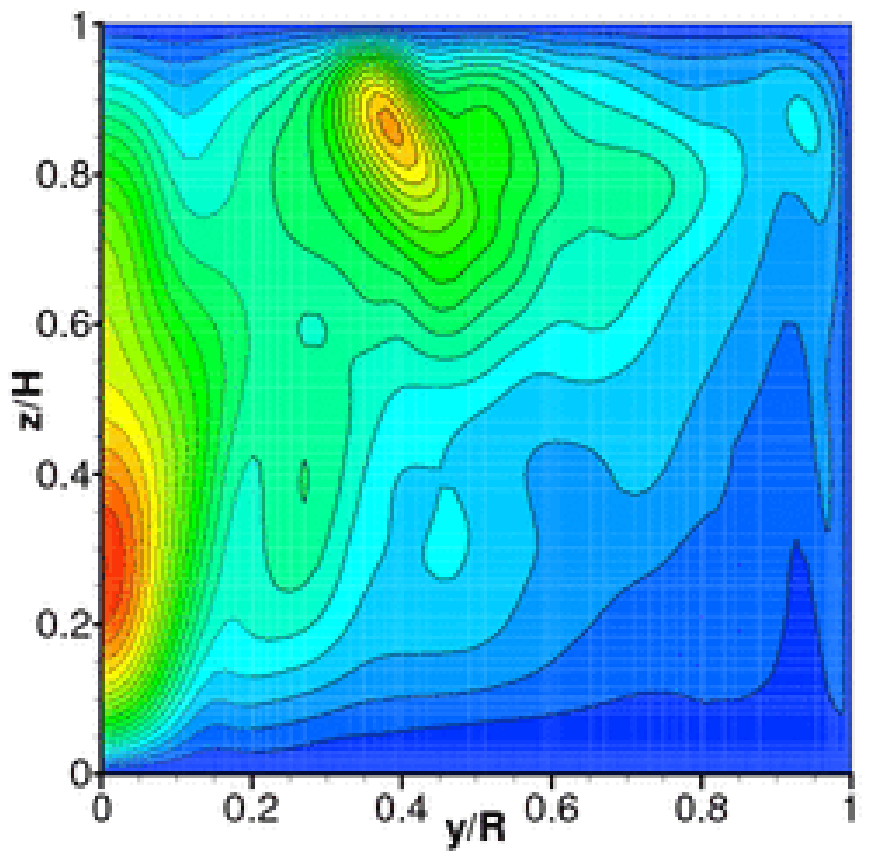}
  \includegraphics[width=0.34\textwidth]{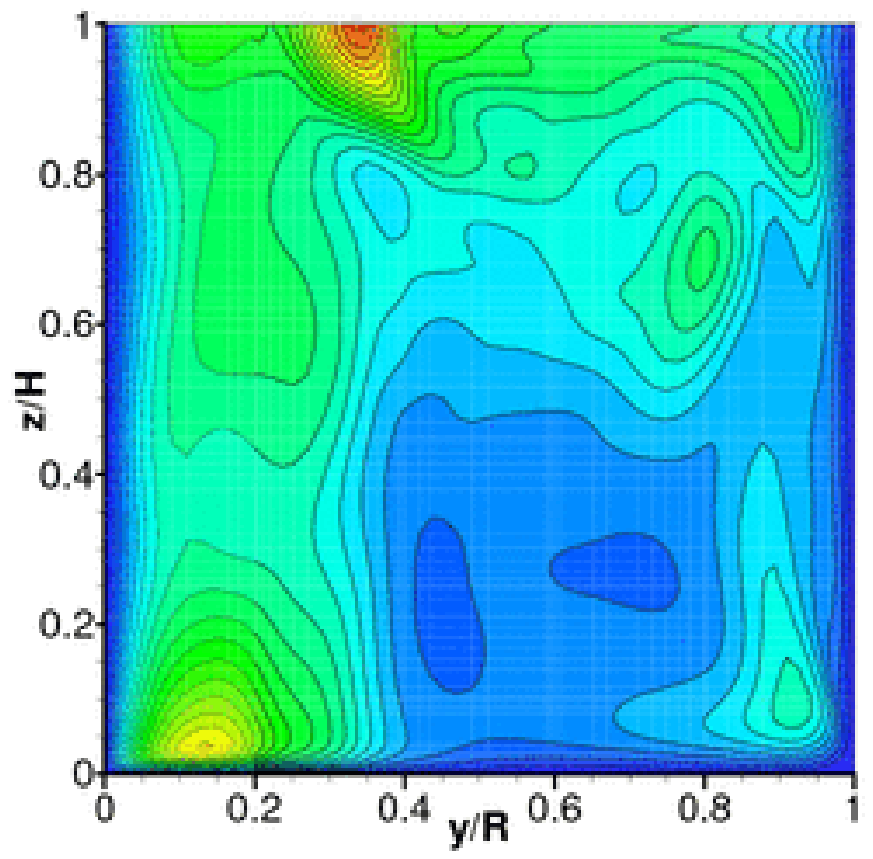}
  \includegraphics[width=0.34\textwidth]{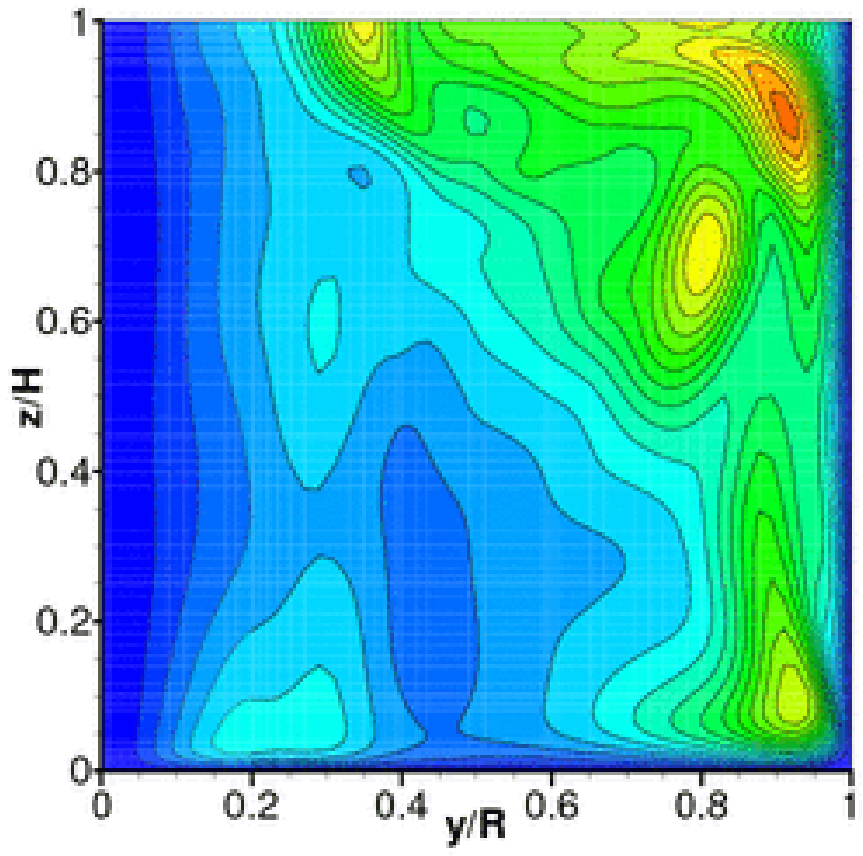}\\
  \caption{Contours in a meridional plane for the case $\Lambda=1$ and
    $\Rey = 6\,000$. Top row: in the meridian plane $y/R=0$; Bottom
    row: in the meridian plane $x/R=0$. From left column to right
    column: rms fluctuations of radial velocity component $u_r$; rms
    fluctuations of the axial velocity component $u_z$; rms
    fluctuations of the azimuthal velocity component $u_\theta$; and
    rms fluctuations of the axial angular momentum
    $\Gamma=ru_\theta$. The 20 contours are uniformly spaced, between
    $0$ and $0.002$ for rms-$u_r$; between $0$ and $0.0024$ for
    rms-$u_z$; between $0$ and $0.004$ for rms-$u_\theta$; and between
    $0$ and $0.001$ for
    rms-$\Gamma$.}\label{fig:rms-ur-uz-ut-gamma-6000}
\end{figure}
\end{landscape}

For the two other radii $r/R=0.08,\ 0.48$, the instantaneous flow
terms are far different from their mean counterparts. As consequence,
the flow in the inner core region appears much more unsteady than the
flow in the outer region of the cavity. In addition, given the high
level of fluctuation in the inner core region of the flow---see
Fig.~\ref{fig:rms-ur-uz-ut-gamma-6000}---it appears irrelevant to
further analyze the results for the mean flow. On the other hand,
variations of the different axial terms for the instantaneous flow
reveals that the axial momentum transfers are more important at high
\Rey\ and are predominant in the top half of the cavity, including
below the free surface. Also, not shown here, the viscous terms are
still insignificant and the two axial convective terms NL$a1$ and
NL$a2$, and the axial pressure gradient dominate the transfers with
other unsteady and nonaxisymmetric terms.

\subsection{Nonaxisymmetric modes in the unsteady transitional flow}\label{sec:modes}
All the previous results dealing with the transitional case $(c)$
reveals a complex flow dynamics due to instabilities developing from a
steady stable flow similar to the steady laminar cases $(a)$ and
$(b)$. The objective of this section is to discuss the physical origin
of this instability based on the results presented earlier and also to
characterize, at least qualitatively, its effect on the flow field.

\begin{landscape}
  \begin{figure}[htbp]
    \centering
    \includegraphics[width=0.4\textwidth]{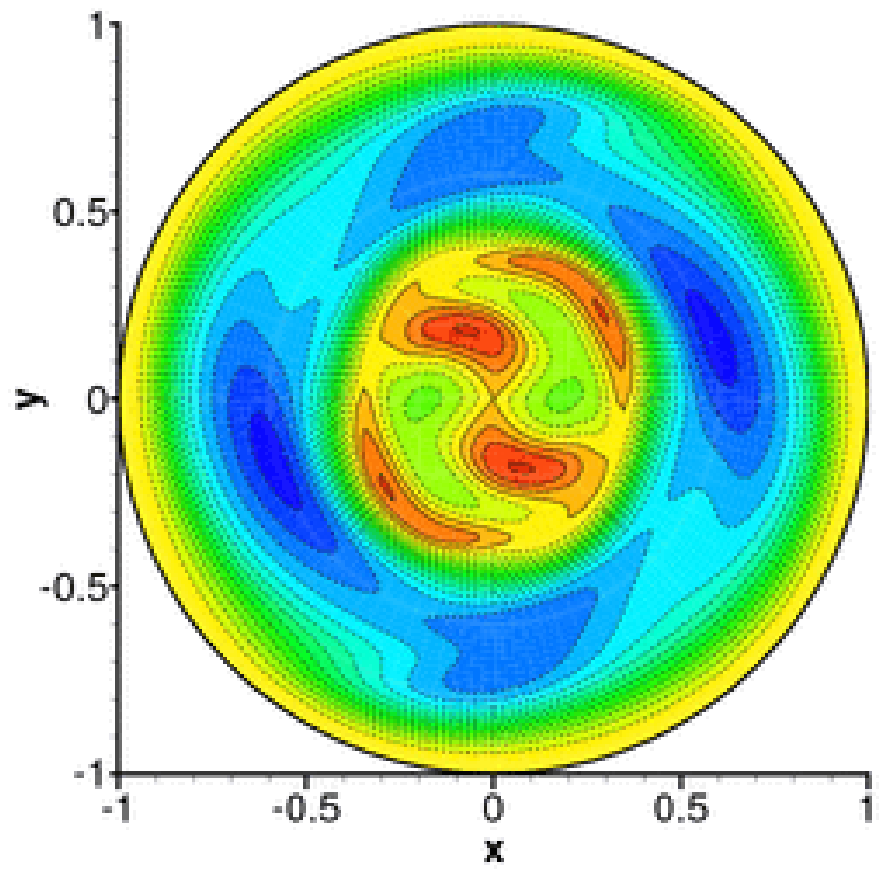}
    \includegraphics[width=0.4\textwidth]{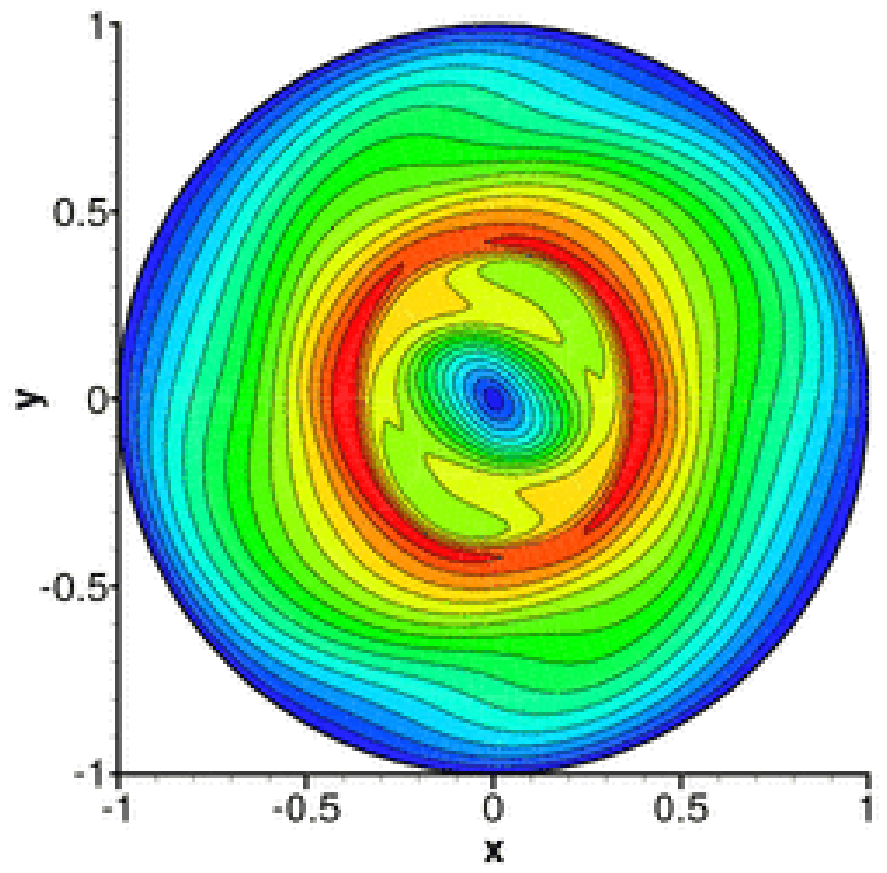}
    \includegraphics[width=0.4\textwidth]{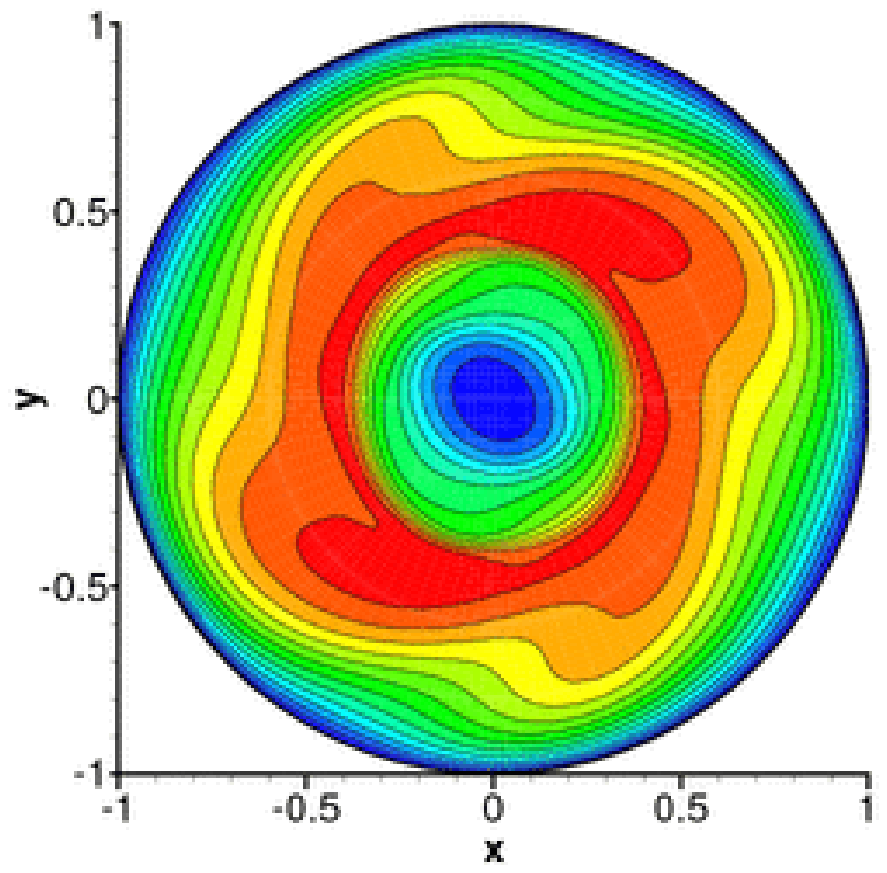}\\
    \includegraphics[width=0.4\textwidth]{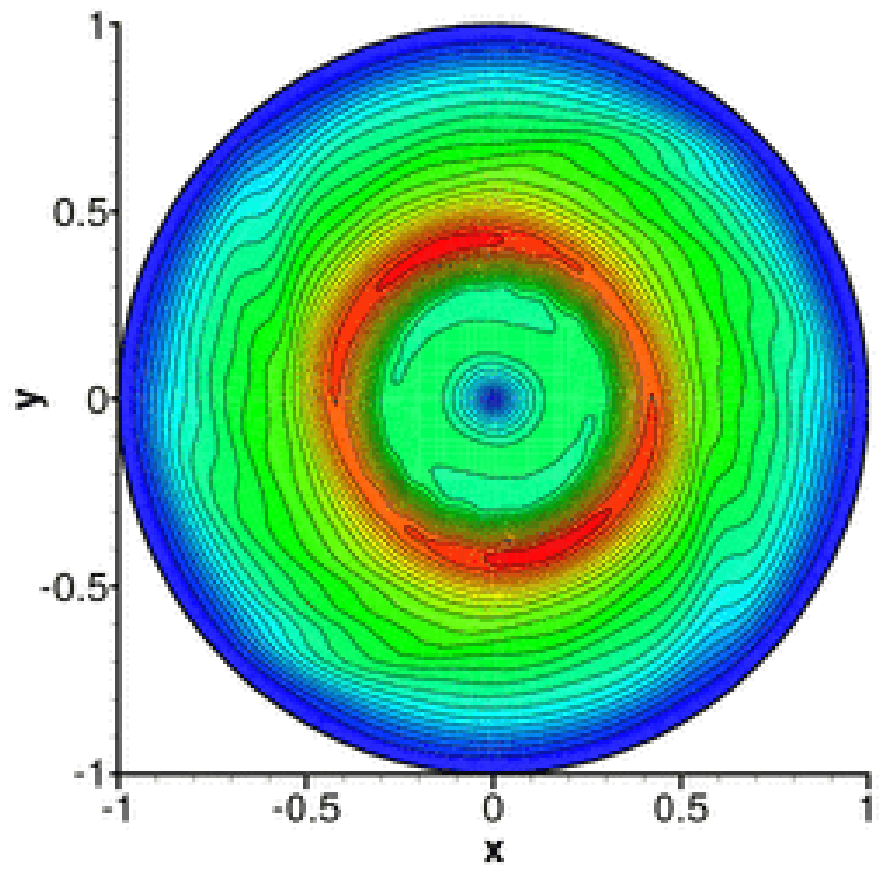}
    \includegraphics[width=0.4\textwidth]{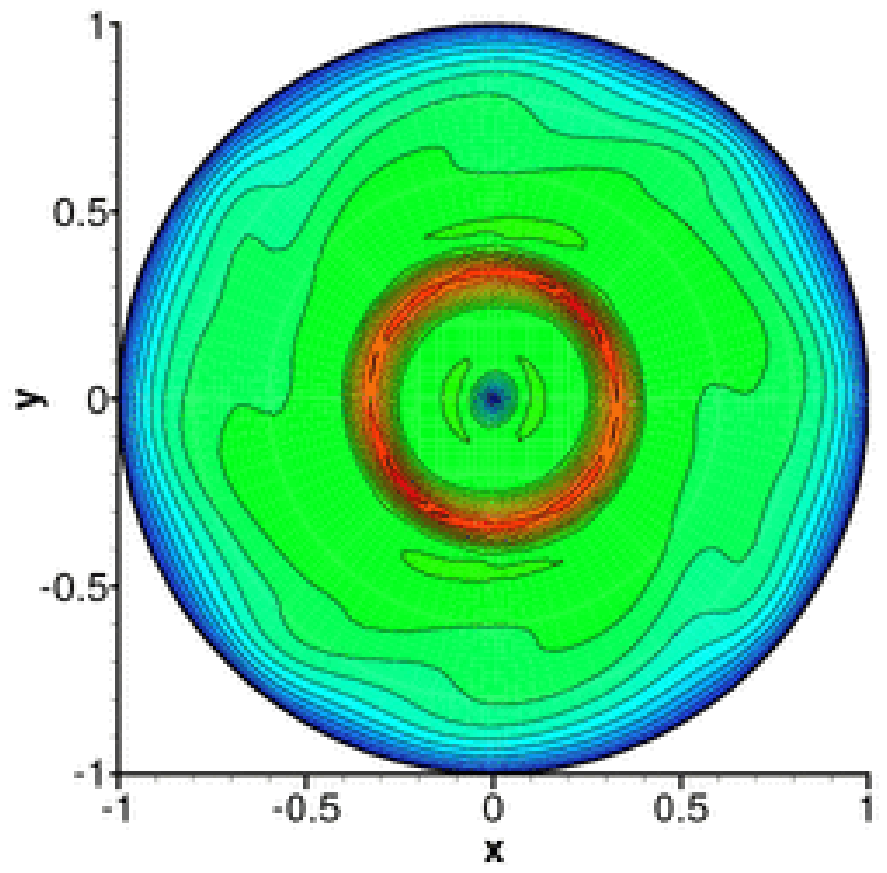}
    \includegraphics[width=0.4\textwidth]{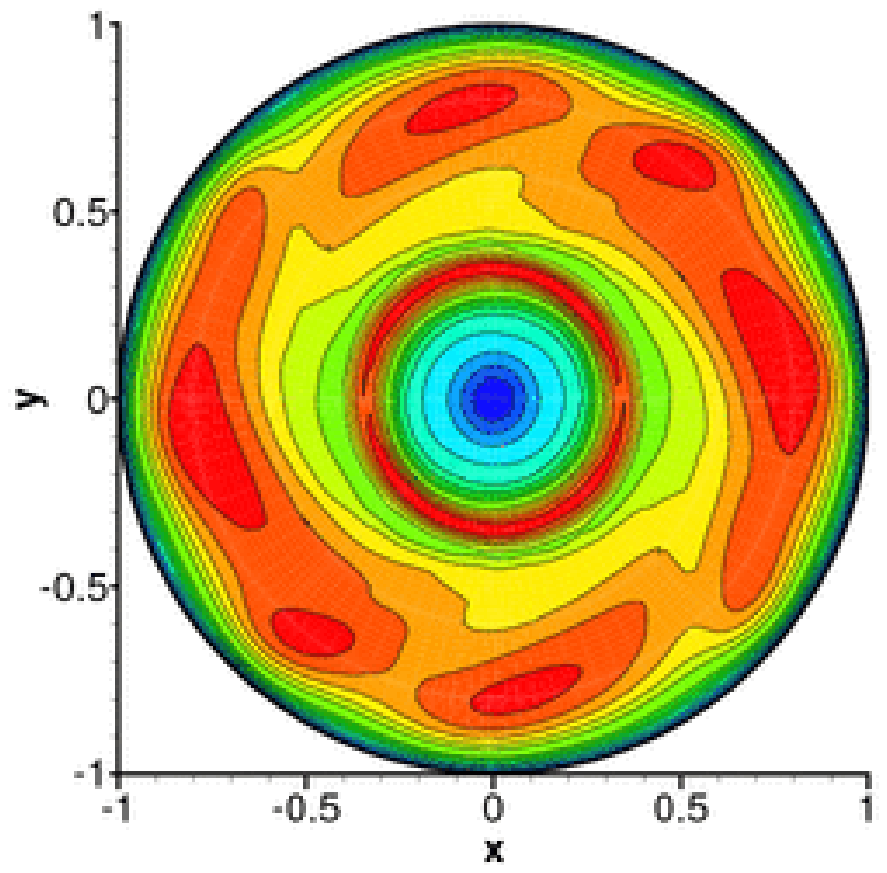}
    \caption{Contours on the free surface $z/H=1$ for the case
      $\Lambda=1$ and $\Rey = 6\,000$. Top row: instantaneous flow;
      Bottom row: rms fluctuations. Left column: radial velocity
      component $u_r$; Central column: azimuthal velocity component
      $u_\theta$; Right column: axial component of the angular
      momentum $\Gamma= r u_\theta$. The 15 contours are uniformly
      spaced, between $-0.13$ and $0.03$ for $u_r$; between $0$ and
      $0.35$ for $u_\theta$; and between $0$ and $0.2$ for
      $\Gamma$. The 20 contours are uniformly spaced, between $0$ and
      $0.002$ for rms-$u_r$; between $0$ and $0.024$ for
      rms-$u_\theta$; and between $0$ and $0.001$ for
      rms-$\Gamma$.}\label{fig:free-surface-6000}
  \end{figure}
\end{landscape}

As mentioned on several occasions in Sec.~\ref{sec:physical} and
Sec.~\ref{sec:one}, the inner core region of the flow $r/R\leq 0.2$ is
globally governed by a quasi-solid-body rotation and in the outer
radial region, a wall-jet along the tubular side-wall drives the
flow. As already discussed, this latter axial wall-jet is turned into
a radial surface jet by the presence of the free surface. It seems
therefore legitimate to consider the interfacial zone---denoted
$\Upsilon$ in the sequel---between the inner core region and the
radially-inward coming jet as prone to developing instabilities. Given
the stress-free condition imposed on the free surface, the effect of
the instabilities should persist all the way to the free surface
itself. Consequently, we consider the variations at the free surface
$z/H=1$ of the radial and azimuthal velocity components---the axial
component vanishes at the free surface because of the kinematic
boundary condition on it---and of the axial angular momentum
$\Gamma=ru_\theta$. These variations for the instantaneous flow are
shown in Fig.~\ref{fig:free-surface-6000} (top row), while the
corresponding rms fluctuations are presented in the row below.

The most significative feature highlighted in these graphs is the
presence of an annular region with $0.3\leq r/R \leq 0.4$, where the
fluctuations of $u_r$, $u_\theta$ and $\Gamma$ are intense. Such
intense fluctuating activity brings the interfacial zone $\Upsilon$ to
light. Focusing now on the rms fluctuations of the axial angular
momentum, one may notice the presence of a second outer annular region
comprised in the interval $0.6 \leq r/R \leq 0.9$, which is
nonaxisymmetric and relatively intense. This outer region of intense
rms fluctuations for $\Gamma$ correspond to the zone where the
jet-like shear layer is turned into the interior by the free
surface. This outer wall jet injects high-angular-momentum fluid
towards the cylinder axis. This radial jet impinges on the cylindrical
core of the fluid that is in solid-body rotation. When increasing the
Reynolds number, the radius of the cylindrical core in solid-body
rotation is reduced, while the intensity of the radial jet is
increased. Above a given value of the Reynolds number, the action of
the impingement of the radial jet on the inner core region starts
developing unstable modes.

\begin{figure}[htbp]
  \centering
  \includegraphics[width=0.4\textwidth]{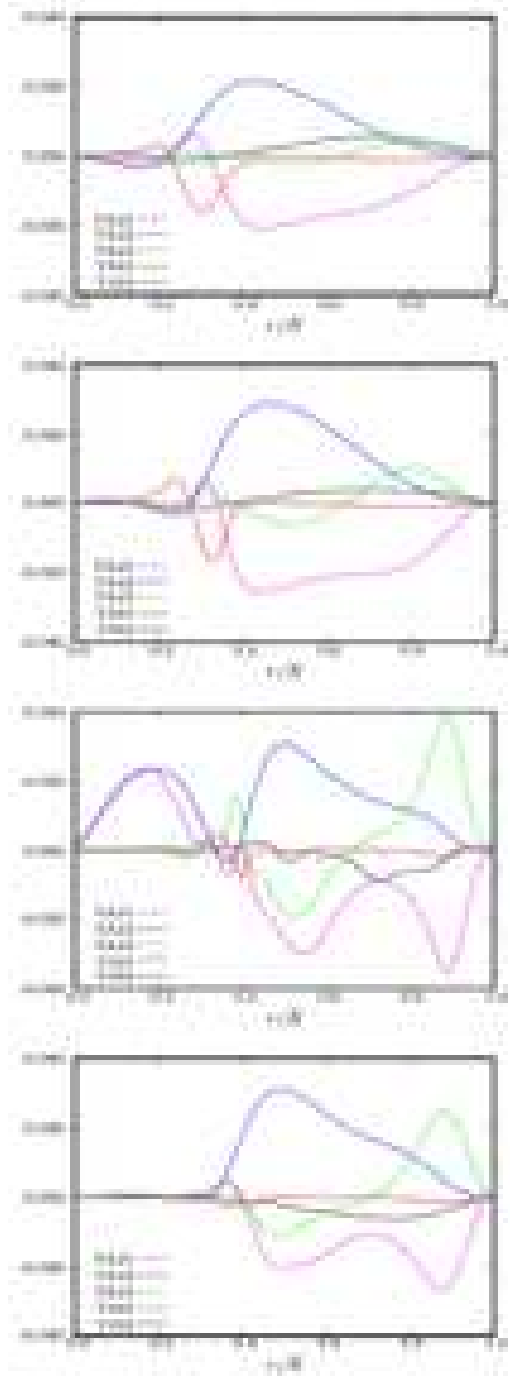}
  \caption{Momentum terms in azimuthal direction plotted along the
    horizontal radial line at $z/H=0.95$. From top to bottom: case
    $(a)$; case $(b)$; case $(c)$ instantaneous; case $(c)$ mean
    flow. The terminology refers to
    Tab.~\ref{tab:name-terms}.}\label{fig:azimuthal-instable}
\end{figure}

The origin of these unstable modes is to be found in the analysis of
the momentum transfers performed in the previous section. Returning on
the radial variations of the five azimuthal terms below the free
surface at $z/H=0.95$, Fig.~\ref{fig:azimuthal-instable} reproduces
these variations for $\Rey=900,\ 1\,500,\ 6\,000$ (instantaneous
flow), $6\,000$ (mean flow) from top to bottom. All these results have
been shown separately before and are now shown together to facilitate
the discussion. The flow in the outer region $r/R>0.5$ has a strong
steady character given the fact that the variations for the mean flow
are fairly close to those of the instantaneous flow. The Coriolis term
NL$t2=-u_ru_\theta /r$ keeps the same radially-outward decreasing
trend. On the other hand, the two other nonlinear terms NL$t1$ and
NL$t3$ develop opposed and equally-intense peaks around
$r/R=0.9$. These opposite peaks have an increasing intensity with the
Reynolds number. In the interfacial zone $\Upsilon$, all the terms
involved present brutal variations and changes of behavior, which give
another characterization of this interfacial zone $\Upsilon$. In the
inner core region $r/R\leq 0.4$, the flow possesses a strong unsteady
character brought to light by the vanishingly small values of the
various terms for the mean flow. Consequently, the unstable azimuthal
modes are to be found into this inner cylindrical region. Indeed, one
may notice that the convective term NL$t1=-u_r\partial u_\theta
/\partial r$ and the Coriolis term NL$t2=-u_r u_\theta /r$ are
negative and have a low magnitude in the laminar cases $(a)$ and
$(b)$, but acquires large positive values at $\Rey =6\,000$. Moreover,
all the three other terms NL$t3$, VS$t1$ and VS$t2$ have very low
magnitudes, and thus cannot counterbalance the azimuthal momentum
injected by NL$t1$ and the Coriolis term NL$t2$. Only an unsteady and
nonaxisymmetric flow can support such azimuthal momentum effects. The
effect of these two destabilizing terms on the flow apparently leads
to the formation of azimuthal rotating waves superimposed to the
stable base flow. The variations at the free surface $z/H=1$, of the
radial and azimuthal velocity components, and of the axial angular
momentum shown in Fig.~\ref{fig:free-surface-6000} (top row), suggest
the conjugate effect of several rotating waves. These rotating waves
correspond to even azimuthal Fourier modes, mainly $n=2$ and $n=4$.

\section{Conclusions} \label{sec:conclusions}
The incompressible flow of a viscous fluid enclosed in a cylindrical
container with an open top flat surface and driven by the constant
rotation of the bottom wall has been thoroughly investigated. The top
surface of the cylindrical cavity is left open with a stress-free
boundary condition imposed on it. No-slip condition is imposed on the
side-wall and also on the rotating bottom end-wall by means of a
regularized angular velocity profile. More specifically, the
stress-free top surface is maintained fixed and flat.

New flow states have been investigated based on a fully
three-dimensional solution of the Navier--Stokes equations for the
free-surface cylindrical swirling flow, without resorting to any
symmetry property unlike all other results available in the
literature. To our knowledge, the present study delivers the most
general available results for this flat-free-surface problem due to
its original mathematical treatment.

Five different cases corresponding to different pairs of governing
parameters $(\Lambda,\Rey)$ have been considered. The Reynolds regime
corresponds to transitional flows with some incursions in the fully
laminar regime. Both steady and unsteady non-oscillatory swirling
flows are considered with a particular emphasis on the case $(\Lambda
=1 , \Rey =6\, 000)$. Of great concern to this study is the question
of space resolution. This is particularly important for the bifurcated
case at $\Rey= 6\,000$. Convergence tests in space and time have been
carried out on this upmost problematic case, and optimal values of the
polynomial degree and time-step have been deduced.

The evolution of the total kinetic energy of this open flow has been
carefully studied for increasing Reynolds numbers and has been
compared to the results for the closed swirling flow. The presence of
the free surface on the top of the cylinder is found to strongly
modify the observed trend: the total kinetic energy is increased with
\Rey\ in the open cylinder case, while the converse is observed in the
closed cylinder case. A physical analysis of the energetic action of
the surface layer below the free surface allows to justify the above
results. A comprehensive physical description of all flow states has
been given with particular emphasis on the vortex breakdown bubbles
and on the structure of the vortex lines. The unsteady case at $\Rey
=6\,000$ has retained more attention, given its unsteady transitional
character. The mean flow and the corresponding rms fluctuations have
been calculated and the results analyzed accordingly. The momentum
transfers in the radial, azimuthal and axial directions have been
studied along various one-dimensional lines. For the transitional case
at $\Rey=6\,000$, the flow in an inner cylindrical core is in
solid-body rotation, while the outer radial layer is dominated by the
jet-like shear layer along the tubular side-wall. This axial wall-jet
is turned into a radial jet of angular momentum, which prevails all
the way up to the free surface. The impingement of this radial jet
onto the inner cylindrical core in solid-body rotation leads to the
development of unstable azimuthal modes. The nonlinear terms, which
includes a Coriolis effect, responsible for the development of these
unstable modes have been found using the azimuthal momentum imbalance
below the free surface. These unstable modes take the form of
even-order azimuthal rotating waves.

\ack
This research is being partially funded by a Swiss National Science Foundation Grant (No. 200020--101707) whose supports are gratefully acknowledged.

The results were obtained on supercomputing facilities on the Pleiades clusters at EPFL--IGM.


 \end{document}